\documentclass[a4paper,12pt]{article}
\usepackage[utf8]{inputenc}
\usepackage[T1]{fontenc}
\usepackage{epsfig,graphicx,xcolor,amsbsy,amssymb,latexsym,amsfonts,amsmath,tcolorbox,setspace,mathrsfs}
\usepackage{pstricks}
\usepackage{color}
\usepackage{soul}
\usepackage{accents}
\usepackage{placeins}
\usepackage{wrapfig,framed,caption}
\usepackage[makeroom]{cancel}
\usepackage{amsfonts}
\usepackage{dsfont}
\usepackage{mmacells}
\mmaDefineMathReplacement{∞}{\infty}

\usepackage[style=numeric,sorting=none,style=numeric-comp,maxnames=10]{biblatex}

\definecolor{shadecolor}{gray}{0.925}

\usepackage{eurosym}
\usepackage[a4paper]{geometry}
\usepackage{graphicx}
\usepackage{tikz}
\usetikzlibrary{cd}
\usetikzlibrary{decorations.pathreplacing,decorations.markings,snakes}
\usetikzlibrary{decorations.pathmorphing}
\usepackage{xcolor}
\usepackage{amsmath,amssymb,amsfonts,pstricks,setspace}
\usepackage[enableskew]{youngtab}
\usepackage{ytableau}

\newenvironment{psmall}
  {\left(\begin{smallmatrix}}
  {\end{smallmatrix}\right)}

\newenvironment{linesmall}[1]
  {\arraycolsep=3pt\scriptsize
   \array{#1}}
  {\endarray}

\numberwithin{equation}{section}

\newcommand{\bea}{\begin{eqnarray}\displaystyle}
\newcommand{\eea}{\end{eqnarray}}

\newcommand{\Qs}{Q_{S}}

\newcommand{\Qt}{Q_{\tau}}
\newcommand{\Qr}{Q_{\rho}}

\newcommand{\A}{\widehat{a}}

\newcommand{\Qa}[1]{Q_{\widehat a_{#1}}}

\usepackage{float}
\newtcolorbox{summary}[2][]{colbacktitle=blue!10!white, colback=yellow!10!white,coltitle=blue!70!black, title={#2},fonttitle=\bfseries,#1}

\title{
\begin{flushright}{\vspace{-2.5cm}\small LYCEN 2023-03\\}\end{flushright}
\vspace{2.3cm}
{\bf Non-perturbative Symmetries of Little Strings \\ and Affine Quiver Algebras}\\[40pt]}

\author{\large \textsc{Baptiste~Filoche\footnote{\tt b.filoche@ip2i.in2p3.fr}}~~,~~\textsc{Stefan~Hohenegger\footnote{\tt s.hohenegger@ipnl.in2p3.fr}}
~,~\,and\,~\textsc{Taro~Kimura\footnote{\tt taro.kimura@u-bourgogne.fr}}
}

\begin{document}

\maketitle
\thispagestyle{empty}
\begin{center}
\renewcommand{\thefootnote}{\fnsymbol{footnote}}\vspace{-0.5cm}
${}^{\footnotemark[1]\,\footnotemark[2]}$ Univ Lyon, Univ Claude Bernard Lyon 1, CNRS/IN2P3, IP2I Lyon, UMR 5822, F-69622, Villeurbanne, France\\[0.2cm]
${}^{\footnotemark[3]}$ Institut de Math\'ematiques de Bourgogne, Universit\'e de Bourgogne, CNRS, France\\[2.5cm]
\end{center}

\begin{abstract}
We consider Little String Theories (LSTs) that are engineered by $N$ parallel M5-branes probing a transverse $\mathbb{Z}_M$ geometry. By exploiting a dual description in terms of F-theory compactified on a toric Calabi-Yau threefold $X_{N,M}$, we establish numerous symmetries that leave the BPS partition function $\mathcal{Z}_{N,M}$ invariant. They furthemore act in a non-perturbative fashion from the point of view of the low energy quiver gauge theory associated with the LST. We present different group theoretical organisations of these symmetries, thereby generalising the results of \cite{Bastian:2018jlf} to the case of generic $M\geq 1$. We also provide a Mathematica package that allows to represent them in terms of matrices that act linearly on the K\"ahler parameters of $X_{N,M}$. From the perspective of dual realisations of the LSTs the symmetries found here act in highly nontrivial ways: as an example, we consider a formulation of $\mathcal{Z}_{N,M}$ in terms of correlation functions of a vertex operator algebra, whose commutation relations are governed by an affine quiver algebra. We show the impact of the symmetry transformations on the latter and discuss invariance of $\mathcal{Z}_{N,M}$ from this perspective for concrete examples.
\end{abstract}

\newpage

\tableofcontents

\section{Introduction}
Symmetries and dualities are among our most powerful tools to study and analyse quantum theories. Of particular interest are those that act non-perturbatively, \emph{i.e.} which mix the coupling constant(s) with the remaining parameters of the theory. Indeed, these provide windows into the strong-coupling regime, which is notoriously difficult to study with other methods (notably perturbative ones). A particularly rich class of examples where such symmetries and dualities can be established (and used for various applications) are field theories that are derived as limits of string theory (or M- and F-theory). The large network of dualities of the latter can be exploited to find different (but equivalent) descriptions of the same field theory that are adapted to different parameter regions. In this way, for example, a large spectrum of supersymmetric gauge theories in different dimensions and with different gauge- and matter content (as well as different amount of supersymmetry) can be covered.

In this paper we discuss a particularly rich class of such theories, namely (orbifolds of) Little String Theories (LSTs) of A-type \cite{Hohenegger:2015btj,Hohenegger:2016eqy,Bastian:2017ing,Bastian:2018dfu}: these theories are special types of quantum theories \cite{Witten:1995zh,Aspinwall:1997ye,Seiberg:1997zk,Intriligator:1997dh,Hanany:1997gh,Brunner:1997gf} (see also \cite{Aharony:1999ks,Kutasov:2001uf} for review-articles) in the sense that they describe point-particle theories at low energies but their UV-completion requires extended (\emph{i.e.} string like) degrees of freedom. Concretely, at low energies they describe supersymmetric circular quiver gauge theories on $\mathbb{R}^4\times T^2$ with $M\in\mathbb{N}$ gauge nodes of type $U(N)$ (with $N\in\mathbb{N}$) and matters in the bifundamental representation (or the adjoint in the case of $M=1$), which is associated with the A-type affine quiver denoted by $\widehat{A}_{M-1}$. See Figure~\ref{Fig:Quiver} down below for a sketch of the quiver. Such theories can be engineered in string- or M-theory in various different ways \cite{Haghighat:2013gba,Haghighat:2013tka,Hohenegger:2013ala,Hohenegger:2015btj,Bastian:2017jje}, which allow to compute important quantities of these theories explicitly (most notably, the full non-perturbative BPS-partition function $\mathcal{Z}_{N,M}$). For example they arise as the world-volume theory of $N$ M5-branes that probe a transverse orbifold of type $A_{M-1}$ \cite{Blum:1997fw,Cecotti:2013mba,Haghighat:2013gba,Haghighat:2013tka,Hohenegger:2013ala}. In the current paper, however, we shall use two further dual descriptions:
\begin{itemize}
\item[\emph{(i)}] {\bf F-theory on a class of toric Calabi-Yau manifolds $X_{N,M}$}\\
As explained in \cite{Haghighat:2013gba,Haghighat:2013tka,Hohenegger:2013ala} (see also \cite{Hohenegger:2015cba,Hohenegger:2015btj,Hohenegger:2016eqy,Bastian:2017ing,Bastian:2017ary,Hohenegger:2016eqy}), the M5-brane world volume theory mentioned above can also be engineered through F-theory compactified on a particular class of Calabi-Yau threefolds $X_{N,M}$ (see \cite{Kanazawa:2016tnt} for the mathematical details). The web diagram of this toric manifold is shown in Figure~\ref{Fig:General Setup} below (and more details are discussed in Section~\ref{SubSect:KahlerParameters}) and is completely determined by (a deformed version of) the M-brane web \cite{Bastian:2017ing,Hohenegger:2016yuv,Bastian:2018fba}. In particular, the K\"ahler parameters of the manifold $X_{N,M}$ are completely determined by the parameters of the M-brane configuration and in turn translate into parameters of the supersymmetric gauge theory (see \cite{Bastian:2017ary} for a convenient choice of basis). Moreover, the non-perturbative BPS-partition function $\mathcal{Z}_{N,M}$ is captured by the topological string partition function of $X_{N,M}$ \cite{Haghighat:2013gba,Haghighat:2013tka,Hohenegger:2013ala}, which can be computed in an algorithmic fashion \cite{Bastian:2017ing} using the (refined) topological vertex \cite{Aganagic:2003db,Iqbal:2007ii,Iqbal:2004ne} (see \cite{Antoniadis:2010iq,Antoniadis:2013bja,Antoniadis:2013mna,Antoniadis:2015spa} for a world-sheet description of the topological string).
\item[\emph{(ii)}] {\bf Affine quiver algebra}\\
Following \cite{Kimura:2015rgi,Kimura:2016dys,Kimura:2017hez} (see \cite{Kimura:2020jxl} for a review), a vertex operator algebra (called W-algebra) can be constructed from the equivariant K-theory on the instanton moduli space of the gauge theory associated with the quiver described above. The vertex operators (called screening charges) are constructed from free field modes whose fundamental commutation relations are governed by a Cartan matrix, which encodes the affine quiver structure and which is deformed by the $\Omega$-background parameters and the bifundamental mass parameters of the gauge theory. In this way, the partition function $\mathcal{Z}_{N,M}$ can be written as a combination of correlation functions of these screening charges, inserted at positions of an elliptic curve, which can be identified with D5-brane positions in a dual string theory picture~\cite{Kimura:2022zsm,Kimura:2023bxy}.
\end{itemize}
The K\"ahler moduli space $\mathbb{K}(X_{N,M})$ of the manifold $X_{N,M}$ used in the first description \emph{(i)}, contains various other regions that describe low energy supersymmetric gauge theories of the form of an affine quiver \cite{Bastian:2017ary}. Moreover, $\mathbb{K}(X_{N,M})$ can be embedded in a larger space (called the extended K\"ahler moduli space $\mathbb{EK}(X_{N,M})$ of $X_{N,M}$) which contains even more such regions, each of which engineering dual supersymmetric gauge theories (\emph{i.e.} all of which sharing the same non-perturbative BPS-partition function $\mathcal{Z}_{N,M}$). Using flop transitions to relate points in $\mathbb{EK}(X_{N,M})$, it was in fact argued in \cite{Bastian:2018dfu} that there exists a web of dual gauge theories described at low energies by quiver gauge theories with $M'$ nodes of the type $U(N')$ for all $(M',N')$ such that $MN=M'N'$ and $\text{gcd}(M,N)=\text{gcd}(M',N')$.\footnote{In the following we denote such a quiver gauge theory with $M$ gauge nodes simply as an $[U(N)]^M$ theory.} For the original $[U(N)]^M$ theory, this web of dualities implies a number of non-perturbative symmetries, which act as linear transformations mixing the coupling constants and Coulomb branch parameters. For $M=1$ (but arbitrary $N$), these symmetries were systematically studied in \cite{Bastian:2018jlf} and shown to form notably a particular dihedral symmetry group. The latter in turn was used \cite{Bastian:2019hpx,Bastian:2019wpx,Hohenegger:2019tii,Hohenegger:2020gio,Hohenegger:2020slq,Filoche:2022qxk} to further study the structure of the BPS-partition function $\mathcal{Z}_{N,M=1}$.

The purpose of the current paper is two-fold: first, we generalise the discussion of \cite{Bastian:2018jlf} to the case $M\geq 1$. We analyse systematically the non-perturbative symmetries that are induced by dualities of the manifold $X_{N,M}$ and write them as linear transformations that act on the ($NM+2$ independent) K\"ahler parameters of $X_{N,M}$. We highlight important algebraic properties of these symmetry transformations and provide the Mathematica package {\tt NPLSTsym}, which expresses them in the form of $(NM+2)\times (NM+2)$ matrices. We also discuss the structure of these symmetry transformations from a group theoretical perspective and provide the generalisation of dihedral group found in \cite{Bastian:2018jlf}. In a second step, we discuss the implications of these non-perturbative symmetries for the affine quiver algebra mentioned in the second dual description \emph{(ii)}. Indeed, since the Cartan matrix that is central to the construction of the (correlation functions of the) screening currents is deformed by the mass parameter of the gauge theory, which can be identified with a specific K\"ahler parameter of $X_{N,M}$, the above mentioned symmetries act in a non-trivial manner on the affine quiver algebra. Concretely, we describe explicitly the modification of the commutator of the screening currents. While for some specific examples we can provide the explicit mechanism how the partition function $\mathcal{Z}_{N,M}$ remains invariant despite this non-trivial morphism of the affine quiver algebra, most other symmetries act in a highly non-trivial fashion that is difficult to predict otherwise.

This paper is organised as follows: Section~\ref{Sect:OverviewPartitionFunction} provides a review of Little String Theories of A-type and the computation of the partition function $\mathcal{Z}_{N,M}$ from the toric manifold $X_{N,M}$. In Section~\ref{Sect:ReviewSyms} we use duality transformation of $X_{N,M}$ (\emph{i.e.} transformations relating different points in $\mathbb{EK}(X_{N,M})$), to establish symmetry transformations that leave the partition function $\mathcal{Z}_{N,M}$ invariant. Section~\ref{Sec:ZLSTVOA} reviews the affine quiver algebra and provides a re-writing of $\mathcal{Z}_{N,M}$. In Section~\ref{Sect:ActionSymmetries} we use this reformulation to establish the action of the symmetries found in Section~\ref{Sect:ReviewSyms} on the affine quiver algebra. Finally, Section~\ref{Sect:Conclusions} contains our conclusions and an outlook for further applications of our results. This paper is supplemented by two Appendices: the first is dedicated to a review of important mathematical functions and notations, while the second one gives a description of the Mathematica package {\tt NPLSTsym}, which automatically computes the transformation matrices for the symmetries established in this work.

\section{Little String Theories and Web Diagrams}\label{Sect:OverviewPartitionFunction}
\subsection{Web Diagrams and K\"ahler Parameters}\label{SubSect:KahlerParameters}
\begin{figure}[ht!]
\centering
\scalebox{0.9}{\parbox{13cm}{\begin{tikzpicture}
\draw[ultra thick] (1.75,1.75) -- (2.25,2.25);
\draw[ultra thick] (2.75,0.75) -- (3.25,1.25);
\draw[ultra thick] (3.75,-0.25) -- (4.25,0.25);
\draw[ultra thick] (0.75,1.75) -- (1.75,1.75) -- (1.75,0.75) -- (2.75,0.75) -- (2.75,-0.25) -- (3.75,-0.25) -- (3.75,-1.25) -- (4.25,-1.25);
\draw[dashed,ultra thick] (4.25,-1.25) -- (4.75,-1.25);
\node[rotate=-45] at (5.5,-1.75) {\Large $\mathbf{\cdots}$};
\draw[dashed,ultra thick] (6.25,-2.25) -- (6.75,-2.25);
\draw[ultra thick] (7.25,-2.25) -- (7.75,-1.75);
\draw[ultra thick] (8.25,-3.25) -- (8.75,-2.75);
\draw[ultra thick] (6.75,-2.25) -- (7.25,-2.25) -- (7.25,-3.25) -- (8.25,-3.25) -- (8.25,-4.25) -- (9.25,-4.25);
\draw[ultra thick] (1,0) -- (1.75,0.75);
\draw[ultra thick] (2,-1) -- (2.75,-0.25);
\draw[ultra thick] (3,-2) -- (3.75,-1.25);
\draw[ultra thick] (7.25,-3.25) -- (6.5,-4);
\draw[ultra thick] (8.25,-4.25) -- (7.5,-5);
\draw[ultra thick] (0,0) -- (1,0) -- (1,-1) -- (2,-1) -- (2,-2) -- (3,-2) -- (3,-3) -- (3.5,-3);
\draw[dashed,ultra thick] (3.5,-3) -- (4,-3);
\node[rotate=-45] at (4.75,-3.5) {\Large $\mathbf{\cdots}$};
\draw[dashed,ultra thick] (5.5,-4) -- (6,-4);
\draw[ultra thick] (6,-4) -- (6.5,-4) -- (6.5,-5) -- (7.5,-5) -- (7.5,-6) -- (8.5,-6);
\draw[ultra thick] (1,-1) -- (0.5,-1.5);
\draw[ultra thick] (2,-2) -- (1.5,-2.5);
\draw[ultra thick] (3,-3) -- (2.5,-3.5);
\draw[ultra thick] (6.5,-5) -- (6,-5.5);
\draw[ultra thick] (7.5,-6) -- (7,-6.5);
\node at (0.5,1.75) {\footnotesize $a_1$}; 
\node at (9.5,-4.25) {\footnotesize $a_1$}; 
\node at (-0.25,0) {\footnotesize $a_2$}; 
\node at (8.75,-6) {\footnotesize $a_2$}; 
\node at (2.4,2.4) {\footnotesize $1$};
\node at (3.4,1.4) {\footnotesize $2$};
\node at (4.4,0.4) {\footnotesize $3$};
\node at (8.3,-1.85) {\footnotesize $N-1$};
\node at (8.9,-3) {\footnotesize $N$};
\draw[dashed,ultra thick] (0.5,-1.5) -- (0.25,-1.75);
\node[rotate=45] at (-0.15,-2.15) {\Large $\mathbf{\cdots}$};
\draw[dashed,ultra thick] (1.5,-2.5) -- (1.25,-2.75);
\node[rotate=45] at (0.85,-3.15) {\Large $\mathbf{\cdots}$};
\draw[dashed,ultra thick] (2.5,-3.5) -- (2.25,-3.75);
\node[rotate=45] at (1.85,-4.15) {\Large $\mathbf{\cdots}$};
\draw[dashed,ultra thick] (6,-5.5) -- (5.75,-5.75);
\node[rotate=45] at (5.35,-6.15) {\Large $\mathbf{\cdots}$};
\draw[dashed,ultra thick] (7,-6.5) -- (6.75,-6.75);
\node[rotate=45] at (6.35,-7.15) {\Large $\mathbf{\cdots}$};
\draw[dashed,ultra thick] (-0.6,-2.6) -- (-0.75,-2.75);
\draw[ultra thick] (-0.75,-2.75) -- (-1.25,-3.25);
\draw[dashed,ultra thick] (0.4,-3.6) -- (0.25,-3.75);
\draw[ultra thick] (0.25,-3.75) -- (-0.25,-4.25);
\draw[dashed,ultra thick] (1.4,-4.6) -- (1.25,-4.75);
\draw[ultra thick] (1.25,-4.75) -- (0.75,-5.25);
\draw[dashed,ultra thick] (4.9,-6.6) -- (4.75,-6.75);
\draw[ultra thick] (4.75,-6.75) -- (4.25,-7.25);
\draw[dashed,ultra thick] (5.9,-7.6) -- (5.75,-7.75);
\draw[ultra thick] (5.75,-7.75) -- (5.25,-8.25);
\draw[ultra thick] (-2.25,-3.25) -- (-1.25,-3.25) -- (-1.25,-4.25) -- (-0.25,-4.25) -- (-0.25,-5.25) -- (0.75,-5.25) -- (0.75,-6.25) -- (1.25,-6.25);
\draw[dashed,ultra thick] (1.25,-6.25) -- (1.75,-6.25);
\node[rotate=-45] at (2.5,-6.75) {\Large $\mathbf{\cdots}$};
\draw[dashed,ultra thick] (3.25,-7.25) -- (3.75,-7.25);
\draw[ultra thick] (3.75,-7.25) -- (4.25,-7.25) -- (4.25,-8.25) -- (5.25,-8.25) -- (5.25,-9.25) -- (6.25,-9.25);
\draw[ultra thick] (-1.25,-4.25) -- (-1.75,-4.75);
\draw[ultra thick] (-0.25,-5.25) -- (-0.75,-5.75);
\draw[ultra thick] (0.75,-6.25) -- (0.25,-6.75);
\draw[ultra thick] (4.25,-8.25) -- (3.75,-8.75);
\draw[ultra thick] (5.25,-9.25) -- (4.75,-9.75);
\node at (-2.6,-3.25) {\footnotesize $a_M$}; 
\node at (6.3,-9.45) {\footnotesize $a_M$}; 
\node at (-1.95,-4.95) {\footnotesize $1$};
\node at (-0.95,-5.95) {\footnotesize $2$};
\node at (0.05,-6.95) {\footnotesize $3$};
\node at (3.5,-9) {\footnotesize $N-1$};
\node at (4.5,-10) {\footnotesize $N$};
\draw[<->,green!50!black] (2,1.9) -- (2.9,1);
\node[green!50!black] at (2.7,1.7) {\footnotesize $\widehat{a}_1^{(1)}$};
\draw[<->,green!50!black] (3,0.9) -- (3.9,0);
\node[green!50!black] at (3.7,0.7) {\footnotesize $\widehat{a}_2^{(1)}$};
\draw[<->,green!50!black] (4,-0.1) -- (4.9,-1);
\node[green!50!black] at (4.7,-0.3) {\footnotesize $\widehat{a}_3^{(1)}$};
\draw[<->,green!50!black] (6.5,-1.1) -- (7.4,-2);
\node[green!50!black] at (7.3,-1.3) {\footnotesize $\widehat{a}_{N-2}^{(1)}$};
\draw[<->,green!50!black] (7.5,-2.1) -- (8.4,-3);
\node[green!50!black] at (8.3,-2.3) {\footnotesize $\widehat{a}_{N-1}^{(1)}$};
\draw[<->,green!50!black] (1.3,0.2) -- (2.2,-0.7);
\node[green!50!black] at (2,0) {\footnotesize $\widehat{a}_1^{(2)}$};
\draw[<->,green!50!black] (2.3,-0.8) -- (3.2,-1.7);
\node[green!50!black] at (3,-1) {\footnotesize $\widehat{a}_2^{(2)}$};
\draw[<->,green!50!black] (3.3,-1.8) -- (4.2,-2.7);
\node[green!50!black] at (4,-2) {\footnotesize $\widehat{a}_3^{(2)}$};
\draw[<->,green!50!black] (5.75,-2.85) -- (6.65,-3.75);
\node[green!50!black] at (6.55,-3.05) {\footnotesize $\widehat{a}_{N-2}^{(2)}$};
\draw[<->,green!50!black] (6.75,-3.85) -- (7.65,-4.75);
\node[green!50!black] at (7.55,-4.05) {\footnotesize $\widehat{a}_{N-1}^{(2)}$};
\draw[<->,green!50!black] (0.75,-1.35) -- (1.65,-2.25);
\node[green!50!black] at (1.45,-1.55) {\footnotesize $\widehat{a}_1^{(3)}$};
\draw[<->,green!50!black] (1.75,-2.35) -- (2.65,-3.25);
\node[green!50!black] at (2.45,-2.55) {\footnotesize $\widehat{a}_2^{(3)}$};
\draw[<->,green!50!black] (2.75,-3.35) -- (3.65,-4.25);
\node[green!50!black] at (3.45,-3.55) {\footnotesize $\widehat{a}_3^{(3)}$};
\draw[<->,green!50!black] (5.25,-4.35) -- (6.15,-5.25);
\node[green!50!black] at (6.05,-4.55) {\footnotesize $\widehat{a}_{N-2}^{(3)}$};
\draw[<->,green!50!black] (6.25,-5.35) -- (7.15,-6.25);
\node[green!50!black] at (7.05,-5.55) {\footnotesize $\widehat{a}_{N-1}^{(3)}$};
\draw[<->,green!50!black] (-1,-3.1) -- (-0.1,-4);
\node[green!50!black] at (0.05,-3.55) {\footnotesize $\widehat{a}_1^{(M)}$};
\draw[<->,green!50!black] (0,-4.1) -- (0.9,-5);
\node[green!50!black] at (0.8,-4.3) {\footnotesize $\widehat{a}_2^{(M)}$};
\draw[<->,green!50!black] (1,-5.1) -- (1.9,-6);
\node[green!50!black] at (1.8,-5.3) {\footnotesize $\widehat{a}_3^{(M)}$};
\draw[<->,green!50!black] (3.5,-6.1) -- (4.4,-7);
\node[green!50!black] at (4.3,-6.3) {\footnotesize $\widehat{a}_{N-2}^{(M)}$};
\draw[<->,green!50!black] (4.5,-7.1) -- (5.4,-8);
\node[green!50!black] at (5.3,-7.3) {\footnotesize $\widehat{a}_{N-1}^{(M)}$};
\draw[dashed] (8.75,-2.75) -- (9.8,-2.75);
\draw[dashed] (6.25,-9.25) -- (9.8,-9.25);
\draw[<->,green!50!black] (9.8,-2.8) -- (9.8,-9.2);
\node[green!50!black] at (9.6,-6) {\footnotesize{\bf $\tau$}};
\draw[dashed] (1.75,1.75) -- (-1.2,-1.2);
\draw[dashed] (-1.25,-4.25) -- (0.35,-2.65);
\draw[<->,green!50!black] (0.3,-2.6) -- (-1.1,-1.2);
\node[green!50!black] at (-0.5,-2.1) {\footnotesize{\bf $S$}};
\draw[dashed] (0.75,1.75) -- (3.05,4.05);
\draw[dashed] (8.75,-2.75) -- (9.3,-2.2);
\draw[<->,green!50!black] (9.25,-2.15) -- (3.05,3.95);
\node[green!50!black] at (6,0.7) {\footnotesize{\bf $\rho$}};
\draw[<->,green!50!black] (0.9,0.1) -- (0.9,1.65);
\node[green!50!black] at (0.7,1.1) {\footnotesize{\bf $\tau_1$}};
\draw[<->,green!50!black] (0.2,-1.65) -- (0.2,-0.1);
\node[green!50!black] at (0,-0.9) {\footnotesize{\bf $\tau_2$}};
\draw[<->,green!50!black] (6,-9.15) -- (6,-7.5);
\node[green!50!black] at (6.5,-8.4) {\footnotesize{\bf $\tau_{M-1}$}};
\draw[blue,fill=blue] (1.75,1.75) circle (0.1cm);
\node[scale=0.7,blue] at (1.45,1.95) {\footnotesize{\bf $(1,1)$}};
\draw[blue,fill=blue] (2.75,0.75) circle (0.1cm);
\node[scale=0.7,blue] at (2.45,0.95) {\footnotesize{\bf $(2,1)$}};
\draw[blue,fill=blue] (3.75,-0.25) circle (0.1cm);
\node[scale=0.7,blue] at (3.45,-0.05) {\footnotesize{\bf $(3,1)$}};
\draw[blue,fill=blue] (7.25,-2.25) circle (0.1cm);
\node[scale=0.7,blue] at (6.65,-2.05) {\footnotesize{\bf $(N-1,1)$}};
\draw[blue,fill=blue] (8.25,-3.25) circle (0.1cm);
\node[scale=0.7,blue] at (7.85,-3.05) {\footnotesize{\bf $(N,1)$}};
\draw[blue,fill=blue] (1,0) circle (0.1cm);
\node[scale=0.7,blue] at (0.65,-0.2) {\footnotesize{\bf $(1,2)$}};
\draw[blue,fill=blue] (2,-1) circle (0.1cm);
\node[scale=0.7,blue] at (1.65,-0.8) {\footnotesize{\bf $(2,2)$}};
\draw[blue,fill=blue] (3,-2) circle (0.1cm);
\node[scale=0.7,blue] at (2.65,-1.8) {\footnotesize{\bf $(3,2)$}};
\draw[blue,fill=blue] (6.5,-4) circle (0.1cm);
\node[scale=0.7,blue] at (5.95,-3.8) {\footnotesize{\bf $(N-1,2)$}};
\draw[blue,fill=blue] (7.5,-5) circle (0.1cm);
\node[scale=0.7,blue] at (7.1,-4.8) {\footnotesize{\bf $(N,2)$}};
\draw[blue,fill=blue] (-1.25,-3.25) circle (0.1cm);
\node[scale=0.7,blue] at (-1.65,-3.05) {\footnotesize{\bf $(1,M)$}};
\draw[blue,fill=blue] (-0.25,-4.25) circle (0.1cm);
\node[scale=0.7,blue] at (-0.65,-4.05) {\footnotesize{\bf $(2,M)$}};
\draw[blue,fill=blue] (0.75,-5.25) circle (0.1cm);
\node[scale=0.7,blue] at (0.35,-5.05) {\footnotesize{\bf $(3,M)$}};
\draw[blue,fill=blue] (4.25,-7.25) circle (0.1cm);
\node[scale=0.7,blue] at (3.7,-7.05) {\footnotesize{\bf $(N-1,M)$}};
\draw[blue,fill=blue] (5.25,-8.25) circle (0.1cm);
\node[scale=0.7,blue] at (4.9,-8.05) {\footnotesize{\bf $(N,M)$}};
\begin{scope}[xshift=8.5cm,yshift=2.5cm]
\draw[dashed,fill=black!5!white] (-1.25,-1.25) -- (1.1,-1.25) -- (1.1,1.15) -- (-1.25,1.15) -- (-1.25,-1.25);
\draw[ultra thick] (-1,0) -- (0,0) -- (0,-1); 
\draw[ultra thick] (0,0) -- (0.75,0.75); 
\draw[blue,fill=blue] (0,0) circle (0.1cm);
\node[scale=0.7,blue] at (-0.2,0.3) {\footnotesize{\bf $(i,j)$}};
\node at (-0.6,-0.35) {\footnotesize $h_i^{(j)}$};
\node at (0.4,-0.6) {\footnotesize $v_i^{(j)}$};
\node at (0.7,0.25) {\footnotesize $m_i^{(j)}$};
\end{scope}
\end{tikzpicture}}}
\caption{\sl General $(N,M)$ web diagram and along with a basis of K\"ahler parameters.}
\label{Fig:General Setup}
\end{figure}
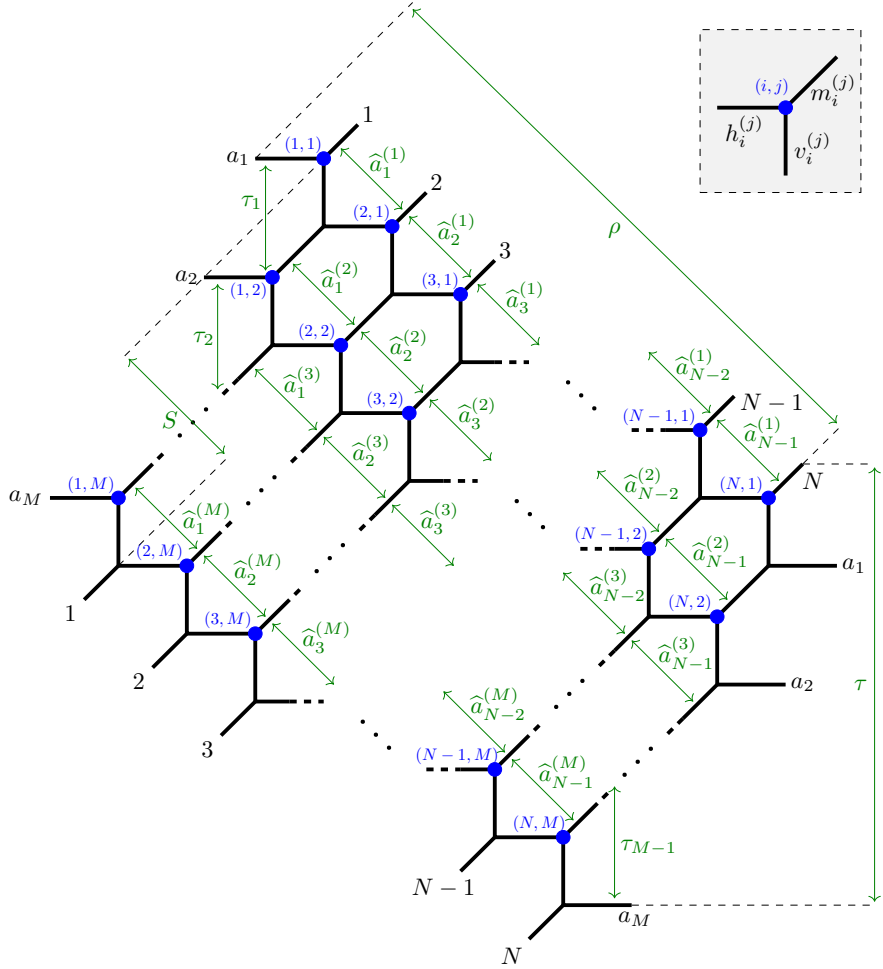

Our starting point is a class of toric Calabi-Yau threefolds $X_{N,M}$ that are labelled by two integer parameters $N,M\in\mathbb{N}$ (see \cite{Haghighat:2013gba,Haghighat:2013tka,Hohenegger:2013ala,Hohenegger:2015btj,Kanazawa:2016tnt}). These manifolds can be described by the web diagram shown in Figure~\ref{Fig:General Setup}. This diagram is doubly periodic, \emph{i.e.} the horizontal legs labelled by the indices $a_{1,\ldots,M}$ and the diagonal legs labelled by $1,\ldots,N$ are respectively identified.\footnote{In the following we shall use lower case letters from the beginning of the alphabet, integers and Roman numerals $\mathbf{I}, \mathbf{II}, \mathbf{III}, \ldots$ to indicate how legs are respectively glued together.}

The lines of the web diagram represent holomorphic curves of the manifold $X_{N,M}$, whose area can be calculated as the integral over the K\"ahler form $\omega$, \emph{i.e.} schematically for a curve $\sigma$ we have the area $A(\sigma)=\int_{\sigma} \omega$. In the following we shall loosely label curves in the diagram through their areas: in Figure~\ref{Fig:General Setup} we have introduced a labelling scheme for the $3NM$ (areas of the) curves $(h_i^{(j)},v_i^{(j)},m_i^{(j)})$, where $i\in\{1,\ldots,N\}$ and $j\in\{1,\ldots,M\}$. These indices are understood to be mod $N$ and $M$ respectively, \emph{e.g.} $h_{N+1}^{(j)}=h_1^{(j)}$ and $h_i^{(M+1)}=h_i^{(1)}$. The curves in Figure~\ref{Fig:General Setup} are labelled according to the $NM$ vertices (shown in blue) they are attached to, following the convention indicated in the grey box. The $3NM$ areas $(h_i^{(j)},v_i^{(j)},m_i^{(j)})$ are not all independent \cite{Haghighat:2013gba,Haghighat:2013tka,Hohenegger:2013ala,Hohenegger:2015cba}: indeed, due to the Calabi-Yau nature of $X_{N,M}$ (see \cite{Kanazawa:2016tnt}) and the double periodicity mentioned above, only $NM+2$ are independent, which can be used to label points in the K\"ahler moduli space of $X_{N,M}$. The choice of these independent parameters is not unique \cite{Bastian:2017ing,Bastian:2017ary,Bastian:2018dfu}, and a particular choice of such a basis is indicated by the green labels in Figure~\ref{Fig:General Setup} for which we shall use the following notation
\begin{align}
\vec{v}=(\vec{a}\,{}^{(1)},\ldots,\vec{a}\,{}^{(M)},\tau_1,\ldots,\tau_{M-1},\tau,S,\rho)^T\in \mathbb{R}^{NM+2}\,,&&\text{with} &&
\vec{a}\,{}^{(k)}=(\widehat{a}_1^{(k)},\ldots,\widehat{a}_{N-1}^{(k)})^T\in\mathbb{R}^{N-1}\,.\label{KaehlerBasis}
\end{align}
This basis of independent parameters is adapted to describe a Little String Theory (LST),

\begin{wrapfigure}{r}{0.22\textwidth}
${}$\\[-0.1cm]
\scalebox{0.7}{\parbox{5.5cm}{\begin{tikzpicture}
\draw [ultra thick,domain=0:200] plot ({2*cos(\x)}, {2*sin(\x)});
\draw [ultra thick,dashed,domain=200:250] plot ({2*cos(\x)}, {2*sin(\x)});
\draw [ultra thick,domain=250:360] plot ({2*cos(\x)}, {2*sin(\x)});
\draw[ultra thick,fill=white] (2,0) circle (0.5cm);
\node at (2,0) {\scriptsize$U(N)$};
\draw[ultra thick,fill=white] ({2*cos(45)}, {2*sin(45)}) circle (0.5cm);
\node at ({2*cos(45)}, {2*sin(45)}) {\scriptsize$U(N)$};
\draw[ultra thick,fill=white] ({2*cos(90)}, {2*sin(90)}) circle (0.5cm);
\node at ({2*cos(90)}, {2*sin(90)}) {\scriptsize$U(N)$};
\draw[ultra thick,fill=white] ({2*cos(135)}, {2*sin(135)}) circle (0.5cm);
\node at ({2*cos(135)}, {2*sin(135)}) {\scriptsize$U(N)$};
\draw[ultra thick,fill=white] ({2*cos(180)}, {2*sin(180)}) circle (0.5cm);
\node at ({2*cos(180)}, {2*sin(180)}) {\scriptsize$U(N)$};
\draw[ultra thick,fill=white] ({2*cos(270)}, {2*sin(270)}) circle (0.5cm);
\node at ({2*cos(270)}, {2*sin(270)}) {\scriptsize$U(N)$};
\draw[ultra thick,fill=white] ({2*cos(315)}, {2*sin(315)}) circle (0.5cm);
\node at ({2*cos(315)}, {2*sin(315)}) {\scriptsize$U(N)$};
\draw [thick,domain=180:270,<->] plot ({2.9*cos(\x)}, {2.9*sin(\x)});
\node[rotate=-45] at (-2.2,-2.2) {\scriptsize $M$ nodes};
\end{tikzpicture}}}
\caption{\sl Quiver with $M$ nodes of groups $U(N)$ and bifundamental matter.}
\label{Fig:Quiver}
${}$\\[-3cm]
\end{wrapfigure}
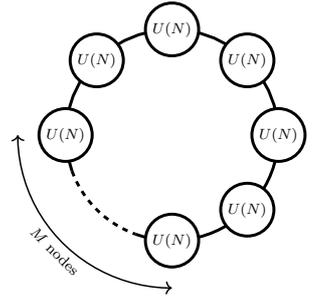

\noindent
which at low energies is given by a quiver gauge theory on $\mathbb{R}^4\times T^2$ and which is described in more detail in Section~\ref{Sec:ZLSTVOA}. This gauge theory has gauge group $[U(N)]^M$ (\emph{i.e.} $M$ nodes of the gauge group $U(N)$) and hypermultiplet matter in the bifundamental representation for $M>1$ and the adjoint representation for $M=1$, as shown in Figure~\ref{Fig:Quiver}:
\begin{itemize}
\item $(\tau_1,\ldots,\tau_{N-1},\tau)$ can be interpreted as coupling constants
\item the sets of parameters $\vec{a}\,{}^{(k)}$ (for $k\in\{1,\ldots,M\}$) play the role of gauge parameters associated with the $k$th $U(N)$ gauge group
\item $\rho$ promotes the Lie algebra $\mathfrak{a}_{N-1}$ associated with each gauge node to the affine algebra $\widehat{\mathfrak{a}}_{N-1}$ 
\item $S$ is a mass deformation of the hypermultiplet matter
\end{itemize}
Geometrically, the K\"ahler moduli space that is parametrised by the basis of K\"ahler parame-

\begin{wrapfigure}{l}{0.14\textwidth}
${}$\\[-0.4cm]
\scalebox{0.52}{\parbox{4.5cm}{\begin{tikzpicture}
\draw[ultra thick] (-2,-4) -- (0.25,2.2);
\draw[ultra thick] (-2,-4) -- (1.5,2);
\draw[ultra thick] (-2,-4) -- (-0.5,1.2);
\draw[fill=white] (-2,-4) -- (0,0) -- (-0.5,1.2) -- (-2,-4);
\draw[ultra thick] (-2,-4) -- (-0.5,1.2);
\draw[ultra thick] (0,0) -- (1,-0.25) -- (2.5,0.8) -- (1.5,2) -- (0.25,2.2) -- (-0.5,1.2) -- (0,0);
\draw[ultra thick,fill=white] (-2,-4) -- (0,0) -- (1,-0.25) -- (-2,-4);
\draw[ultra thick] (-2,-4) -- (0,0);
\draw[ultra thick] (-2,-4) -- (1,-0.25);
\draw[ultra thick] (-2,-4) -- (2.5,0.8);
\draw[dashed] (-2,-4) -- (1.5,2);
\draw[dashed] (-2,-4) -- (0.25,2.2);
\end{tikzpicture}}}
\caption{\sl K\"ahler cone of $X_{N,M}$.}
\label{Fig:ConeSingle}
${}$\\[-2cm]
\end{wrapfigure}

\noindent
ters~(\ref{KaehlerBasis}) takes the structure of a cone, which is schematically depicted in Figure~\ref{Fig:ConeSingle}. More concretely, this \emph{K\"ahler cone} $\mathbb{K}(X_{N,M})$ represents the regime
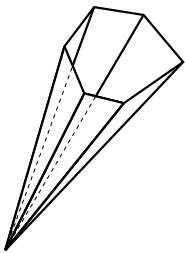
\begin{align}
&\int_{X_{N,M}}\!\!\!\!\omega\wedge\omega\wedge\omega>0\,,&&\int_{\sigma_2}\omega\wedge\omega>0\,,&&\int_{\sigma}\omega>0\,,&&\forall \sigma,\sigma_2\in X_{N,M}\,,\label{InteriorCone}
\end{align}
where $\sigma$ denotes holomorphic curves and $\sigma_2$ two-dimensional complex submanifolds of $X_{N,M}$. The walls of the cone correspond to equalities in the relations (\ref{InteriorCone}), \emph{i.e.} loci where the size of a rational curve with normal bundle $\mathcal{O}(1)\oplus\mathcal{O}(1)$ shrinks to zero \cite{Hohenegger:2016yuv}. As we shall discuss in the next Subsection, certain structures beyond these walls can also be described as Calabi-Yau manifolds.

\subsection{Extended Moduli Space of $X_{N,M}$}
As mentioned before, the walls of the K\"ahler cone $\mathbb{K}(X_{N,M})$ correspond to degenerations of

\begin{wrapfigure}{r}{0.2\textwidth}
${}$\\[-0.5cm]
\centering
\scalebox{0.55}{\parbox{4.5cm}{\begin{tikzpicture}
\draw[ultra thick] (-2,-4) -- (0.25,2.2);
\draw[ultra thick] (-2,-4) -- (1.5,2);
\draw[ultra thick] (-2,-4) -- (-0.5,1.2);
\draw[ultra thick] (-2,-4) -- (-1,1.7);
\draw[ultra thick] (-2,-4) -- (-1.6,1.5);
\draw[fill=yellow!90!green] (-2,-4) -- (0,0) -- (-0.5,1.2) -- (-2,-4);
\draw[ultra thick] (-2,-4) -- (-0.5,1.2);
\draw[ultra thick] (0,0) -- (1,-0.25) -- (2.5,0.8) -- (1.5,2) -- (0.25,2.2) -- (-0.5,1.2) -- (0,0);
\draw[ultra thick] (0,0) -- (-0.5,1.2) -- (-1,1.7) -- (-1.6,1.5) -- (-2,0.5) -- (-1.2,-0.4) -- (0,0);
\draw[ultra thick,fill=white] (-2,-4) -- (0,0) -- (1,-0.25) -- (-2,-4);
\draw[ultra thick] (-2,-4) -- (0,0);
\draw[ultra thick] (-2,-4) -- (1,-0.25);
\draw[ultra thick] (-2,-4) -- (2.5,0.8);
\draw[dashed] (-2,-4) -- (1.5,2);
\draw[dashed] (-2,-4) -- (0.25,2.2);
\draw[ultra thick,fill=white] (-2,-4) -- (-1.995,0.495) -- (-1.2,-0.4) -- (-2,-4);
\draw[ultra thick,fill=white] (-2,-4) -- (-1.2,-0.4) -- (0,0) -- (-2,-4);
\draw[ultra thick] (-2,-4) -- (-1.2,-0.4);
\draw[ultra thick] (-2,-4) -- (-2,0.5);
\draw[dashed] (-2,-4) -- (-0.5,1.2);
\draw[dashed] (-2,-4) -- (-1,1.7);
\draw[dashed] (-2,-4) -- (-1.6,1.5);
\draw[dashed] (-2,-4) -- (-0.5,1.2);
\draw[dashed] (-2,-4) -- (0.25,2.2);
\end{tikzpicture}}}
\caption{\sl Adjacent K\"ahler cones in $\mathbb{EK}(X_{N,M})$.}
\label{Fig:FlopTransition}
${}$\\[-1.1cm]
\end{wrapfigure}
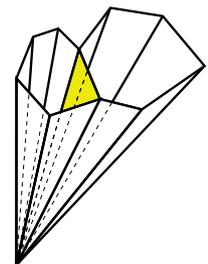

\noindent
rational curves. Such a degeneration can be resolved to produce a new rational curve, which is part of a new Calabi-Yau manifold (which shares the same Hodge numbers as $X_{N,M}$ but has different triple intersection numbers). As is schematically sketched in Figure~\ref{Fig:FlopTransition}, the moduli space of the latter is not part of $\mathbb{K}(X_{N,M})$, but lies outside of it and is part of the so-called \emph{extended moduli space of }$X_{N,M}$, which we shall denote $\mathbb{EK}(X_{N,M})$: indeed with respect to the K\"ahler form $\omega$ of $X_{N,M}$, the area of the new curve is negative. Instead, the new manifold is characterised by its own K\"ahler cone, which is attached to $\mathbb{K}(X_{N,M})$ along one of its walls (shown in yellow in Figure~\ref{Fig:FlopTransition}) and the passage from one cone to the other is called a \emph{flop transition}.  There are numerous different families of Calabi-Yau manifolds that are related in this way through flop (or conifold) transitions, \emph{e.g.}\cite{Reid1987,Green:1988bp,Green:1988wa,Candelas:1988di,Candelas:1989ug,doi:10.1142/1410,Avram:1997rs,Candelas:2007ac,Candelas:2010ve}. In \cite{Hohenegger:2016yuv} it was argued that $\mathbb{K}(X_{N',M'})\subset \mathbb{EK}(X_{N,M})$ (which we denote as the duality $X_{N',M'}\sim X_{N,M}$)~if 
\begin{align}
&N'M'=NM\,,&&\text{and}&&\text{gcd}(N',M')=\text{gcd}(N,M)\,.\label{DualityConditions}
\end{align} 
Concretely, under these conditions, it was argued in \cite{Hohenegger:2016yuv} that there exists a series of flop trans-

\noindent
itions such that the web diagram of $X_{N,M}$ can be transformed into the web diagram of $X_{N',M'}$. The flop transition of an individual curve and its impact on the K\"ahler parameters of the geometry is shown through the following example:

\begin{align}
\scalebox{1}{\parbox{4cm}{\begin{tikzpicture}
\draw[ultra thick] (1.75,0.75) -- (1.75,1.75);
\draw[ultra thick] (0,0) -- (1,0) -- (1.75,0.75) -- (2.75,0.75) -- (3.5,1.5);
\draw[ultra thick,red] (1,0) -- (1.75,0.75);
\draw[ultra thick] (0.75,1.75) -- (1.75,1.75) -- (2.5,2.5);
\draw[ultra thick] (1,0) -- (1,-1);
\draw[ultra thick] (2.75,0.75) -- (2.75,-0.25);
\node at (0.5,0.35) {\footnotesize $h_{i_1}^{(j_1)}$};
\node at (0.6,-0.6) {\footnotesize $v_{i_1}^{(j_1)}$};
\node at (1.25,0.75) {\footnotesize $m_{i_1}^{(j_1)}$};
\node at (1.25,2.1) {\footnotesize $h_{i_2}^{(j_2)}$};
\node at (2.2,1.4) {\footnotesize $v_{i_2}^{(j_2)}$};
\node at (2.6,2.05) {\footnotesize $m_{i_2}^{(j_2)}$};
\node at (3.6,1.05) {\footnotesize $m_{i_3}^{(j_3)}$};
\node at (3.2,0.3) {\footnotesize $v_{i_3}^{(j_3)}$};
\node at (2.2,0.4) {\footnotesize $h_{i_3}^{(j_3)}$};
\draw[blue,fill=blue] (1,0) circle (0.1cm);
\node[scale=0.7,blue] at (1.5,-0.1) {\footnotesize{\bf $(i_1,j_1)$}};
\draw[blue,fill=blue] (1.75,1.75) circle (0.1cm);
\node[scale=0.7,blue] at (1.3,1.5) {\footnotesize{\bf $(i_2,j_2)$}};
\draw[blue,fill=blue] (2.75,0.75) circle (0.1cm);
\node[scale=0.7,blue] at (2.35,0.95) {\footnotesize{\bf $(i_3,j_3)$}};
\end{tikzpicture}}}
&&
\scalebox{1}{\parbox{4cm}{\begin{tikzpicture}
\draw[ultra thick] (0.75,1.75) -- (1.75,1.75) -- (2.5,2.5);
\draw[ultra thick] (1.75,0.75) -- (1.75,1.75);
\draw[ultra thick] (0.75,0.75) -- (1.75,0.75) -- (2.5,0) -- (3.5,0) -- (4.25,0.75);
\draw[ultra thick] (2.5,0) -- (2.5,-1);
\draw[ultra thick] (3.5,0) -- (3.5,-1);
\draw[ultra thick,red] (1.75,0.75) -- (2.5,0);
\node at (1.25,0.4) {\footnotesize $\tilde{h}_{i_1}^{(j_1)}$};
\node at (2.1,-0.65) {\footnotesize $\tilde{v}_{i_1}^{(j_1)}$};
\node at (1.7,0) {\footnotesize $-m_{i_1}^{(j_1)}$};
\node at (1.25,2.1) {\footnotesize $h_{i_2}^{(j_2)}$};
\node at (2.2,1.25) {\footnotesize $\tilde{v}_{i_2}^{(j_2)}$};
\node at (2.55,2) {\footnotesize $m_{i_2}^{(j_2)}$};
\node at (3.1,0.35) {\footnotesize $\tilde{h}_{i_3}^{(j_3)}$};
\node at (4.4,0.25) {\footnotesize $m_{i_3}^{(j_3)}$};
\node at (4,-0.65) {\footnotesize $v_{i_3}^{(j_3)}$};
\draw[orange,fill=yellow!90!green] (1.75,0.75) circle (0.1cm);
\node[scale=0.7,orange] at (1.3,1) {\footnotesize{\bf $(i_1,j_1)$}};
\draw[blue,fill=blue] (1.75,1.75) circle (0.1cm);
\node[scale=0.7,blue] at (1.3,1.5) {\footnotesize{\bf $(i_2,j_2)$}};
\draw[blue,fill=blue] (3.5,0) circle (0.1cm);
\node[scale=0.7,blue] at (3.95,-0.2) {\footnotesize{\bf $(i_3,j_3)$}};
\end{tikzpicture}}}
&&
\begin{array}{l}
\tilde{v}_{i_1}^{(j_1)}=v_{i_1}^{(j_1)}+m_{i_1}^{(j_1)}\,,\\[2pt]
\tilde{v}_{i_2}^{(j_2)}=v_{i_2}^{(j_2)}+m_{i_1}^{(j_1)}\,,\\[2pt]
\tilde{h}_{i_1}^{(j_1)}=h_{i_1}^{(j_1)}+m_{i_1}^{(j_1)}\,,\\[2pt]
\tilde{h}_{i_3}^{(j_3)}=h_{i_3}^{(j_3)}+m_{i_1}^{(j_1)}\,.\label{FlopGenericDef}
\end{array}
\end{align}
Passing from the diagram on the left to the one on the right, the curve indicated in red has undergone a flop transition: the area of the same curve is now negative (with respect to the old K\"ahler form), while the area of all other curves that intersect the red curve are shifted as indicated.\footnote{All curves with vanishing intersection number with $m_{i_1}^{(j_1)}$ remain unchanged.} Notice, with regards to our labelling scheme (based on vertices), the flop transition changes the nature of the vertex directly attached to the red line (which for this reason we now colour in yellow) and shifts the lines of two adjacent vertices. Using flop transitions of this form (as well as other symmetry transformations of the web diagram), the precise duality map between $X_{N,M}$ and $X_{N',M'}$ (under the conditions (\ref{DualityConditions})) was given in \cite{Hohenegger:2016yuv}. Furthermore, it was shown in \cite{Bastian:2017ing} for $\text{gcd}(N,M)=1$ that the partition functions of the gauge theories engineered by $X_{N,M}$ and $X_{N',M'}$ for ($NM=N'M'$) are identical. Using different techniques, further evidence for the duality $X_{N,M}\sim X_{N',M'}$ was given in \cite{Haghighat:2018gqf}.

\subsection{Shifted Web Diagrams}
Before discussing the LSTs (and supersymmetric gauge theories) that are engineered  by $X_{N,M}$, 

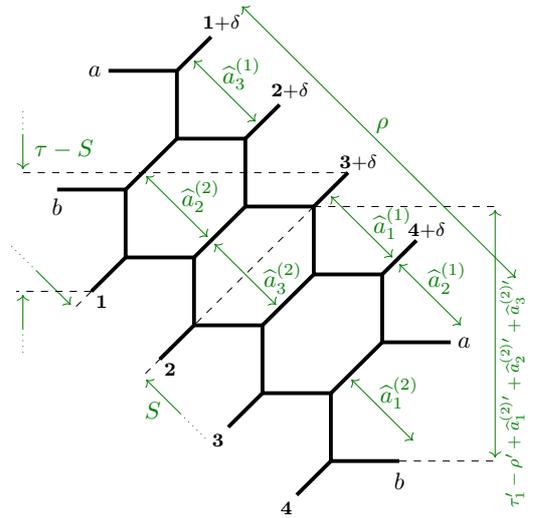
\begin{wrapfigure}{r}{0.41\textwidth}
${}$\\[-0.5cm]
\scalebox{0.9}{\parbox{7.5cm}{\begin{tikzpicture}
\draw[ultra thick] (1.75,1.75) -- (2.25,2.25);
\draw[ultra thick] (2.75,0.75) -- (3.25,1.25);
\draw[ultra thick] (3.75,-0.25) -- (4.25,0.25);
\draw[ultra thick] (4.75,-1.25) -- (5.25,-0.75);
\draw[ultra thick] (0.75,1.75) -- (1.75,1.75) -- (1.75,0.75) -- (2.75,0.75) -- (2.75,-0.25) -- (3.75,-0.25) -- (3.75,-1.25) -- (4.75,-1.25) -- (4.75,-2.25) -- (5.75,-2.25);
\draw[ultra thick] (1,0) -- (1.75,0.75);
\draw[ultra thick] (2,-1) -- (2.75,-0.25);
\draw[ultra thick] (3,-2) -- (3.75,-1.25);
\draw[ultra thick] (4,-3) -- (4.75,-2.25);
\draw[ultra thick] (0,0) -- (1,0) -- (1,-1) -- (2,-1) -- (2,-2) -- (3,-2) -- (3,-3) -- (4,-3) -- (4,-4) -- (5,-4);
\draw[ultra thick] (1,-1) -- (0.5,-1.5);
\draw[ultra thick] (2,-2) -- (1.5,-2.5);
\draw[ultra thick] (3,-3) -- (2.5,-3.5);
\draw[ultra thick] (4,-4) -- (3.5,-4.5);
\node at (0.55,1.75) {\footnotesize $a$}; 
\node at (5.95,-2.25) {\footnotesize $a$}; 
\node at (0,-0.25) {\footnotesize $b$}; 
\node at (5,-4.3) {\footnotesize $b$}; 
\node at (2.4,2.45) {\scriptsize {\bf 1$+\delta$}};
\node at (3.4,1.45) {\scriptsize {\bf 2$+\delta$}};
\node at (4.4,0.4) {\scriptsize {\bf 3$+\delta$}};
\node at (5.4,-0.6) {\scriptsize {\bf 4$+\delta$}};
\node at (0.65,-1.65) {\scriptsize {\bf 1}};
\node at (1.65,-2.65) {\scriptsize {\bf 2}};
\node at (2.35,-3.7) {\scriptsize {\bf 3}};
\node at (3.35,-4.7) {\scriptsize {\bf 4}};
\draw[<->,green!50!black] (2,1.9) -- (2.9,1);
\node[green!50!black] at (2.7,1.7) {\footnotesize $\widehat{a}_3^{(1)}$};
\draw[<->,green!50!black] (4,-0.1) -- (4.9,-1);
\node[green!50!black] at (4.9,-0.5) {\footnotesize $\widehat{a}_1^{(1)}$};
\draw[<->,green!50!black] (5,-1.1) -- (5.9,-2);
\node[green!50!black] at (5.7,-1.3) {\footnotesize $\widehat{a}_2^{(1)}$};
\draw[<->,green!50!black] (1.3,0.2) -- (2.2,-0.7);
\node[green!50!black] at (2.1,-0.1) {\footnotesize $\widehat{a}_2^{(2)}$};
\draw[<->,green!50!black] (2.3,-0.8) -- (3.2,-1.7);
\node[green!50!black] at (3.3,-1.3) {\footnotesize $\widehat{a}_3^{(2)}$};
\draw[<->,green!50!black] (4.3,-2.8) -- (5.2,-3.7);
\node[green!50!black] at (5,-3) {\footnotesize $\widehat{a}_1^{(2)}$};
\draw[<->,green!50!black] (2.7,2.7) -- (6.7,-1.3);
\node[green!50!black] at (4.75,0.95) {\footnotesize{\bf $\rho$}};
\draw[dashed] (3.75,-0.25) -- (1.25,-2.75);
\draw[dashed] (0.5,-1.5) -- (0.25,-1.75);
\draw[<-,green!50!black] (0.2,-1.7) -- (-0.3,-1.2);
\draw[dotted,green!50!black]  (-0.3,-1.2) -- (-0.7,-0.8);
\draw[<-,green!50!black] (1.3,-2.8) -- (1.8,-3.3);
\draw[dotted,green!50!black] (1.8,-3.3) -- (2.2,-3.7);
\node[green!50!black] at (1.4,-3.25) {\footnotesize{\bf $S$}};
\draw[dashed] (3.75,-0.25) -- (6.4,-0.25);
\draw[dashed] (5,-4) -- (6.4,-4);
\draw[<->,green!50!black] (6.4,-0.3) -- (6.4,-3.95);
\node[rotate=90,green!50!black,scale=0.9] at (6.65,-3) {\scriptsize{\bf $\tau'_1-\rho'+\widehat{a}_1^{(2)\prime}+\widehat{a}_2^{(2)\prime}+\widehat{a}_3^{(2)\prime}$}};
\draw[dashed] (4.25,0.25) -- (-0.6,0.25);
\draw[dashed] (0.5,-1.5) -- (-0.6,-1.5);
\draw[<-,green!50!black] (-0.5,0.3) -- (-0.5,0.8);
\draw[dotted,green!50!black] (-0.5,0.8) -- (-0.5,1.2);
\draw[<-,green!50!black] (-0.5,-1.55) -- (-0.5,-2.05);
\draw[dotted,green!50!black] (-0.5,-2.05) -- (-0.5,-2.45);
\node[green!50!black] at (0.1,0.6) {\footnotesize{\bf $\tau-S$}};
\end{tikzpicture}}}
\caption{\emph{Shifted web diagram $(N,M)=(4,2)$ and $\delta=2$ with a basis of K\"ahler parameters.}}
\label{Fig:Shifted}
${}$\\[-2cm]
\end{wrapfigure}

\noindent
we comment on so-called \emph{shifted web diagrams} \cite{Bastian:2018dfu,Bastian:2018fba}: certain points of $\mathbb{EK}(X_{N,M})$ can be represented by shifted versions of the diagram in Figure~\ref{Fig:General Setup}, for which the external legs are identified after a cyclic rotation. An example is shown in Figure~\ref{Fig:Shifted}, where the 4 diagonal legs are glued after a cyclic shift by $\delta=2$ (while the 2 horizontal legs are identified without shift): this is denoted by the labels $\mathbf{i}+\delta\in \{1,\ldots,N\}$ which are understood modulo $N$ (\emph{e.g.} $\mathbf{3}+\delta=\mathbf{1}$ for $\delta=2$). The Figure also shows a basis of independent K\"ahler parameters $(\widehat{a}_1^{(1)},\ldots,\widehat{a}_3^{(2)},\tau'_1,\tau',S',\rho')$, where the definition of the couplings is more involved than for $\delta=0$. For generic $\delta$ a general formula for $\tau$ can be found by summing the areas of all diagonal lines of the diagram
\begin{align}
\tau &= S + \frac{1}{N} \sum_{i=1}^N \sum_{j=1}^M m_i^{(j)}\,.
\end{align}
Similarly, the $\{\tau_1,\ldots,\tau_{M-1}\}$ can be related to the sum over all $m_i^{(j+1)}$ in the following manner
\begin{align}
    \tau_j &= \frac{1}{N} \sum_{i=1}^N m_i^{(j+1)} + \frac{1}{M} S - \sum_{i=1}^{N-1} \frac{i(N-i)}{2N} \left(\widehat{a}_i^{(j)}-2\widehat{a}_i^{(j+1)}+\widehat{a}_i^{(j+2)}\right)\,.
\end{align}
\subsection{Little String Partition Function} \label{subsec:partitionfunction}
The manifolds $X_{N,M}$ described above, geometrically engineer six-dimensional Little String Theories (LSTs), which at low energies are described by supersymmetric gauge theories on $\mathbb{R}^4_{\epsilon_1,\epsilon_2}\times T^2$ with different gauge and matter content \cite{Hohenegger:2015btj}. The parameters $\epsilon_{1,2}$ are geometric deformation parameters which are required to render the topological string partition function $\mathcal{Z}_{N,M}$ well defined, as we shall discuss in the following. Concretely, it was argued in \cite{Bastian:2017ary} that $\mathbb{K}(X_{N,M})$ contains three regions that allow to describe weakly coupled quiver gauge theories with gauge groups
\begin{align}
&G_{\text{hor}}=[U(M)]^N\,,&&\text{or}&&G_{\text{vert}}=[U(N)]^M\,,&&\text{or} &&G_{\text{diag}}=\left[U\left(\tfrac{NM}{k}\right)\right]^k\,,&&\text{with} &&k=\text{gcd}(N,M)\,,\label{GaugeTheoriesDualities}
\end{align}
and matter in the bifundamental representation (or the adjoint in the case where there is only one gauge node present). The quiver representing $G_{\text{vert}}$ is shown in Figure~\ref{Fig:Quiver} and, as mentioned before, the concrete parametrisation of $\mathbb{K}(X_{N,M})$ we have chosen in this paper is adapted to describe this theory. 

The three theories (\ref{GaugeTheoriesDualities}) are dual to each other in the sense that their full non-perturbative BPS partition functions are identical \cite{Bastian:2017ary,Bastian:2018dfu} and are captured by the topological string partition function of $X_{N,M}$ (see \cite{Haghighat:2013gba,Haghighat:2013tka,Hohenegger:2013ala}). Instanton expansions can be calculated efficiently in each case

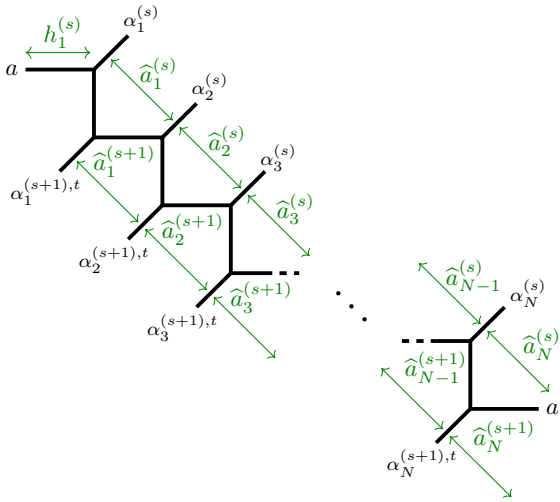
\begin{wrapfigure}{l}{0.46\textwidth}
${}$\\[-0.3cm]
\scalebox{0.9}{\parbox{8.4cm}{\begin{tikzpicture}
\draw[ultra thick] (1.75,1.75) -- (2.25,2.25);
\draw[ultra thick] (2.75,0.75) -- (3.25,1.25);
\draw[ultra thick] (3.75,-0.25) -- (4.25,0.25);
\draw[ultra thick] (0.75,1.75) -- (1.75,1.75) -- (1.75,0.75) -- (2.75,0.75) -- (2.75,-0.25) -- (3.75,-0.25) -- (3.75,-1.25) -- (4.25,-1.25); 
\draw[dashed,ultra thick] (4.25,-1.25) -- (4.75,-1.25);
\node[rotate=-45] at (5.5,-1.75) {\Large $\mathbf{\cdots}$};
\draw[dashed,ultra thick] (6.25,-2.25) -- (6.75,-2.25);
\draw[ultra thick] (7.25,-2.25) -- (7.75,-1.75);
\draw[ultra thick] (6.75,-2.25) -- (7.25,-2.25) -- (7.25,-3.25) -- (8.25,-3.25);
\draw[ultra thick] (1.25,0.25) -- (1.75,0.75);
\draw[ultra thick] (2.25,-0.75) -- (2.75,-0.25);
\draw[ultra thick] (3.25,-1.75) -- (3.75,-1.25);
\draw[ultra thick] (6.75,-3.75) -- (7.25,-3.25);
\node at (0.55,1.75) {\footnotesize $a$}; 
\node at (8.45,-3.25) {\footnotesize $a$}; 
\node at (2.45,2.5) {\scriptsize $\alpha_1^{(s)}$};
\node at (3.45,1.5) {\scriptsize $\alpha_2^{(s)}$};
\node at (4.45,0.45) {\scriptsize $\alpha_3^{(s)}$};
\node at (8.1,-1.55) {\scriptsize $\alpha_N^{(s)}$};
\node at (1.05,-0.05) {\scriptsize $\alpha_1^{(s+1),t}$};
\node at (2.05,-1.05) {\scriptsize $\alpha_2^{(s+1),t}$};
\node at (3.05,-2.05) {\scriptsize $\alpha_3^{(s+1),t}$};
\node at (6.55,-4.05) {\scriptsize $\alpha_N^{(s+1),t}$};
\draw[<->,green!50!black] (2,1.9) -- (2.9,1);
\node[green!50!black] at (2.7,1.7) {\footnotesize $\widehat{a}_1^{(s)}$};
\draw[<->,green!50!black] (3,0.9) -- (3.9,0);
\node[green!50!black] at (3.7,0.7) {\footnotesize $\widehat{a}_2^{(s)}$};
\draw[<->,green!50!black] (4,-0.1) -- (4.9,-1);
\node[green!50!black] at (4.7,-0.3) {\footnotesize $\widehat{a}_3^{(s)}$};
\draw[<->,green!50!black] (6.5,-1.1) -- (7.4,-2);
\node[green!50!black] at (7.3,-1.3) {\footnotesize $\widehat{a}_{N-1}^{(s)}$};
\draw[<->,green!50!black] (7.5,-2.1) -- (8.4,-3);
\node[green!50!black] at (8.3,-2.3) {\footnotesize $\widehat{a}_{N}^{(s)}$};
\draw[<->,green!50!black] (1.5,0.4) -- (2.4,-0.5);
\node[green!50!black] at (2.2,0.4) {\footnotesize $\widehat{a}_1^{(s+1)}$};
\draw[<->,green!50!black] (2.5,-0.6) -- (3.4,-1.5);
\node[green!50!black] at (3.2,-0.6) {\footnotesize $\widehat{a}_2^{(s+1)}$};
\draw[<->,green!50!black] (3.5,-1.6) -- (4.4,-2.5);
\node[green!50!black] at (4.2,-1.6) {\footnotesize $\widehat{a}_3^{(s+1)}$};
\draw[<->,green!50!black] (5.95,-2.65) -- (6.85,-3.55);
\node[green!50!black] at (6.75,-2.65) {\footnotesize $\widehat{a}_{N-1}^{(s+1)}$};
\draw[<->,green!50!black] (6.95,-3.65) -- (7.85,-4.55);
\node[green!50!black] at (7.75,-3.65) {\footnotesize $\widehat{a}_{N}^{(s+1)}$};
\draw[<->,green!50!black] (0.75,2) -- (1.7,2);
\node[green!50!black]  at (1.3,2.3) {\footnotesize $h_1^{(s)}$};
\end{tikzpicture}}}
\caption{\emph{Schematic depiction of the building block in (\ref{WbuildingBlock}).}}
\label{Fig:BuildingBlock}
${}$\\[-1.1cm]
\end{wrapfigure}

\noindent
 using the (refined) topological vertex formalism \cite{Aganagic:2003db,Iqbal:2007ii}: to this end \cite{Bastian:2017ing}, the web-diagram in Figure~\ref{Fig:General Setup} can be decomposed into $M$ fundamental building blocks (with $s\in\{1,\ldots,M\}$), as shown in Figure~\ref{Fig:BuildingBlock}. This building block is parametrised by the K\"ahler parameters $(h_1^{(s)},\widehat{a}_{1,\ldots,N},\widehat{b}_{1,\ldots,N})$: here $h_1^{(s)}$ can be expressed in the basis (\ref{KaehlerBasis})\footnote{The K\"ahler parameters entering the basis~\eqref{KaehlerBasis} are real since they correspond to the area of certain holomorphic curves in $X_{N,M}$. For generality, we now consider complex K\"ahler parameters, this ensures the convergence of the theta functions as defined in~\eqref{eq:thetafunction} and of the summation over the partitions performed in~\eqref{DefPartitionFunction}.}, while $\widehat{a}_N^{(s)}$ and $\widehat{a}_N^{(s+1)}$ are defined through
\begin{align}
\rho=\sum_{i=1}^N\widehat{a}_i^{(s)}=\sum_{i=1}^N\widehat{a}_i^{(s+1)}\,.
\end{align}
Furthermore, $\alpha^{(s)}_{1,\ldots,N}$ and $\alpha^{(s+1)}_{1,\ldots,N}$ denote integer partitions (with $t$ denoting the transposition) associated with the free legs. The web diagram in Figure~\ref{Fig:General Setup} is obtained by 'gluing' $M$ of the strips \ref{Fig:BuildingBlock} together by summing over these partitions (weighted by the K\"ahler parameters $m_i^{(j)}$). The contribution of an individual building block to the partition function was found in \cite{Bastian:2017ing} (see also \cite{Hohenegger:2013ala,Hohenegger:2016eqy}): 
\begin{align}
W^{\alpha_1^{(s)}\ldots \alpha_N^{(s)}}_{\alpha_1^{(s+1)}\ldots\alpha_N^{(s+1)}}=W_N(\emptyset)\cdot\hat{Z}\cdot\prod_{i,j=1}^N\frac{\mathcal{J}_{\alpha_i^{(s)}\alpha_i^{(s+1)}}(\widehat{Q}^{(s)}_{i,i-j};q,t)\,\mathcal{J}_{\alpha_j^{(s+1)}\alpha_i^{(s)}}\left((\widehat{Q}^{(s)}_{i,i-j})^{-1} Q_\rho;q,t\right)}{\mathcal{J}_{\alpha_i^{(s)}\alpha_j^{(s)}}(\overline{Q}^{(s)}_{i,i-j}\sqrt{q/t};q,t)\mathcal{J}_{\alpha_j^{(s+1)}\alpha_i^{(s+1)}}(\dot{Q}^{(s)}_{i,j-1}\sqrt{t/q};q,t)}\,,\label{WbuildingBlock}
\end{align}
with the definitions\footnote{For the definitions of $W_N(\emptyset)$ and $\hat{Z}$, which shall not play a role in our work, we refer the reader to \cite{Bastian:2017ing}. Furthermore, in this work we shall use the notation $Q_x:=e^{2\pi i x}\,\, \forall x\in\mathbb{C}$, such that \emph{e.g.} $Q_{\widehat{a}_i^{(j)}}=e^{2\pi i \widehat{a}_i^{(j)}}$ as well as the shorthand $q=e^{2\pi i \epsilon_1}$ and $t=e^{-2 \pi i \epsilon_2}$.} 
\begin{align}\label{eq:Qvariables}
&\widehat{Q}^{(s)}_{i,j}=Q_{h_i^{(s)}}\,\prod_{r=1}^i Q_{\widehat{a}_r^{(s)}}\,(Q_{\widehat{a}_r^{(s+1)}})^{-1}\,\prod_{k=1}^{j-1}\,Q_{\widehat{a}_{i-k}^{(s)}}\,,&&\dot{Q}_{i,j}=\prod_{k=1}^j Q_{\widehat{a}_{i+k}^{(s+1)}}\,,&&\overline{Q}_{i,j}=\left\{\begin{array}{lcl} 1 & \text{if} & j=N\,,\\ \prod_{k=1}^jQ_{\widehat{a}_{i-k}^{(s)}} & \text{if} & j\neq N\,. \end{array}\right.
\end{align}
The functions $\mathcal{J}_{\mu\nu}$, which depend on two integer partitions $\mu$, $\nu$, are defined in (\ref{Jfunction}). Combining the building blocks (\ref{WbuildingBlock}), the non-perturbative BPS partition function can be written in the form \cite{Hohenegger:2015cba,Hohenegger:2015btj,Hohenegger:2016eqy,Hohenegger:2016yuv,Ahmed:2017hfr,Bastian:2017ing,Bastian:2017ary,Bastian:2018dfu}
\begin{align}
&\mathcal{Z}_{N,M}\big(\vec{v};\epsilon_{1,2}\big)= W_N^M(\emptyset) \sum_{\vec \alpha^{(1)},\ldots,\vec \alpha^{(M)}} \Qt^{\sum_{i=1}^N \left|\alpha_i^{(1)}\right|} \prod_{j=1}^{M-1} Q_{\tau_j}^{\sum_{i=1}^N \left|\alpha_i^{(j+1)}\right|-\left|\alpha_i^{(1)}\right|}\, \mathcal{P}_{\vec \alpha^{(1)},\ldots,\vec \alpha^{(M)}}\, ,\label{DefPartitionFunction}
\end{align}
where we have defined
\begin{align}\label{Pexternalpartition}
    \mathcal{P}_{\vec \alpha^{(1)},\ldots,\vec \alpha^{(M)}} = \prod_{j=1}^M \prod_{1 \leq k \leq l \leq N} \frac{\vartheta_{\alpha_k^{(j+1)}\alpha_l^{(j)}}((\widehat Q_{k,k-l}^{(j)})^{-1};\rho)\vartheta_{\alpha_k^{(j)}\alpha_l^{(j+1)}}(\widehat Q_{k,k-l}^{(j)};\rho)}{\vartheta_{\alpha_l^{(j)} \alpha_k^{(j)}}(Q_{\widehat b_l^{(j)}}/Q_{\widehat b_{-k}^{(j)}}\sqrt{q/t};\rho)\vartheta_{\alpha_k^{(j+1)}\alpha_l^{(j+1)}}(Q_{\widehat b_{1-l}^{(j+1)}}/Q_{\widehat b_k^{(j+1)}}\sqrt{t/q};\rho)}\,.
\end{align}
For later convenience, we have introduced the notation $\vec \alpha^{(j)} = \{ \alpha_1^{(j)}, \ldots, \alpha_N^{(j)} \}$ and $\widehat b_i^{(j)}$ combinations of $\widehat a_i^{(j)}$:
\begin{align}
&\widehat{b}_1^{(j)}=0\,,&&\text{and} &&\widehat{b}_i^{(j)}=\sum_{n=1}^{i-1}\widehat{a}_n^{(j)}\,,&& \begin{array}{l}\forall i=2,\ldots,N\,, \\ \forall j=1,\ldots,M\,.\end{array}\label{DefBParameters}
\end{align}
The formalism described above can also be used to compute the topological string partition functions of Calabi-Yau manifolds associated with shifted web-diagrams as in Figure~\ref{Fig:Shifted}. Indeed, as discussed in \cite{Bastian:2018dfu,Bastian:2018fba} the fundamental building blocks (\ref{WbuildingBlock}) simply need to be glued together in a way to reflect the shifting of the external legs.




\section{Non-perturbative Symmetries}\label{Sect:ReviewSyms}
In the previous Section we have seen that Little String Orbifolds can be constructed from toric Calabi-Yau manifolds, which among others act as blue prints to compute the partition function $\mathcal{Z}_{N,M}$. This geometric engineering also allows us to translate dualities of the Calabi-Yau geometries $X_{N,M}$ into symmetries of the corresponding LST theory. As was shown in \cite{Bastian:2018jlf} in the case $M=1$, (some of) these symmetries act inherently non-perturbatively from the perspective of low energy gauge theories and are therefore exceedingly difficult to understand with other means. In this Section we first provide examples of such theories in the case of $M=1$ (thus reviewing and re-formulating some of the findings of \cite{Bastian:2018jlf}) and then provide a general discussion for the case $M\geq 1$ in Section~\ref{Sect:GenSymTrans}. 
\subsection{Similarity Transformations and the Example $(N,1)$}
The basic strategy for finding non-trivial symmetries for the partition function in \cite{Bastian:2018jlf} is to study self-similarity transformation of the web diagram, which leave its general form (nearly) invariant but act in a non-trivial fashion on the elements of a basis of the K\"ahler parameters. Conceptually, from the perspective of the extended K\"ahler moduli space of the Calabi-Yau manifold $X_{N,M}$, there are two different types of such transformations $\mathfrak{s}$:
\begin{itemize}
\item[\emph{(i)}] \emph{symmetries relating points in the same a K\"ahler cone:} these can be established by finding different presentations of the web diagram which are again of the form of a $(N,M)$ web diagram. Examples of such alternative presentations are
\begin{itemize} 
\item[$\bullet$] \emph{cut-and-reglue}: Since the web diagrams are understood to be drawn on a torus, they need to be cut open to be drawn on a plane as in Figure~\ref{Fig:General Setup}. Different choices of these cuts can lead to different presentations of the web diagram, which still describe the same point in the moduli space of $X_{N,M}$. However, if both presentations are similar to one another (\emph{i.e.} are of the form of an '$(N,M)$ web'), they can be used to deduce actual symmetries between inequivalent points of the moduli space. A simple example is given in the form of the following two diagrams\footnote{Although the labelling of the vertices is sufficient, in order to better exhibit the various transformations, we have also retained the labelling of individual lines. Furthermore, although for $M=1$ the superscript on the labelling $(h_i^{(j)},v_i^{(j)},m_i^{(j)})$ is redundant, for consistency we are using the same notation as in previous Sections (we have, however, somewhat streamlined the notation of basis elements compared to (\ref{KaehlerBasis})).}
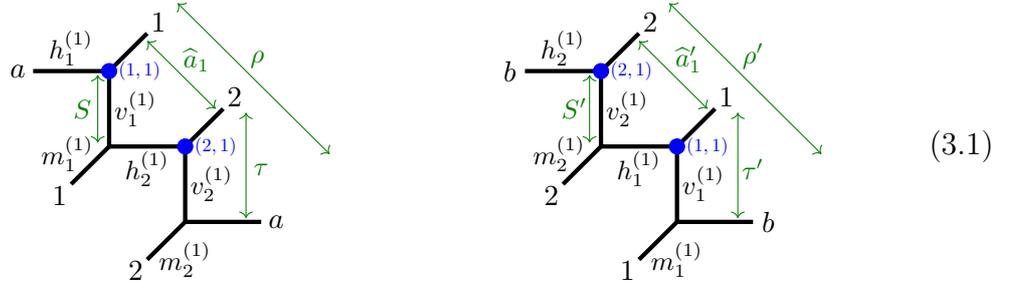
\begin{align}
&\parbox{4.5cm}{\begin{tikzpicture}
\draw[ultra thick] (0,0) -- (1,0) -- (1,-1) -- (2,-1) -- (2,-2) -- (3,-2);
\node at (-0.2,0) {\small $a$};
\node at (3.2,-2) {\small $a$};
\draw[ultra thick] (1,0) -- (1.5,0.5);
\node at (1.65,0.65) {\small $1$};
\draw[ultra thick] (2,-1) -- (2.5,-0.5);
\node at (2.65,-0.35) {\small $2$};
\draw[ultra thick] (1,-1) -- (0.5,-1.5);
\node at (0.35,-1.65) {\small $1$};
\draw[ultra thick] (2,-2) -- (1.5,-2.5);
\node at (1.35,-2.65) {\small $2$};
\node at (0.5,0.3) {\footnotesize $h_1^{(1)}$};
\node at (1.5,-1.3) {\footnotesize $h_2^{(1)}$};
\node at (1.35,-0.5) {\footnotesize $v_1^{(1)}$};
\node at (2.35,-1.5) {\footnotesize $v_2^{(1)}$};
\node at (0.45,-1.1) {\footnotesize $m_1^{(1)}$};
\node at (2,-2.5) {\footnotesize $m_2^{(1)}$};
\draw[blue,fill=blue] (1,0) circle (0.1cm);
\node[scale=0.7,blue] at (1.4,0) {\footnotesize{\bf $(1,1)$}};
\draw[blue,fill=blue] (2,-1) circle (0.1cm);
\node[scale=0.7,blue] at (2.4,-1) {\footnotesize{\bf $(2,1)$}};
\draw[<->,green!50!black] (1.5,0.4) -- (2.4,-0.5);
\node[green!50!black] at (2.15,0.15) {\footnotesize $\widehat{a}_1$};
\draw[<->,green!50!black] (2.8,-0.55) -- (2.8,-1.95);
\node[green!50!black] at (3,-1.3) {\footnotesize $\tau$};
\draw[<->,green!50!black] (0.85,-0.95) -- (0.85,-0.05);
\node[green!50!black] at (0.65,-0.5) {\footnotesize $S$};
\draw[<->,green!50!black] (1.9,0.9) -- (3.9,-1.1);
\node[green!50!black] at (2.95,0.15) {\footnotesize{\bf $\rho$}};
\end{tikzpicture}}
&&
\parbox{4.6cm}{\begin{tikzpicture}
\draw[ultra thick] (0,0) -- (1,0) -- (1,-1) -- (2,-1) -- (2,-2) -- (3,-2);
\node at (-0.2,0) {\small $b$};
\node at (3.2,-2) {\small $b$};
\draw[ultra thick] (1,0) -- (1.5,0.5);
\node at (1.65,0.65) {\small $2$};
\draw[ultra thick] (2,-1) -- (2.5,-0.5);
\node at (2.65,-0.35) {\small $1$};
\draw[ultra thick] (1,-1) -- (0.5,-1.5);
\node at (0.35,-1.65) {\small $2$};
\draw[ultra thick] (2,-2) -- (1.5,-2.5);
\node at (1.35,-2.65) {\small $1$};
\node at (0.5,0.3) {\footnotesize $h_2^{(1)}$};
\node at (1.5,-1.3) {\footnotesize $h_1^{(1)}$};
\node at (1.35,-0.5) {\footnotesize $v_2^{(1)}$};
\node at (2.35,-1.5) {\footnotesize $v_1^{(1)}$};
\node at (0.45,-1.1) {\footnotesize $m_2^{(1)}$};
\node at (2,-2.5) {\footnotesize $m_1^{(1)}$};
\draw[blue,fill=blue] (1,0) circle (0.1cm);
\node[scale=0.7,blue] at (1.4,0) {\footnotesize{\bf $(2,1)$}};
\draw[blue,fill=blue] (2,-1) circle (0.1cm);
\node[scale=0.7,blue] at (2.4,-1) {\footnotesize{\bf $(1,1)$}};
\draw[<->,green!50!black] (1.5,0.4) -- (2.4,-0.5);
\node[green!50!black] at (2.15,0.15) {\footnotesize $\widehat{a}'_1$};
\draw[<->,green!50!black] (2.8,-0.55) -- (2.8,-1.95);
\node[green!50!black] at (3,-1.3) {\footnotesize $\tau'$};
\draw[<->,green!50!black] (0.85,-0.95) -- (0.85,-0.05);
\node[green!50!black] at (0.65,-0.5) {\footnotesize $S'$};
\draw[<->,green!50!black] (1.9,0.9) -- (3.9,-1.1);
\node[green!50!black] at (3,0.2) {\footnotesize{\bf $\rho'$}};\end{tikzpicture}}
\end{align}
Both diagrams represent the same point in the K\"ahler cone of the Calabi-Yau geometry $X_{2,1}$: visually, they can be related by cutting the left diagram along the line labelled $h_2^{(1)}$ and regluing it along the line labeled $h_1^{(1)}$. From the perspective of the vertices, this transformation corresponds to the shift $(i,j)\to (i+1,j)$. However, the sets of K\"ahler parameters $(\widehat{a}_1,\tau,S,\rho)$ and $(\widehat{a}'_1,\tau',S',\rho')$ constitute two different (but equivalent) bases of the K\"ahler moduli space. The equivalence of the two diagrams above therefore implies a (linear) symmetry transformation 
\begin{align}
&\mathfrak{s}:\,\begin{psmall} \widehat{a}_1\\\tau \\ S \\\rho\end{psmall}\longrightarrow \begin{psmall} \widehat{a}'_1\\\tau' \\ S' \\ \rho'\end{psmall}=\mathcal{S}\cdot\begin{psmall} \widehat{a}_1\\ \tau \\ S \\ \rho\end{psmall}\,,&&\text{with} && \mathcal{S}= \begin{psmall}-1 & 0 & 0 & 1 \\ 0 & 1 & 0 & 0 \\ 0 & 0 & 1 & 0 \\ 0 & 0 & 0 & 1\end{psmall} \,.
\end{align}
$\mathfrak{s}$ maps two \emph{inequivalent} points in $\mathbb{K}(X_{2,1})$, however, in a way which leaves the full partition function $\mathcal{Z}_{2,1}$ invariant.
\item[$\bullet$] \emph{rotations:} Simple re-orientations of the web diagram do not change the associated Calabi-Yau geometry and can thus be used to construct various similarity transformations. A simple example is given by the two diagrams below\footnote{Here we choose an example in the context of $X_{3,1}$ rather than $X_{2,1}$, since the equivalent of the transformation (\ref{TrafoRotationToy}) for $(N,M)=(2,1)$ is trivial.}
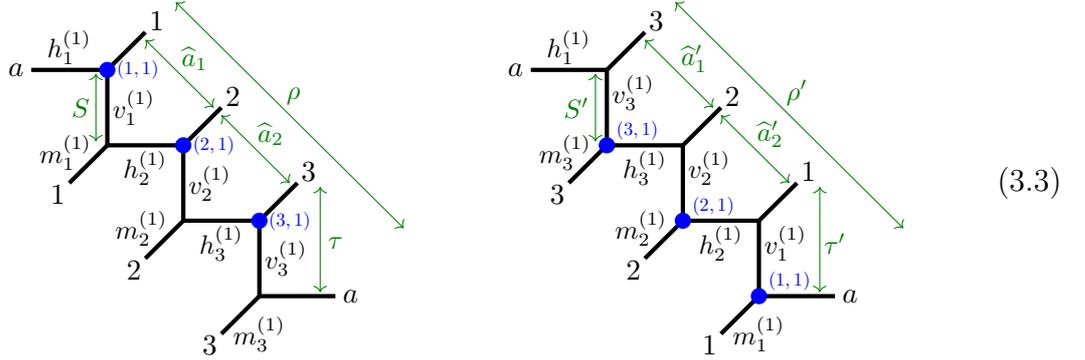
\begin{align}
&\parbox{5.5cm}{\begin{tikzpicture}
\draw[ultra thick] (0,0) -- (1,0) -- (1,-1) -- (2,-1) -- (2,-2) -- (3,-2) -- (3,-3) -- (4,-3);
\node at (-0.2,0) {\small $a$};
\node at (4.2,-3) {\small $a$};
\draw[ultra thick] (1,0) -- (1.5,0.5);
\node at (1.65,0.65) {\small $1$};
\draw[ultra thick] (2,-1) -- (2.5,-0.5);
\node at (2.65,-0.35) {\small $2$};
\draw[ultra thick] (3,-2) -- (3.5,-1.5);
\node at (3.65,-1.35) {\small $3$};
\draw[ultra thick] (1,-1) -- (0.5,-1.5);
\node at (0.35,-1.65) {\small $1$};
\draw[ultra thick] (2,-2) -- (1.5,-2.5);
\node at (1.35,-2.65) {\small $2$};
\draw[ultra thick] (3,-3) -- (2.5,-3.5);
\node at (2.35,-3.65) {\small $3$};
\draw[blue,fill=blue] (1,0) circle (0.1cm);
\node[scale=0.7,blue] at (1.4,0) {\footnotesize{\bf $(1,1)$}};
\draw[blue,fill=blue] (2,-1) circle (0.1cm);
\node[scale=0.7,blue] at (2.4,-1) {\footnotesize{\bf $(2,1)$}};
\draw[blue,fill=blue] (3,-2) circle (0.1cm);
\node[scale=0.7,blue] at (3.4,-2) {\footnotesize{\bf $(3,1)$}};
\node at (0.5,0.3) {\footnotesize $h_1^{(1)}$};
\node at (1.5,-1.3) {\footnotesize $h_2^{(1)}$};
\node at (2.5,-2.3) {\footnotesize $h_3^{(1)}$};
\node at (1.35,-0.5) {\footnotesize $v_1^{(1)}$};
\node at (2.35,-1.5) {\footnotesize $v_2^{(1)}$};
\node at (3.35,-2.5) {\footnotesize $v_3^{(1)}$};
\node at (0.45,-1.1) {\footnotesize $m_1^{(1)}$};
\node at (1.45,-2.1) {\footnotesize $m_2^{(1)}$};
\node at (3,-3.5) {\footnotesize $m_3^{(1)}$};
\draw[<->,green!50!black] (1.5,0.4) -- (2.4,-0.5);
\node[green!50!black] at (2.15,0.15) {\footnotesize $\widehat{a}_1$};
\draw[<->,green!50!black] (2.5,-0.6) -- (3.4,-1.5);
\node[green!50!black] at (3.15,-0.85) {\footnotesize $\widehat{a}_2$};
\draw[<->,green!50!black] (3.8,-1.55) -- (3.8,-2.95);
\node[green!50!black] at (4,-2.3) {\footnotesize $\tau$};
\draw[<->,green!50!black] (0.85,-0.95) -- (0.85,-0.05);
\node[green!50!black] at (0.65,-0.5) {\footnotesize $S$};
\draw[<->,green!50!black] (1.9,0.9) -- (4.9,-2.1);
\node[green!50!black] at (3.45,-0.35) {\footnotesize{\bf $\rho$}};
\end{tikzpicture}}
&&
\parbox{5.5cm}{\begin{tikzpicture}
\draw[ultra thick] (0,0) -- (1,0) -- (1,-1) -- (2,-1) -- (2,-2) -- (3,-2) -- (3,-3) -- (4,-3);
\node at (-0.2,0) {\small $a$};
\node at (4.2,-3) {\small $a$};
\draw[ultra thick] (1,0) -- (1.5,0.5);
\node at (1.65,0.65) {\small $3$};
\draw[ultra thick] (2,-1) -- (2.5,-0.5);
\node at (2.65,-0.35) {\small $2$};
\draw[ultra thick] (3,-2) -- (3.5,-1.5);
\node at (3.65,-1.35) {\small $1$};
\draw[ultra thick] (1,-1) -- (0.5,-1.5);
\node at (0.35,-1.65) {\small $3$};
\draw[ultra thick] (2,-2) -- (1.5,-2.5);
\node at (1.35,-2.65) {\small $2$};
\draw[ultra thick] (3,-3) -- (2.5,-3.5);
\node at (2.35,-3.65) {\small $1$};
\draw[blue,fill=blue] (1,-1) circle (0.1cm);
\node[scale=0.7,blue] at (1.4,-0.8) {\footnotesize{\bf $(3,1)$}};
\draw[blue,fill=blue] (2,-2) circle (0.1cm);
\node[scale=0.7,blue] at (2.4,-1.8) {\footnotesize{\bf $(2,1)$}};
\draw[blue,fill=blue] (3,-3) circle (0.1cm);
\node[scale=0.7,blue] at (3.4,-2.8) {\footnotesize{\bf $(1,1)$}};
\node at (0.5,0.3) {\footnotesize $h_1^{(1)}$};
\node at (1.5,-1.3) {\footnotesize $h_3^{(1)}$};
\node at (2.5,-2.3) {\footnotesize $h_2^{(1)}$};
\node at (1.35,-0.3) {\footnotesize $v_3^{(1)}$};
\node at (2.35,-1.3) {\footnotesize $v_2^{(1)}$};
\node at (3.35,-2.3) {\footnotesize $v_1^{(1)}$};
\node at (0.45,-1.1) {\footnotesize $m_3^{(1)}$};
\node at (1.45,-2.1) {\footnotesize $m_2^{(1)}$};
\node at (3,-3.5) {\footnotesize $m_1^{(1)}$};
\draw[<->,green!50!black] (1.5,0.4) -- (2.4,-0.5);
\node[green!50!black] at (2.15,0.15) {\footnotesize $\widehat{a}'_1$};
\draw[<->,green!50!black] (2.5,-0.6) -- (3.4,-1.5);
\node[green!50!black] at (3.15,-0.85) {\footnotesize $\widehat{a}'_2$};
\draw[<->,green!50!black] (3.8,-1.55) -- (3.8,-2.95);
\node[green!50!black] at (4,-2.3) {\footnotesize $\tau'$};
\draw[<->,green!50!black] (0.85,-0.95) -- (0.85,-0.05);
\node[green!50!black] at (0.6,-0.5) {\footnotesize $S'$};
\draw[<->,green!50!black] (1.9,0.9) -- (4.9,-2.1);
\node[green!50!black] at (3.5,-0.3) {\footnotesize{\bf $\rho'$}};
\end{tikzpicture}}\end{align}
Visually, the right diagram is obtained from the left one by a rotation by 180 degrees\footnote{Notice also that the vertices are simply rotated, since the cyclic order of $(h_i^{(1)},v_i^{(1)},m_i^{(1)})$ compared to the left diagram remains preserved.} and therefore a priori describes the same point in the moduli space of $X_{3,1}$. However, $(\widehat{a}_1,\widehat{a}_2,\tau,S,\rho)$ and $(\widehat{a}'_1,\widehat{a}'_2,\tau',S',\rho')$ constitute two different basis of this moduli space, which are related by a linear transformation
\begin{align}
&\mathcal{S}:\,\begin{psmall} \widehat{a}_1\\ \widehat{a}_2 \\ \tau \\ S \\ \rho\end{psmall}\longrightarrow \begin{psmall} \widehat{a}'_1\\ \widehat{a}'_2\\ \tau' \\ S' \\ \rho'\end{psmall}=\mathcal{S}\cdot\begin{psmall} \widehat{a}_1\\ \widehat{a}_2 \\ \tau \\ S \\ \rho\end{psmall}\,,&&\text{with} && \mathcal{S}= \begin{psmall} -1 & -1 & 0 & 0 & 1 \\ 0 & 1 & 0 & 0 & 0 \\ 0 & 0 & 1 & 0 & 0 \\ 0 & 0 & 0 & 1 & 0 \\ 0 & 0 & 0 & 0 & 1  \end{psmall} \,.\label{TrafoRotationToy}
\end{align}
The transformation $\mathfrak{s}$ maps two inequivalent points in $\mathbb{K}(X_{3,1})$ which leaves $\mathcal{Z}_{3,1}$ invariant.
\item[$\bullet$]  \emph{re-orientation:} Since the web diagrams are drawn on a torus, we can use $SL(2,\mathbb{Z})$ transformations to simultaneously change the orientations of all lines. A simple example how to use this freedom to deduce a similarity transformation is shown in the following Figures
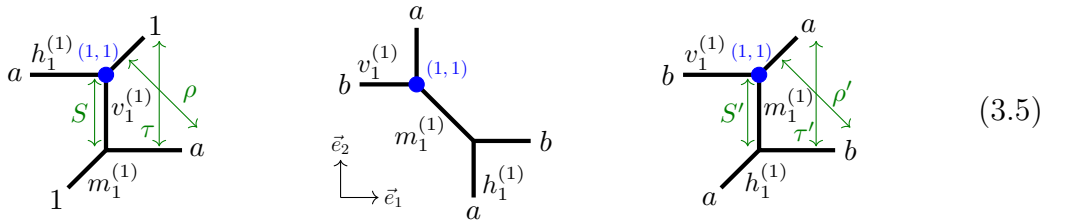
\begin{align}
&\parbox{3cm}{\begin{tikzpicture}
\draw[ultra thick] (0,0) -- (1,0) -- (1,-1) -- (2,-1);
\node at (-0.2,0) {\small $a$};
\node at (2.2,-1) {\small $a$};
\draw[ultra thick] (1,0) -- (1.5,0.5);
\node at (1.65,0.65) {\small $1$};
\draw[ultra thick] (1,-1) -- (0.5,-1.5);
\node at (0.35,-1.65) {\small $1$};
\draw[blue,fill=blue] (1,0) circle (0.1cm);
\node[scale=0.7,blue] at (0.9,0.3) {\footnotesize{\bf $(1,1)$}};
\node at (0.3,0.3) {\footnotesize $h_1^{(1)}$};
\node at (1.35,-0.4) {\footnotesize $v_1^{(1)}$};
\node at (1.1,-1.4) {\footnotesize $m_1^{(1)}$};
\draw[<->,green!50!black] (1.3,0.2) -- (2.2,-0.7);
\node[green!50!black] at (2.1,-0.25) {\footnotesize $\rho$};
\draw[<->,green!50!black] (1.7,0.45) -- (1.7,-0.95);
\node[green!50!black] at (1.55,-0.75) {\footnotesize $\tau$};
\draw[<->,green!50!black] (0.85,-0.95) -- (0.85,-0.05);
\node[green!50!black] at (0.65,-0.5) {\footnotesize $S$};
\end{tikzpicture}}
&&\parbox{3cm}{\begin{tikzpicture}
\draw[ultra thick] (0.25,0) -- (1,0) -- (1.75,-0.75) -- (2.5,-0.75);
\node at (0.05,0) {\small $b$};
\node at (2.7,-0.75) {\small $b$};
\draw[ultra thick] (1,0) -- (1,0.75);
\node at (1,0.95) {\small $a$};
\draw[ultra thick] (1.75,-0.75) -- (1.75,-1.5);
\node at (1.75,-1.7) {\small $a$};
\draw[blue,fill=blue] (1,0) circle (0.1cm);
\node[scale=0.7,blue] at (1.4,0.2) {\footnotesize{\bf $(1,1)$}};
\node at (0.5,0.3) {\footnotesize $v_1^{(1)}$};
\node at (1.05,-0.65) {\footnotesize $m_1^{(1)}$};
\node at (2.15,-1.3) {\footnotesize $h_1^{(1)}$};
\draw[->] (0,-1.5) -- (0.5,-1.5);
\draw[->] (0,-1.5) -- (0,-1);
\node[scale=0.7] at (0.7,-1.5) {$\vec{e}_1$};
\node[scale=0.7] at (0,-0.8) {$\vec{e}_2$};
\end{tikzpicture}}
&\parbox{3cm}{\begin{tikzpicture}
\draw[ultra thick] (0,0) -- (1,0) -- (1,-1) -- (2,-1);
\node at (-0.2,0) {\small $b$};
\node at (2.2,-1) {\small $b$};
\draw[ultra thick] (1,0) -- (1.5,0.5);
\node at (1.65,0.65) {\small $a$};
\draw[ultra thick] (1,-1) -- (0.5,-1.5);
\node at (0.35,-1.65) {\small $a$};
\draw[blue,fill=blue] (1,0) circle (0.1cm);
\node[scale=0.7,blue] at (0.9,0.3) {\footnotesize{\bf $(1,1)$}};
\node at (0.3,0.3) {\footnotesize $v_1^{(1)}$};
\node at (1.4,-0.4) {\footnotesize $m_1^{(1)}$};
\node at (1.1,-1.4) {\footnotesize $h_1^{(1)}$};
\draw[<->,green!50!black] (1.3,0.2) -- (2.2,-0.7);
\node[green!50!black] at (2.1,-0.25) {\footnotesize $\rho'$};
\draw[<->,green!50!black] (1.75,0.45) -- (1.75,-0.95);
\node[green!50!black] at (1.6,-0.75) {\footnotesize $\tau'$};
\draw[<->,green!50!black] (0.85,-0.95) -- (0.85,-0.05);
\node[green!50!black] at (0.65,-0.5) {\footnotesize $S'$};
\end{tikzpicture}}
\end{align}
All three diagrams represent the same point in the moduli space of $X_{1,1}$. Visually, the middle diagram can be obtained from the leftmost diagram through rotations and cutting-and-regluing procedures as explained above. The rightmost diagram is obtained by re-orienting all lines of the diagram. This corresponds to an $SL(2,\mathbb{Z})$ transformation of the form $\begin{psmall} 1 & 1\\ 0 & 1\end{psmall}$ which acts through matrix multiplication on the unit vectors $(\vec{e}_1,\vec{e}_2)$ along which the lines of the diagram are oriented. Notice that the orientation of the vertex remains unchanged by this procedure. The parameters $(\tau,S,\rho)$ and $(\tau',S',\rho')$, however, are two different bases of the K\"ahler moduli space of $X_{1,1}$. The equivalence of the left- and rightmost diagram therefore suggests the symmetry transformation
\begin{align}
&\mathfrak{s}:\,\begin{psmall} \tau \\ S \\ \rho\end{psmall}\longrightarrow \begin{psmall} \tau' \\ S' \\ \rho'\end{psmall}=\mathcal{S}\cdot\begin{psmall} \tau \\ S \\ \rho\end{psmall}\,,&&\text{with} && \mathcal{S}= \begin{psmall}1 & -2 & 1 \\ 1 & -1 & 0 \\ 1 & 0 & 0\end{psmall} \,.
\end{align}
which leaves the partition function $\mathcal{Z}_{1,1}$ invariant.
\item[$\bullet$] \emph{mirroring:} Another possibility is to mirror the web diagram , which changes the orientation of the vertices. A simple example is given by the sequence of the following diagrams 
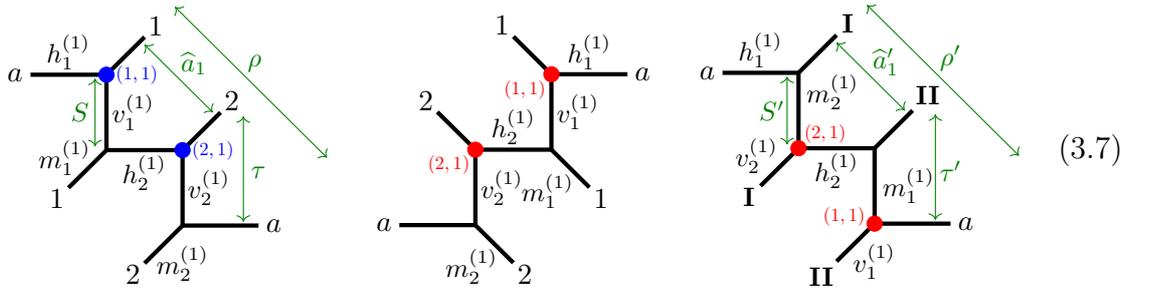
\begin{align}
&\parbox{4.5cm}{\begin{tikzpicture}
\draw[ultra thick] (0,0) -- (1,0) -- (1,-1) -- (2,-1) -- (2,-2) -- (3,-2);
\node at (-0.2,0) {\small $a$};
\node at (3.2,-2) {\small $a$};
\draw[ultra thick] (1,0) -- (1.5,0.5);
\node at (1.65,0.65) {\small $1$};
\draw[ultra thick] (2,-1) -- (2.5,-0.5);
\node at (2.65,-0.35) {\small $2$};
\draw[ultra thick] (1,-1) -- (0.5,-1.5);
\node at (0.35,-1.65) {\small $1$};
\draw[ultra thick] (2,-2) -- (1.5,-2.5);
\node at (1.35,-2.65) {\small $2$};
\node at (0.5,0.3) {\footnotesize $h_1^{(1)}$};
\node at (1.5,-1.3) {\footnotesize $h_2^{(1)}$};
\node at (1.35,-0.5) {\footnotesize $v_1^{(1)}$};
\node at (2.35,-1.5) {\footnotesize $v_2^{(1)}$};
\node at (0.45,-1.1) {\footnotesize $m_1^{(1)}$};
\node at (2,-2.5) {\footnotesize $m_2^{(1)}$};
\draw[blue,fill=blue] (1,0) circle (0.1cm);
\node[scale=0.7,blue] at (1.4,0) {\footnotesize{\bf $(1,1)$}};
\draw[blue,fill=blue] (2,-1) circle (0.1cm);
\node[scale=0.7,blue] at (2.4,-1) {\footnotesize{\bf $(2,1)$}};
\draw[<->,green!50!black] (1.5,0.4) -- (2.4,-0.5);
\node[green!50!black] at (2.15,0.15) {\footnotesize $\widehat{a}_1$};
\draw[<->,green!50!black] (2.8,-0.55) -- (2.8,-1.95);
\node[green!50!black] at (3,-1.3) {\footnotesize $\tau$};
\draw[<->,green!50!black] (0.85,-0.95) -- (0.85,-0.05);
\node[green!50!black] at (0.65,-0.5) {\footnotesize $S$};
\draw[<->,green!50!black] (1.9,0.9) -- (3.9,-1.1);
\node[green!50!black] at (2.95,0.15) {\footnotesize{\bf $\rho$}};
\end{tikzpicture}}
&&
\parbox{3.9cm}{\begin{tikzpicture}
\draw[ultra thick] (1,1) -- (0.5,1.5);
\node at (0.35,1.65) {\small $2$};
\draw[ultra thick] (2,2) -- (1.5,2.5);
\node at (1.35,2.65) {\small $1$};
\draw[ultra thick] (0,0) -- (1,0) -- (1,1) -- (2,1) -- (2,2) -- (3,2);
\draw[ultra thick] (1,0) -- (1.5,-0.5);
\node at (1.65,-0.65) {\small $2$};
\draw[ultra thick] (2,1) -- (2.5,0.5);
\node at (2.65,0.35) {\small $1$};
\node at (-0.2,0) {\small $a$};
\node at (3.2,2) {\small $a$};
\draw[red,fill=red] (1,1) circle (0.1cm);
\node[scale=0.7,red] at (0.65,0.8) {\footnotesize{\bf $(2,1)$}};
\draw[red,fill=red] (2,2) circle (0.1cm);
\node[scale=0.7,red] at (1.65,1.8) {\footnotesize{\bf $(1,1)$}};
\node at (2.5,2.3) {\footnotesize $h_1^{(1)}$};
\node at (1.5,1.3) {\footnotesize $h_2^{(1)}$};
\node at (2.35,1.5) {\footnotesize $v_1^{(1)}$};
\node at (1.35,0.5) {\footnotesize $v_2^{(1)}$};
\node at (1.95,0.45) {\footnotesize $m_1^{(1)}$};
\node at (0.95,-0.55) {\footnotesize $m_2^{(1)}$};
\end{tikzpicture}}
&&
\parbox{4.6cm}{\begin{tikzpicture}
\draw[ultra thick] (0,0) -- (1,0) -- (1,-1) -- (2,-1) -- (2,-2) -- (3,-2);
\node at (-0.2,0) {\small $a$};
\node at (3.2,-2) {\small $a$};
\draw[ultra thick] (1,0) -- (1.5,0.5);
\node at (1.65,0.65) {\small {\footnotesize {\bf I}}};
\draw[ultra thick] (2,-1) -- (2.5,-0.5);
\node at (2.7,-0.3) {\footnotesize {\bf II}};
\draw[ultra thick] (1,-1) -- (0.5,-1.5);
\node at (0.35,-1.65) {\small {\footnotesize {\bf I}}};
\draw[ultra thick] (2,-2) -- (1.5,-2.5);
\node at (1.3,-2.7){\footnotesize {\bf II}};
\node at (0.5,0.3) {\footnotesize $h_1^{(1)}$};
\node at (1.5,-1.3) {\footnotesize $h_2^{(1)}$};
\node at (1.45,-0.25) {\footnotesize $m_2^{(1)}$};
\node at (2.45,-1.5) {\footnotesize $m_1^{(1)}$};
\node at (0.45,-1.1) {\footnotesize $v_2^{(1)}$};
\node at (2,-2.5) {\footnotesize $v_1^{(1)}$};
\draw[red,fill=red] (1,-1) circle (0.1cm);
\node[scale=0.7,red] at (1.35,-0.8) {\footnotesize{\bf $(2,1)$}};
\draw[red,fill=red] (2,-2) circle (0.1cm);
\node[scale=0.7,red] at (1.6,-1.9) {\footnotesize{\bf $(1,1)$}};
\draw[<->,green!50!black] (1.5,0.4) -- (2.4,-0.5);
\node[green!50!black] at (2.15,0.15) {\footnotesize $\widehat{a}'_1$};
\draw[<->,green!50!black] (2.8,-0.55) -- (2.8,-1.95);
\node[green!50!black] at (3,-1.3) {\footnotesize $\tau'$};
\draw[<->,green!50!black] (0.85,-0.95) -- (0.85,-0.05);
\node[green!50!black] at (0.65,-0.5) {\footnotesize $S'$};
\draw[<->,green!50!black] (1.9,0.9) -- (3.9,-1.1);
\node[green!50!black] at (3,0.2) {\footnotesize{\bf $\rho'$}};\end{tikzpicture}}
\label{MirroringExample}
\end{align}
Visually, the middle diagram is obtained by mirroring the leftmost diagram along a vertical axis. Notice that this changes the orientation of the vertices, which we shall indicate by changing their colour from blue to red:
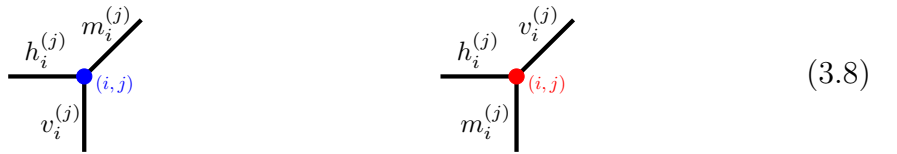
\begin{align}
&\scalebox{1}{\parbox{2cm}{\begin{tikzpicture}
\draw[ultra thick] (-1,0) -- (0,0) -- (0,-1); 
\draw[ultra thick] (0,0) -- (0.75,0.75); 
\draw[blue,fill=blue] (0,0) circle (0.1cm);
\node[scale=0.7,blue] at (0.4,-0.1) {\footnotesize{\bf $(i,j)$}};
\node at (-0.5,0.35) {\footnotesize $h_i^{(j)}$};
\node at (-0.3,-0.6) {\footnotesize $v_i^{(j)}$};
\node at (0.3,0.7) {\footnotesize $m_i^{(j)}$};
\end{tikzpicture}}}
&&
\scalebox{1}{\parbox{2cm}{\begin{tikzpicture}
\draw[ultra thick] (-1,0) -- (0,0) -- (0,-1); 
\draw[ultra thick] (0,0) -- (0.75,0.75); 
\draw[red,fill=red] (0,0) circle (0.1cm);
\node[scale=0.7,red] at (0.4,-0.1) {\footnotesize{\bf $(i,j)$}};
\node at (-0.5,0.35) {\footnotesize $h_i^{(j)}$};
\node at (-0.4,-0.6) {\footnotesize $m_i^{(j)}$};
\node at (0.3,0.7) {\footnotesize $v_i^{(j)}$};
\end{tikzpicture}}}\label{OrientationVertices}
\end{align}
The rightmost diagram in (\ref{MirroringExample}) is obtained by a re-orientation of the middle one. The leftmost and rightmost diagrams in (\ref{MirroringExample}) still represent the same point in $\mathbb{K}(X_{2,1})$, however, $(\widehat{a}_1,\tau,S,\rho)$ and $(\widehat{a}'_1,\tau',S',\rho')$ constitute two different (but equivalent) bases of the K\"ahler moduli space. The equivalence of the leftmost and rightmost diagram in (\ref{MirroringExample}) therefore implies a (linear) symmetry transformation 
\begin{align}
&\mathfrak{s}:\,\begin{psmall} \widehat{a}_1\\\tau \\ S \\\rho\end{psmall}\longrightarrow \begin{psmall} \widehat{a}'_1\\\tau' \\ S' \\ \rho'\end{psmall}=\mathcal{S}\cdot\begin{psmall} \widehat{a}_1\\ \tau \\ S \\ \rho\end{psmall}\,,&&\text{with} && \mathcal{S}= \begin{psmall} 1 & 1 & -2 & 0 \\ 0 & 1 & 0 & 0 \\ 0 & 1 & -1 & 0 \\ 0 & 2 & -4 & 1\end{psmall} \,.
\end{align}
$\mathfrak{s}$ maps two \emph{inequivalent} points in $\mathbb{K}(X_{2,1})$, however, in a way which leaves the full partition function $\mathcal{Z}_{2,1}$ invariant.   
\end{itemize}

\item[\emph{(ii)}] \emph{symmetries relating points in different K\"ahler cones:} There are two different ways of using the extended K\"ahler moduli space of $X_{N,M}$ to establish new symmetries.
\begin{itemize}
\item[$\bullet$] \emph{finding an alternative presentation of the web diagram which allows for a similarity transformation to a point in the extended moduli space of $X_{N,M}$:} A simple example of this type is given by the two diagrams below
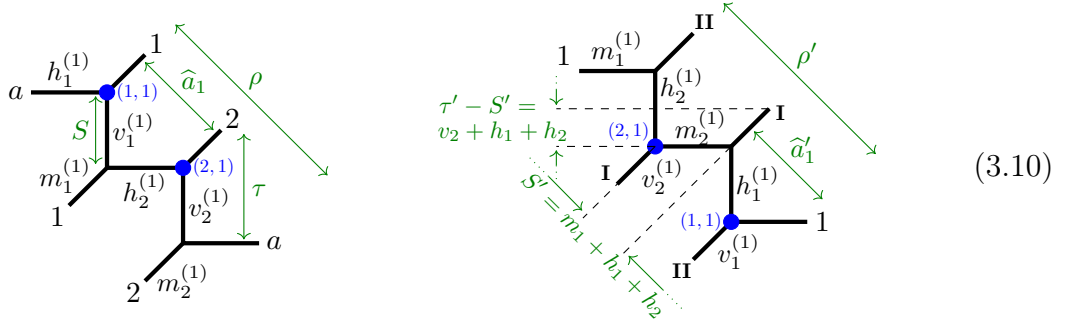
\begin{align}
&\parbox{4.5cm}{\begin{tikzpicture}
\draw[ultra thick] (0,0) -- (1,0) -- (1,-1) -- (2,-1) -- (2,-2) -- (3,-2);
\node at (-0.2,0) {\small $a$};
\node at (3.2,-2) {\small $a$};
\draw[ultra thick] (1,0) -- (1.5,0.5);
\node at (1.65,0.65) {\small $1$};
\draw[ultra thick] (2,-1) -- (2.5,-0.5);
\node at (2.65,-0.35) {\small $2$};
\draw[ultra thick] (1,-1) -- (0.5,-1.5);
\node at (0.35,-1.65) {\small $1$};
\draw[ultra thick] (2,-2) -- (1.5,-2.5);
\node at (1.35,-2.65) {\small $2$};
\node at (0.5,0.3) {\footnotesize $h_1^{(1)}$};
\node at (1.5,-1.3) {\footnotesize $h_2^{(1)}$};
\node at (1.35,-0.5) {\footnotesize $v_1^{(1)}$};
\node at (2.35,-1.5) {\footnotesize $v_2^{(1)}$};
\node at (0.45,-1.1) {\footnotesize $m_1^{(1)}$};
\node at (2,-2.5) {\footnotesize $m_2^{(1)}$};
\draw[blue,fill=blue] (1,0) circle (0.1cm);
\node[scale=0.7,blue] at (1.4,0) {\footnotesize{\bf $(1,1)$}};
\draw[blue,fill=blue] (2,-1) circle (0.1cm);
\node[scale=0.7,blue] at (2.4,-1) {\footnotesize{\bf $(2,1)$}};
\draw[<->,green!50!black] (1.5,0.4) -- (2.4,-0.5);
\node[green!50!black] at (2.15,0.15) {\footnotesize $\widehat{a}_1$};
\draw[<->,green!50!black] (2.8,-0.55) -- (2.8,-1.95);
\node[green!50!black] at (3,-1.3) {\footnotesize $\tau$};
\draw[<->,green!50!black] (0.85,-0.95) -- (0.85,-0.05);
\node[green!50!black] at (0.65,-0.5) {\footnotesize $S$};
\draw[<->,green!50!black] (1.9,0.9) -- (3.9,-1.1);
\node[green!50!black] at (2.95,0.15) {\footnotesize{\bf $\rho$}};
\end{tikzpicture}}
&&
\parbox{6cm}{\begin{tikzpicture}
\draw[ultra thick] (0,0) -- (1,0) -- (1,-1) -- (2,-1) -- (2,-2) -- (3,-2);
\node at (-0.2,0.2) {\small $1$};
\node at (3.2,-2) {\small $1$};
\draw[ultra thick] (1,0) -- (1.5,0.5);
\node[scale=0.8] at (1.65,0.65) {\footnotesize {\bf II}};
\draw[ultra thick] (2,-1) -- (2.5,-0.5);
\node[scale=0.8] at (2.65,-0.55) {\footnotesize {\bf I}};
\draw[ultra thick] (1,-1) -- (0.5,-1.5);
\node[scale=0.8] at (0.35,-1.35) {\footnotesize {\bf I}};
\draw[ultra thick] (2,-2) -- (1.5,-2.5);
\node[scale=0.8] at (1.35,-2.65) {\footnotesize {\bf II}};
\draw[blue,fill=blue] (1,-1) circle (0.1cm);
\node[scale=0.7,blue] at (0.65,-0.8) {\footnotesize{\bf $(2,1)$}};
\draw[blue,fill=blue] (2,-2) circle (0.1cm);
\node[scale=0.7,blue] at (1.6,-2) {\footnotesize{\bf $(1,1)$}};
\node at (0.5,0.3) {\footnotesize $m_1^{(1)}$};
\node at (1.6,-0.75) {\footnotesize $m_2^{(1)}$};
\node at (1.35,-0.2) {\footnotesize $h_2^{(1)}$};
\node at (2.35,-1.5) {\footnotesize $h_1^{(1)}$};
\node at (1.1,-1.4) {\footnotesize $v_2^{(1)}$};
\node at (2.1,-2.4) {\footnotesize $v_1^{(1)}$};
\draw[<->,green!50!black] (2.3,-0.8) -- (3.2,-1.7);
\node[green!50!black] at (2.95,-1.05) {\footnotesize $\widehat{a}'_1$};
%
\draw[dashed] (1,-1) -- (0.05,-1.95);
\draw[green!50!black,<-] (0.04,-1.85) -- (-0.4,-1.4);
\draw[green!50!black,dotted] (-0.65,-1.15) -- (-0.4,-1.4);
\draw[dashed] (2,-1) -- (0.55,-2.45);
\draw[green!50!black,<-] (0.65,-2.45) -- (1.1,-2.9);
\draw[green!50!black,dotted] (1.1,-2.9) -- (1.35,-3.15);
\node[rotate=-45,green!50!black,scale=0.9] at (0.2,-2.3) {\footnotesize $S'=m_1+h_1+h_2$};
\draw[dashed] (1,-1) -- (-0.3,-1);
\draw[dashed] (2.5,-0.5) -- (-0.3,-0.5);
\draw[green!50!black,<-] (-0.3,-0.45) -- (-0.3,-0.25);
\draw[green!50!black,dotted] (-0.3,-0.25) -- (-0.3,0);
\draw[green!50!black,<-] (-0.3,-1.05) -- (-0.3,-1.25);
\draw[green!50!black,dotted] (-0.3,-1.25) -- (-0.3,-1.5);
\node[green!50!black,scale=0.9] at (-1.2,-0.45) {\footnotesize $\tau'-S'=$};
\node[green!50!black,scale=0.9] at (-1,-0.8) {\footnotesize $v_2+h_1+h_2$};
\draw[<->,green!50!black] (1.9,0.9) -- (3.9,-1.1);
\node[green!50!black] at (3,0.2) {\footnotesize{\bf $\rho'$}};\end{tikzpicture}}
\end{align}
Both diagrams describe the same point in $\mathbb{K}(X_{2,1})$. However, the right diagram can also be interpreted as a shifted web diagram and thus represents a point in a different K\"ahler cone of the extended moduli space $\mathbb{EK}(X_{2,1})$, parametrised by the basis $(\widehat{a}_1',\tau',S',\rho')$ as indicated in the Figure. The equivalence of the two diagrams therefore implies the following symmetry transformation
\begin{align}
&\mathfrak{s}:\,\begin{psmall} \widehat{a}_1\\\tau \\ S \\ \rho\end{psmall}\longrightarrow \begin{psmall} \widehat{a}'_1\\\tau' \\ S' \\ \rho'\end{psmall}=\mathcal{S}\cdot\begin{psmall} \widehat{a}_1\\\tau \\ S \\ \rho\end{psmall}\,,&&\text{with} && \mathcal{S}= \begin{psmall}-1 & 1 & -2 & 1 \\ 0 & 1 & -4 & 2 \\ 0 & 1 & -3 & 1 \\ 0 & 2 & -4 & 1\end{psmall} \,.
\end{align}
which leaves the partition function $\mathcal{Z}_{2,1}$ invariant.
\item[$\bullet$] \emph{flop transitions:} Instead of searching for an alternative presentation of the web diagram, we may consider a (flop) transition to another K\"ahler cone in $\mathbb{EK}(X_{N,M})$. A simple example of this strategy is showcased by the diagrams below
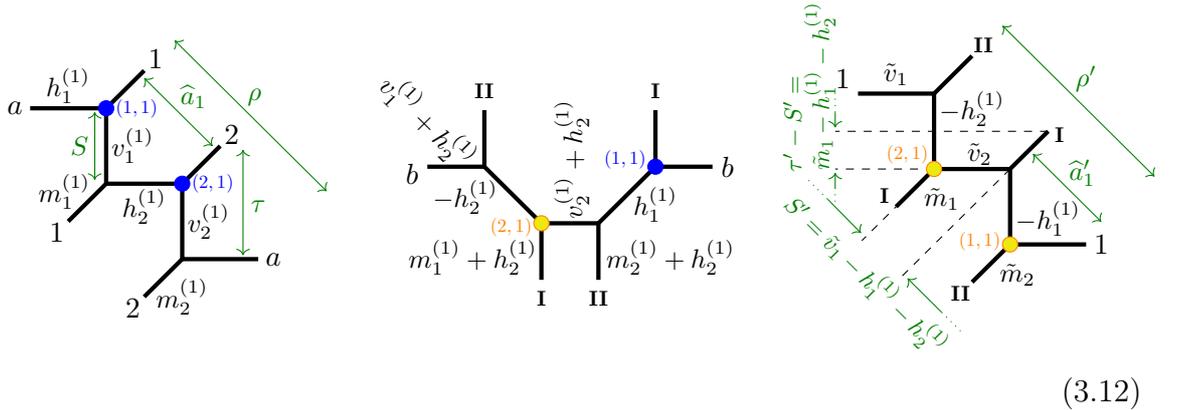
\begin{align}
&\parbox{4.4cm}{\begin{tikzpicture}
\draw[ultra thick] (0,0) -- (1,0) -- (1,-1) -- (2,-1) -- (2,-2) -- (3,-2);
\node at (-0.2,0) {\small $a$};
\node at (3.2,-2) {\small $a$};
\draw[ultra thick] (1,0) -- (1.5,0.5);
\node at (1.65,0.65) {\small $1$};
\draw[ultra thick] (2,-1) -- (2.5,-0.5);
\node at (2.65,-0.35) {\small $2$};
\draw[ultra thick] (1,-1) -- (0.5,-1.5);
\node at (0.35,-1.65) {\small $1$};
\draw[ultra thick] (2,-2) -- (1.5,-2.5);
\node at (1.35,-2.65) {\small $2$};
\node at (0.5,0.3) {\footnotesize $h_1^{(1)}$};
\node at (1.5,-1.3) {\footnotesize $h_2^{(1)}$};
\node at (1.35,-0.5) {\footnotesize $v_1^{(1)}$};
\node at (2.35,-1.5) {\footnotesize $v_2^{(1)}$};
\node at (0.45,-1.1) {\footnotesize $m_1^{(1)}$};
\node at (2,-2.5) {\footnotesize $m_2^{(1)}$};
\draw[blue,fill=blue] (1,0) circle (0.1cm);
\node[scale=0.7,blue] at (1.4,0) {\footnotesize{\bf $(1,1)$}};
\draw[blue,fill=blue] (2,-1) circle (0.1cm);
\node[scale=0.7,blue] at (2.4,-1) {\footnotesize{\bf $(2,1)$}};
\draw[<->,green!50!black] (1.5,0.4) -- (2.4,-0.5);
\node[green!50!black] at (2.15,0.15) {\footnotesize $\widehat{a}_1$};
\draw[<->,green!50!black] (2.8,-0.55) -- (2.8,-1.95);
\node[green!50!black] at (3,-1.3) {\footnotesize $\tau$};
\draw[<->,green!50!black] (0.85,-0.95) -- (0.85,-0.05);
\node[green!50!black] at (0.65,-0.5) {\footnotesize $S$};
\draw[<->,green!50!black] (1.9,0.9) -- (3.9,-1.1);
\node[green!50!black] at (2.95,0.15) {\footnotesize{\bf $\rho$}};
\end{tikzpicture}}
&&
\parbox{5cm}{\begin{tikzpicture}
\draw[ultra thick] (0,0) -- (0.75,0) -- (1.5,-0.75) -- (2.25,-0.75) -- (3,0) -- (3.75,0);
\node at (-0.2,-0.1) {\small $b$};
\node at (3.95,0) {\small $b$};
\draw[ultra thick] (0.75,0) -- (0.75,0.75);
\node[scale=0.8] at (0.75,1) {\footnotesize {\bf II}};
\draw[ultra thick] (3,0) -- (3,0.75);
\node[scale=0.8] at (3,1) {\footnotesize {\bf I}};
\draw[ultra thick] (1.5,-0.75) -- (1.5,-1.5);
\node[scale=0.8] at (1.5,-1.75) {\footnotesize {\bf I}};
\draw[ultra thick] (2.25,-0.75) -- (2.25,-1.5);
\node[scale=0.8] at (2.25,-1.75) {\footnotesize {\bf II}};
\node[rotate=-45] at (0,0.6) {\footnotesize $v_1^{(1)}+h_2^{(1)}$};
\node[rotate=90] at (1.9,0.1) {\footnotesize $v_2^{(1)}+h_2^{(1)}$};
\node at (0.5,-0.4) {\footnotesize $-h_2^{(1)}$};
\node at (3,-0.5) {\footnotesize $h_1^{(1)}$};
\node at (0.6,-1.2) {\footnotesize $m_1^{(1)}+h_2^{(1)}$};
\node at (3.2,-1.2) {\footnotesize $m_2^{(1)}+h_2^{(1)}$};
\draw[blue,fill=blue] (3,0) circle (0.1cm);
\node[scale=0.7,blue] at (2.6,0.1) {\footnotesize{\bf $(1,1)$}};
\draw[orange,fill=yellow!90!green] (1.5,-0.75) circle (0.1cm);
\node[scale=0.7,orange] at (1.1,-0.8) {\footnotesize{\bf $(2,1)$}};
\end{tikzpicture}}
&&
\parbox{5.1cm}{\begin{tikzpicture}
\draw[ultra thick] (0,0) -- (1,0) -- (1,-1) -- (2,-1) -- (2,-2) -- (3,-2);
\node at (-0.2,0.2) {\small $1$};
\node at (3.2,-2) {\small $1$};
\draw[ultra thick] (1,0) -- (1.5,0.5);
\node[scale=0.8] at (1.65,0.65) {\footnotesize {\bf II}};
\draw[ultra thick] (2,-1) -- (2.5,-0.5);
\node[scale=0.8] at (2.65,-0.55) {\footnotesize {\bf I}};
\draw[ultra thick] (1,-1) -- (0.5,-1.5);
\node[scale=0.8] at (0.35,-1.35) {\footnotesize {\bf I}};
\draw[ultra thick] (2,-2) -- (1.5,-2.5);
\node[scale=0.8] at (1.35,-2.65) {\footnotesize {\bf II}};
\node at (0.5,0.25) {\footnotesize $\tilde{v}_1$};
\node at (1.6,-0.8) {\footnotesize $\tilde{v}_2$};
\node at (1.5,-0.2) {\footnotesize $-h_2^{(1)}$};
\node at (2.5,-1.65) {\footnotesize $-h_1^{(1)}$};
\node at (1.1,-1.4) {\footnotesize $\tilde{m}_1$};
\node at (2.1,-2.4) {\footnotesize $\tilde{m}_2$};
\draw[<->,green!50!black] (2.3,-0.8) -- (3.2,-1.7);
\node[green!50!black] at (2.95,-1.05) {\footnotesize $\widehat{a}'_1$};
%
\draw[dashed] (1,-1) -- (0.05,-1.95);
\draw[green!50!black,<-] (0.04,-1.85) -- (-0.4,-1.4);
\draw[green!50!black,dotted] (-0.65,-1.15) -- (-0.4,-1.4);
\draw[dashed] (2,-1) -- (0.55,-2.45);
\draw[green!50!black,<-] (0.65,-2.45) -- (1.1,-2.9);
\draw[green!50!black,dotted] (1.1,-2.9) -- (1.35,-3.15);
\node[rotate=-45,green!50!black,scale=0.9] at (0.1,-2.4) {\footnotesize $S'=\tilde{v}_1-h_1^{(1)}-h_2^{(1)}$};
\draw[dashed] (1,-1) -- (-0.3,-1);
\draw[dashed] (2.5,-0.5) -- (-0.3,-0.5);
\draw[green!50!black,<-] (-0.3,-0.45) -- (-0.3,-0.25);
\draw[green!50!black,dotted] (-0.3,-0.25) -- (-0.3,0);
\draw[green!50!black,<-] (-0.3,-1.05) -- (-0.3,-1.25);
\draw[green!50!black,dotted] (-0.3,-1.25) -- (-0.3,-1.5);
\node[green!50!black,scale=0.9,rotate=90] at (-0.85,-0.4) {\footnotesize $\tau'-S'=$};
\node[green!50!black,scale=0.9,rotate=90] at (-0.5,0.1) {\footnotesize $\tilde{m}_1-h^{(1)}_1-h^{(1)}_2$};
\draw[<->,green!50!black] (1.9,0.9) -- (3.9,-1.1);
\node[green!50!black] at (3,0.2) {\footnotesize{\bf $\rho'$}};
\draw[orange,fill=yellow!90!green] (1,-1) circle (0.1cm);
\node[scale=0.7,orange] at (0.65,-0.8) {\footnotesize{\bf $(2,1)$}};
\draw[orange,fill=yellow!90!green] (2,-2) circle (0.1cm);
\node[scale=0.7,orange] at (1.6,-1.95) {\footnotesize{\bf $(1,1)$}};
\end{tikzpicture}}\label{FlopTransition}
\end{align}
with the shorthand notation
\begin{align}
&\tilde{v}_i=v_i^{(1)}+h_1^{(1)}+h_2^{(1)}\,,&&\tilde{m}_i=m_i^{(1)}+h_1^{(1)}+h_2^{(1)}\,,&&\forall i\in\{1,2\}\,.\label{ExampleFlopAreas}
\end{align}
Starting from the leftmost diagram in (\ref{FlopTransition}), the middle diagram represents the geometry after a flop of the line labelled $h_2^{(1)}$, while the rightmost diagram after flopping the two lines labelled $h_{1,2}^{(1)}$. After each flop, the vertices (and areas (\ref{ExampleFlopAreas})) are modified according to (\ref{FlopGenericDef}). The rightmost diagram can be interpreted as a shifted web diagram in a different K\"ahler cone of $\mathbb{EK}(X_{2,1})$, parametrised by the basis $(\widehat{a}_1',\tau',S',\rho')$. The latter is trivially related to the original basis
\begin{align}
&\mathfrak{s}:\,\begin{psmall} \widehat{a}_1\\\tau \\ S \\ \rho\end{psmall}\longrightarrow \begin{psmall} \widehat{a}'_1\\\tau' \\ S' \\ \rho'\\\end{psmall}=\begin{psmall} \widehat{a}_1\\\tau \\ S \\\rho\end{psmall}\,.
\end{align}
In fact, the choice of basis for the shifted web diagrams is such that this specific transformation acts as the unit transformation and thus leaves $\mathcal{Z}_{2,1}$ invariant.
\end{itemize}
\end{itemize}
Combining these various similarity transformations, we can find various different symmetries of the partition function $\mathcal{Z}_{N,M}$. In \cite{Bastian:2018jlf} the case $(N,1)$ was studied in detail and the symmetry transformations worked out systematically. In this way it was found that the free energy associated with the partition function $\mathcal{Z}_{N,1}$ is invariant under the symmetry group 
\begin{align}
&\widetilde{\mathbb{G}}(N)\cong \mathbb{G}(N)\times \widetilde{S}_N\,,&&\text{with} &&\mathbb{G}(N)\cong\text{Dih}_n\,,&&\text{for}&&n=\left\{\begin{array}{lcl}3 & \text{for} & N=1\,,\\2 & \text{for} & N=2\,,\\3 & \text{for} & N=3\,,\\\infty & \text{for} & N\geq 4\,. \end{array}\right.\label{DihedralGroup}
\end{align}
Concretely, $\mathbb{G}(N)$ is generated by two involutions for which the following form was found in~\cite{Bastian:2018jlf} 
\begin{align}
&\mathbb{G}(N)\cong\left\langle\left\{\mathcal{G}_2(N),\mathcal{G}'_2(N)|\left(\mathcal{G}_2(N)\right)^2=\left(\mathcal{G}_2(N)\right)^2=\left(\mathcal{G}_2(N)\cdot \mathcal{G}'_2(N)\right)^n=\mathbbm{1}\right\}\right\rangle\,,\label{M1Generators}
\end{align}
with $n$ as in (\ref{DihedralGroup}) and where a concrete matrix representation of $\mathcal{G}_2(N)$ and $\mathcal{G}'_2(N)$ was given.

\subsection{Generalisation for $M>1$}\label{Sect:GenSymTrans}
The similarity transformations of the previous subsection, which allow us to discover non-trivial

\begin{wrapfigure}{l}{0.45\textwidth}
${}$\\[-0.5cm]
\scalebox{0.99}{\parbox{7.1cm}{\begin{tikzpicture}
\draw[ultra thick] (1.75,1.75) -- (2.25,2.25);
\draw[ultra thick] (2.75,0.75) -- (3.25,1.25);
\draw[ultra thick] (3.75,-0.25) -- (4.25,0.25);
\draw[ultra thick] (4.75,-1.25) -- (5.25,-0.75);
\draw[ultra thick] (0.75,1.75) -- (1.75,1.75) -- (1.75,0.75) -- (2.75,0.75) -- (2.75,-0.25) -- (3.75,-0.25) -- (3.75,-1.25) -- (4.75,-1.25) -- (4.75,-2.25) -- (5.75,-2.25);
\draw[ultra thick] (1,0) -- (1.75,0.75);
\draw[ultra thick] (2,-1) -- (2.75,-0.25);
\draw[ultra thick] (3,-2) -- (3.75,-1.25);
\draw[ultra thick] (4,-3) -- (4.75,-2.25);
\draw[ultra thick] (0,0) -- (1,0) -- (1,-1) -- (2,-1) -- (2,-2) -- (3,-2) -- (3,-3) -- (4,-3) -- (4,-4) -- (5,-4);
\draw[ultra thick] (1,-1) -- (0.5,-1.5);
\draw[ultra thick] (2,-2) -- (1.5,-2.5);
\draw[ultra thick] (3,-3) -- (2.5,-3.5);
\draw[ultra thick] (4,-4) -- (3.5,-4.5);
\node at (0.55,1.75) {\footnotesize $a$}; 
\node at (5.95,-2.25) {\footnotesize $a$}; 
\node at (-0.05,0.25) {\footnotesize $b$}; 
\node at (5,-4.3) {\footnotesize $b$}; 
\node at (2.4,2.4) {\footnotesize $1$};
\node at (3.4,1.4) {\footnotesize $2$};
\node at (4.4,0.4) {\footnotesize $3$};
\node at (5.4,-0.6) {\footnotesize $4$};
\node at (0.35,-1.65) {\footnotesize $1$};
\node at (1.35,-2.65) {\footnotesize $2$};
\node at (2.35,-3.65) {\footnotesize $3$};
\node at (3.35,-4.65) {\footnotesize $4$};
\draw[<->,green!50!black] (2,1.9) -- (2.9,1);
\node[green!50!black] at (2.7,1.7) {\footnotesize $\widehat{a}_1^{(1)}$};
\draw[<->,green!50!black] (3,0.9) -- (3.9,0);
\node[green!50!black] at (3.7,0.7) {\footnotesize $\widehat{a}_2^{(1)}$};
\draw[<->,green!50!black] (4,-0.1) -- (4.9,-1);
\node[green!50!black] at (4.7,-0.3) {\footnotesize $\widehat{a}_3^{(1)}$};
\draw[<->,green!50!black] (1.3,0.2) -- (2.2,-0.7);
\node[green!50!black] at (2,0) {\footnotesize $\widehat{a}_1^{(2)}$};
\draw[<->,green!50!black] (2.3,-0.8) -- (3.2,-1.7);
\node[green!50!black] at (3,-1) {\footnotesize $\widehat{a}_2^{(2)}$};
\draw[<->,green!50!black] (3.3,-1.8) -- (4.2,-2.7);
\node[green!50!black] at (4,-2) {\footnotesize $\widehat{a}_3^{(2)}$};
\draw[dashed] (1.75,1.75) -- (-0.325,-0.325);
\draw[dashed] (5,-4) -- (5.5,-4);
\draw[<->,green!50!black] (2.7,2.7) -- (6.7,-1.3);
\node[green!50!black] at (4.75,0.95) {\footnotesize{\bf $\rho$}};
\draw[<->,green!50!black] (-0.25,-0.35) -- (0.65,-1.25);
\node[green!50!black] at (0.4,-0.65) {\footnotesize{\bf $S$}};
\draw[<->,green!50!black] (0.9,0.1) -- (0.9,1.65);
\node[green!50!black] at (0.65,1.3) {\footnotesize{\bf $\tau_1$}};
\draw[<->,green!50!black] (5.45,-0.85) -- (5.45,-3.9);
\node[green!50!black] at (5.25,-3) {\footnotesize{\bf $\tau$}};
\draw[blue,fill=blue] (1.75,1.75) circle (0.1cm);
\node[scale=0.7,blue] at (1.45,1.95) {\footnotesize{\bf $(1,1)$}};
\draw[blue,fill=blue] (2.75,0.75) circle (0.1cm);
\node[scale=0.7,blue] at (2.45,0.95) {\footnotesize{\bf $(2,1)$}};
\draw[blue,fill=blue] (3.75,-0.25) circle (0.1cm);
\node[scale=0.7,blue] at (3.45,-0.05) {\footnotesize{\bf $(3,1)$}};
\draw[blue,fill=blue] (4.75,-1.25) circle (0.1cm);
\node[scale=0.7,blue] at (4.45,-1.05) {\footnotesize{\bf $(4,1)$}};
\draw[blue,fill=blue] (1,0) circle (0.1cm);
\node[scale=0.7,blue] at (0.65,-0.2) {\footnotesize{\bf $(1,2)$}};
\draw[blue,fill=blue] (2,-1) circle (0.1cm);
\node[scale=0.7,blue] at (1.65,-1.2) {\footnotesize{\bf $(2,2)$}};
\draw[blue,fill=blue] (3,-2) circle (0.1cm);
\node[scale=0.7,blue] at (2.65,-2.2) {\footnotesize{\bf $(3,2)$}};
\draw[blue,fill=blue] (4,-3) circle (0.1cm);
\node[scale=0.7,blue] at (3.65,-3.2) {\footnotesize{\bf $(4,2)$}};
\end{tikzpicture}}}
\caption{\emph{Web diagram $(N,M)=(4,2)$.}}
\label{Fig:Web42}
${}$\\[-1.5cm]
\end{wrapfigure}
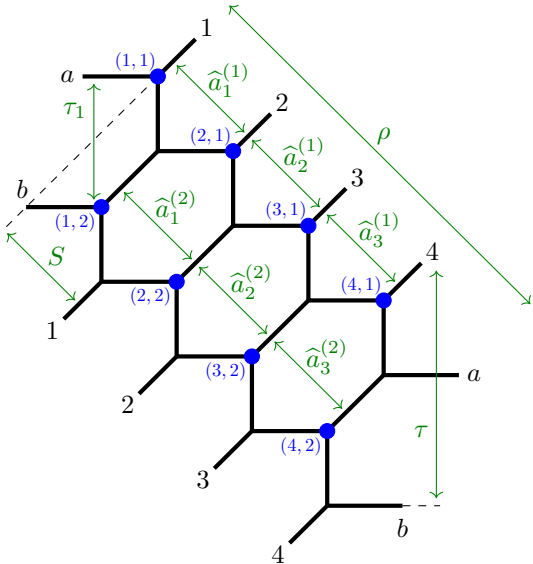

\noindent
symmetries of the LST partition function, can be generalised to the orbifold case $(N,M)$. As explained above, using the duality discovered in \cite{Hohenegger:2016yuv}, it is sufficient to consider cases with $N/M\in\mathbb{N}$. In the following we shall give an overview of various similarity transformations (following a similar organisation as in the previous Subsection) and exhibit them explicitly for the example $(4,2)$. For the reader's convenience, the webdiagram in this case is shown in Figure~\ref{Fig:Web42}: here we have also indicated a basis of the K\"ahler parameters (green lines) as well as a labelling of individual vertices, which shall allows us to keep track of individual curves throughout the various similarity transformations. In the following we shall furthermore explicitly determine the general form (\emph{i.e.} which is valid for the general case $(N,M)$) of the symmetry transformation $\mathfrak{s}$: as in the previous Subsection, $\mathfrak{s}$ acts on the basis of K\"ahler parameters (as showcased in Figure~\ref{Fig:General Setup} for the generic case $(N,M)$ and in Figure~\ref{Fig:Web42} for the case $(4,2)$) as a linear map 
\begin{align}
&\mathfrak{s}\,:\hspace{0.2cm}\vec{v}\longrightarrow \mathcal{S}\cdot \vec{v}\,,&&\text{with} &&\begin{array}{l}\vec{v}=(\vec{a}^{(1)},\ldots,\vec{a}^{(M)},\tau_1,\ldots,\tau_{M-1},\tau,S,\rho)^T\in \mathbb{R}^{NM+2}\,,\\
\vec{a}^{(k)}=(\widehat{a}_1^{(k)},\ldots,\widehat{a}_{N-1}^{(k)})^T\in\mathbb{R}^{N-1}\hspace{0.2cm}\forall k\in\{1,\ldots,M\}\,.\end{array}\label{VectorKaehlerModuli}
\end{align}
Here $\mathcal{S}$ is a $(NM+2)\times (NM+2)$ matrix with rational entries, which can be written in block form
\begin{align}
&\mathcal{S}=
\left(\begin{linesmall}{@{}ccc@{}}&&\\[-6pt]
\mathcal{S}_{\widehat{a}\widehat{a}} & \mathcal{S}_{\widehat{a}\tau} & \mathcal{S}_{\widehat{a}\omega}  \\[2pt]&&\\[-5pt] \mathcal{S}_{\tau\widehat{a}} & \mathcal{S}_{\tau\tau} & \mathcal{S}_{\tau\omega}\\[2pt] &&\\[-5pt] \mathcal{S}_{\omega\widehat{a}} & \mathcal{S}_{\omega\tau} & \mathcal{S}_{\omega\omega}
\end{linesmall}\right)\,,&&\text{with} &&\begin{array}{lll}\mathcal{S}_{\widehat{a}\widehat{a}}\in \mathbb{M}_{M(N-1),M(N-1)}\,, & \mathcal{S}_{\widehat{a}\tau}\in \mathbb{M}_{M(N-1),M-1}\,,& \mathcal{S}_{\widehat{a}\omega}\in \mathbb{M}_{M(N-1),3}\,, \\ \mathcal{S}_{\tau\widehat{a}}\in \mathbb{M}_{M-1,M(N-1)}\,, & \mathcal{S}_{\tau\tau}\in \mathbb{M}_{M-1,M-1}\,,& \mathcal{S}_{\tau\omega}\in \mathbb{M}_{M-1,3}\,, \\ \mathcal{S}_{\omega\widehat{a}}\in \mathbb{M}_{3,M(N-1)}\,, & \mathcal{S}_{\omega\tau}\in \mathbb{M}_{\omega,M-1}\,,& \mathcal{S}_{\omega\omega}\in \mathbb{M}_{3,3}\,,   \end{array}\nonumber
\end{align}
where $\mathbb{M}_{n,m}$ is the space of $n\times m$ matrices with rational entries. These matrices have been obtained by studying examples with low values of $(N,M)$ and generalised to higher values. We have checked them explicitly up to $(N,M)=(10,10)$. The matrices $\mathcal{S}$ can also be obtained with the Mathematica package {\tt NPLSTsym}, which is briefly described in Appendix~\ref{App:MathematicaPackage}.

\noindent
Generalising the similarity transformations discussed in the previous Subsection, we can organise them in the following manner 
\begin{itemize}
\item[\emph{(i)}] \emph{Symmetries relating points in the same K\"ahler cone of $X_{N,M}$}\\ 
As in the case of $(N,1)$, these correspond to simple cutting and re-gluing operations or re-orientations of the web diagram. Here we particularly mention three 
\begin{itemize}
\item cut and re-glue along horizontal lines $\mathfrak{c}^{\text{hor}}$:\\
The transformed web diagram for the case $(4,2)$ and the associated general symmetry transformation $\mathcal{C}^{\text{hor}}$ are as follows
\begin{align}
&\scalebox{0.72}{\parbox{7.1cm}{\begin{tikzpicture}
\draw[ultra thick] (1.75,1.75) -- (2.25,2.25);
\draw[ultra thick] (2.75,0.75) -- (3.25,1.25);
\draw[ultra thick] (3.75,-0.25) -- (4.25,0.25);
\draw[ultra thick] (4.75,-1.25) -- (5.25,-0.75);
\draw[ultra thick] (0.75,1.75) -- (1.75,1.75) -- (1.75,0.75) -- (2.75,0.75) -- (2.75,-0.25) -- (3.75,-0.25) -- (3.75,-1.25) -- (4.75,-1.25) -- (4.75,-2.25) -- (5.75,-2.25);
\draw[ultra thick] (1,0) -- (1.75,0.75);
\draw[ultra thick] (2,-1) -- (2.75,-0.25);
\draw[ultra thick] (3,-2) -- (3.75,-1.25);
\draw[ultra thick] (4,-3) -- (4.75,-2.25);
\draw[ultra thick] (0,0) -- (1,0) -- (1,-1) -- (2,-1) -- (2,-2) -- (3,-2) -- (3,-3) -- (4,-3) -- (4,-4) -- (5,-4);
\draw[ultra thick] (1,-1) -- (0.5,-1.5);
\draw[ultra thick] (2,-2) -- (1.5,-2.5);
\draw[ultra thick] (3,-3) -- (2.5,-3.5);
\draw[ultra thick] (4,-4) -- (3.5,-4.5);
\node at (0.55,1.75) {\footnotesize $a$}; 
\node at (5.95,-2.25) {\footnotesize $a$}; 
\node at (-0.05,0.25) {\footnotesize $b$}; 
\node at (5,-4.3) {\footnotesize $b$}; 
\node at (2.4,2.4) {\footnotesize $2$};
\node at (3.4,1.4) {\footnotesize $3$};
\node at (4.4,0.4) {\footnotesize $4$};
\node at (5.4,-0.6) {\footnotesize $1$};
\node at (0.35,-1.65) {\footnotesize $2$};
\node at (1.35,-2.65) {\footnotesize $3$};
\node at (2.35,-3.65) {\footnotesize $4$};
\node at (3.35,-4.65) {\footnotesize $1$};
\draw[<->,green!50!black] (2,1.9) -- (2.9,1);
\node[green!50!black] at (2.7,1.7) {\footnotesize $\widehat{a}_1^{(1)\prime}$};
\draw[<->,green!50!black] (3,0.9) -- (3.9,0);
\node[green!50!black] at (3.7,0.7) {\footnotesize $\widehat{a}_2^{(1)\prime}$};
\draw[<->,green!50!black] (4,-0.1) -- (4.9,-1);
\node[green!50!black] at (4.7,-0.3) {\footnotesize $\widehat{a}_3^{(1)\prime}$};
\draw[<->,green!50!black] (1.3,0.2) -- (2.2,-0.7);
\node[green!50!black] at (2,0) {\footnotesize $\widehat{a}_1^{(2)\prime}$};
\draw[<->,green!50!black] (2.3,-0.8) -- (3.2,-1.7);
\node[green!50!black] at (3,-1) {\footnotesize $\widehat{a}_2^{(2)\prime}$};
\draw[<->,green!50!black] (3.3,-1.8) -- (4.2,-2.7);
\node[green!50!black] at (4,-2) {\footnotesize $\widehat{a}_3^{(2)\prime}$};
\draw[dashed] (1.75,1.75) -- (-0.325,-0.325);
\draw[dashed] (5,-4) -- (5.5,-4);
\draw[<->,green!50!black] (2.7,2.7) -- (6.7,-1.3);
\node[green!50!black] at (4.75,0.95) {\footnotesize{\bf $\rho'$}};
\draw[<->,green!50!black] (-0.25,-0.35) -- (0.65,-1.25);
\node[green!50!black] at (0.4,-0.65) {\footnotesize{\bf $S'$}};
\draw[<->,green!50!black] (0.9,0.1) -- (0.9,1.65);
\node[green!50!black] at (0.65,1.3) {\footnotesize{\bf $\tau'_1$}};
\draw[<->,green!50!black] (5.45,-0.85) -- (5.45,-3.9);
\node[green!50!black] at (5.25,-3) {\footnotesize{\bf $\tau'$}};
\draw[blue,fill=blue] (1.75,1.75) circle (0.1cm);
\node[scale=0.7,blue] at (1.45,1.95) {\footnotesize{\bf $(2,1)$}};
\draw[blue,fill=blue] (2.75,0.75) circle (0.1cm);
\node[scale=0.7,blue] at (2.45,0.95) {\footnotesize{\bf $(3,1)$}};
\draw[blue,fill=blue] (3.75,-0.25) circle (0.1cm);
\node[scale=0.7,blue] at (3.45,-0.05) {\footnotesize{\bf $(4,1)$}};
\draw[blue,fill=blue] (4.75,-1.25) circle (0.1cm);
\node[scale=0.7,blue] at (4.45,-1.05) {\footnotesize{\bf $(1,1)$}};
\draw[blue,fill=blue] (1,0) circle (0.1cm);
\node[scale=0.7,blue] at (0.65,-0.2) {\footnotesize{\bf $(2,2)$}};
\draw[blue,fill=blue] (2,-1) circle (0.1cm);
\node[scale=0.7,blue] at (1.65,-1.2) {\footnotesize{\bf $(3,2)$}};
\draw[blue,fill=blue] (3,-2) circle (0.1cm);
\node[scale=0.7,blue] at (2.65,-2.2) {\footnotesize{\bf $(4,2)$}};
\draw[blue,fill=blue] (4,-3) circle (0.1cm);
\node[scale=0.7,blue] at (3.65,-3.2) {\footnotesize{\bf $(1,2)$}};
\end{tikzpicture}}}
&&\mathcal{C}^{\text{hor}}=\left(\begin{linesmall}{@{}cccccccc|c|c@{}}&&&&&&&&\\[-6pt]\mathcal{P}_{N} & \mathbf{0} & \mathbf{0} & \mathbf{0} & \cdots & \mathbf{0} & \mathbf{0} & \mathbf{0} & \mathbf{0} & \mathcal{E}_3  \\[2pt]
&&&&&&&&\\[-6pt]\mathbf{0} &\mathcal{P}_{N} & \mathbf{0} & \mathbf{0} & \cdots & \mathbf{0} & \mathbf{0} & \mathbf{0} &  \mathbf{0} & \mathcal{E}_3  \\[2pt]
&&&&&&&&\\[-6pt]\mathbf{0}  & \mathbf{0} &\mathcal{P}_{N} & \mathbf{0} & \cdots & \mathbf{0} & \mathbf{0} & \mathbf{0} & \mathbf{0} & \mathcal{E}_3  \\[2pt]
&&&&&&&&\\[-6pt]\mathbf{0}  & \mathbf{0} & \mathbf{0} &\mathcal{P}_{N} & \cdots & \mathbf{0} & \mathbf{0} & \mathbf{0} & \mathbf{0} & \mathcal{E}_3  \\[2pt]
&&&&&&&&\\[-6pt]\vdots  & \vdots &\vdots &\vdots & \ddots  & \vdots & \vdots & \vdots & \vdots  & \vdots \\[2pt]
&&&&&&&&\\[-6pt]\mathbf{0}  & \mathbf{0} & \mathbf{0} & \mathbf{0} & \cdots & \mathcal{P}_{N} & \mathbf{0} & \mathbf{0}  & \mathbf{0} & \mathcal{E}_3  \\[2pt]
&&&&&&&&\\[-6pt]\mathbf{0}  & \mathbf{0} & \mathbf{0} & \mathbf{0} & \cdots  & \mathbf{0} & \mathcal{P}_{N} & \mathbf{0}  & \mathbf{0} & \mathcal{E}_3  \\[2pt]
&&&&&&&&\\[-6pt]\mathbf{0}  & \mathbf{0} & \mathbf{0} & \mathbf{0} & \cdots  & \mathbf{0}  & \mathbf{0} & \mathcal{P}_{N}  & \mathbf{0} & \mathcal{E}_3  \\[2pt]\hline
&&&&&&&&\\[-5pt]\mathcal{L}  & -2\mathcal{L} &\mathcal{L} & 0 & \cdots & 0 &  0 & 0 & & \mathbf{0}  \\[2pt]
&&&&&&&&\\[-5pt]0 & \mathcal{L}  & -2\mathcal{L} &\mathcal{L} & \cdots & 0 &  0 & 0 & & \mathbf{0}  \\[2pt]
&&&&&&&&\\[-12pt]\vdots & \vdots  & \ddots &\ddots & \ddots & \vdots & \vdots & \vdots  & \mathbbm{1}  & \vdots  \\[2pt]
&&&&&&&&\\[-5pt]0 &0  & 0 & 0 & \cdots &  \mathcal{L} &  -2\mathcal{L} & \mathcal{L} & & \mathbf{0}  \\[2pt]
&&&&&&&&\\[-5pt]\mathcal{L} &0  & 0 & 0 & \cdots & 0 &  \mathcal{L}  &  -2\mathcal{L} &  & \mathbf{0}  \\[2pt]\hline
&&&&&&&&\\[-5pt] \mathbf{0} & \mathbf{0}  & \mathbf{0} &\mathbf{0} & \cdots & \mathbf{0} & \mathbf{0} & \mathbf{0} & \mathbf{0} & \mathbbm{1}   
\end{linesmall}\right)\,,
\end{align}
while for $M=2$ we have $\mathcal{C}^{\text{hor}}_{\tau\widehat{a}}=(2\mathcal{L}\,,-2\mathcal{L})$. Furthermore, we have used the shorthand matrices
\begin{align}
&\mathcal{P}_{N}=\begin{psmall} 0 & 1 & 0 & 0 & \cdots & 0 \\ 0 & 0 & 1 & 0 & \cdots & 0 \\ 0 & 0 & 0 & 1 & \cdots & 0 \\[-3pt] \vdots & \vdots & \vdots & \vdots & \ddots & \vdots \\[2pt] 0 & 0 & 0 & 0 & \cdots & 1 \\ -1 & -1 & -1 & -1 & \cdots & -1 \end{psmall}\in\mathbb{M}_{N-1,N-1}\,,&&\mathcal{E}_3=\begin{psmall} 0 & 0 & 0 \\[-3pt] \vdots & \vdots & \vdots \\[2pt] 0 & 0 & 0 \\ 0 & 0 & 1 \end{psmall}\in\mathbb{M}_{N-1,3}\,,\nonumber\\
&\mathcal{L}=-\begin{psmall} \frac{N-1}{N} & \frac{N-2}{N} & \cdots & \frac{2}{N} & \frac{1}{N} \end{psmall}\in \mathbb{M}_{1,N-1}\,.\label{DefMatrices}
\end{align}
We note that the symmetry matrix $\mathcal{C}^{\text{hor}}$ satisfies $(\mathcal{C}^{\text{hor}})^N=\mathbbm{1}\,,$ and $\text{det}(\mathcal{C}^{\text{hor}})=1$. This transformation corresponds to a map between alternative parametrisation of a given web diagram. 
\item cut and re-glue along diagonal lines $\mathfrak{c}^{\text{ver}}$:\\
The transformed web diagram for the case $(4,2)$ and the associated general symmetry transformation $\mathcal{C}^{\text{ver}}$ are as follows
\begin{align}
&\scalebox{0.72}{\parbox{7.1cm}{\begin{tikzpicture}
\draw[ultra thick] (1.75,1.75) -- (2.25,2.25);
\draw[ultra thick] (2.75,0.75) -- (3.25,1.25);
\draw[ultra thick] (3.75,-0.25) -- (4.25,0.25);
\draw[ultra thick] (4.75,-1.25) -- (5.25,-0.75);
\draw[ultra thick] (0.75,1.75) -- (1.75,1.75) -- (1.75,0.75) -- (2.75,0.75) -- (2.75,-0.25) -- (3.75,-0.25) -- (3.75,-1.25) -- (4.75,-1.25) -- (4.75,-2.25) -- (5.75,-2.25);
\draw[ultra thick] (1,0) -- (1.75,0.75);
\draw[ultra thick] (2,-1) -- (2.75,-0.25);
\draw[ultra thick] (3,-2) -- (3.75,-1.25);
\draw[ultra thick] (4,-3) -- (4.75,-2.25);
\draw[ultra thick] (0,0) -- (1,0) -- (1,-1) -- (2,-1) -- (2,-2) -- (3,-2) -- (3,-3) -- (4,-3) -- (4,-4) -- (5,-4);
\draw[ultra thick] (1,-1) -- (0.5,-1.5);
\draw[ultra thick] (2,-2) -- (1.5,-2.5);
\draw[ultra thick] (3,-3) -- (2.5,-3.5);
\draw[ultra thick] (4,-4) -- (3.5,-4.5);
\node at (0.55,1.75) {\footnotesize $b$}; 
\node at (5.95,-2.25) {\footnotesize $b$}; 
\node at (-0.05,0.25) {\footnotesize $a$}; 
\node at (5,-4.3) {\footnotesize $a$}; 
\node at (2.4,2.4) {\footnotesize $1$};
\node at (3.4,1.4) {\footnotesize $2$};
\node at (4.4,0.4) {\footnotesize $3$};
\node at (5.4,-0.6) {\footnotesize $4$};
\node at (0.35,-1.65) {\footnotesize $1$};
\node at (1.35,-2.65) {\footnotesize $2$};
\node at (2.35,-3.65) {\footnotesize $3$};
\node at (3.35,-4.65) {\footnotesize $4$};
\draw[<->,green!50!black] (2,1.9) -- (2.9,1);
\node[green!50!black] at (2.7,1.7) {\footnotesize $\widehat{a}_1^{(1)\prime}$};
\draw[<->,green!50!black] (3,0.9) -- (3.9,0);
\node[green!50!black] at (3.7,0.7) {\footnotesize $\widehat{a}_2^{(1)\prime}$};
\draw[<->,green!50!black] (4,-0.1) -- (4.9,-1);
\node[green!50!black] at (4.7,-0.3) {\footnotesize $\widehat{a}_3^{(1)\prime}$};
\draw[<->,green!50!black] (1.3,0.2) -- (2.2,-0.7);
\node[green!50!black] at (2,0) {\footnotesize $\widehat{a}_1^{(2)\prime}$};
\draw[<->,green!50!black] (2.3,-0.8) -- (3.2,-1.7);
\node[green!50!black] at (3,-1) {\footnotesize $\widehat{a}_2^{(2)\prime}$};
\draw[<->,green!50!black] (3.3,-1.8) -- (4.2,-2.7);
\node[green!50!black] at (4,-2) {\footnotesize $\widehat{a}_3^{(2)\prime}$};
\draw[dashed] (1.75,1.75) -- (-0.325,-0.325);
\draw[dashed] (5,-4) -- (5.5,-4);
\draw[<->,green!50!black] (2.7,2.7) -- (6.7,-1.3);
\node[green!50!black] at (4.75,0.95) {\footnotesize{\bf $\rho'$}};
\draw[<->,green!50!black] (-0.25,-0.35) -- (0.65,-1.25);
\node[green!50!black] at (0.4,-0.65) {\footnotesize{\bf $S'$}};
\draw[<->,green!50!black] (0.9,0.1) -- (0.9,1.65);
\node[green!50!black] at (0.65,1.3) {\footnotesize{\bf $\tau'_1$}};
\draw[<->,green!50!black] (5.45,-0.85) -- (5.45,-3.9);
\node[green!50!black] at (5.25,-3) {\footnotesize{\bf $\tau'$}};
\draw[blue,fill=blue] (1.75,1.75) circle (0.1cm);
\node[scale=0.7,blue] at (1.45,1.95) {\footnotesize{\bf $(1,2)$}};
\draw[blue,fill=blue] (2.75,0.75) circle (0.1cm);
\node[scale=0.7,blue] at (2.45,0.95) {\footnotesize{\bf $(2,2)$}};
\draw[blue,fill=blue] (3.75,-0.25) circle (0.1cm);
\node[scale=0.7,blue] at (3.45,-0.05) {\footnotesize{\bf $(3,2)$}};
\draw[blue,fill=blue] (4.75,-1.25) circle (0.1cm);
\node[scale=0.7,blue] at (4.45,-1.05) {\footnotesize{\bf $(4,2)$}};
\draw[blue,fill=blue] (1,0) circle (0.1cm);
\node[scale=0.7,blue] at (0.65,-0.2) {\footnotesize{\bf $(1,1)$}};
\draw[blue,fill=blue] (2,-1) circle (0.1cm);
\node[scale=0.7,blue] at (1.65,-1.2) {\footnotesize{\bf $(2,1)$}};
\draw[blue,fill=blue] (3,-2) circle (0.1cm);
\node[scale=0.7,blue] at (2.65,-2.2) {\footnotesize{\bf $(3,1)$}};
\draw[blue,fill=blue] (4,-3) circle (0.1cm);
\node[scale=0.7,blue] at (3.65,-3.2) {\footnotesize{\bf $(4,1)$}};
\end{tikzpicture}}}
&&\mathcal{C}^{\text{ver}}=\left(\begin{linesmall}{@{}ccccc|c|c@{}}&&&&&&\\[-6pt]
\mathbf{0} & \mathbbm{1} & \mathbf{0} & \cdots & \mathbf{0} & \mathbf{0} & \mathbf{0}\\[2pt]
&&&&&&\\[-6pt]
\mathbf{0} & \mathbf{0} & \mathbbm{1} & \cdots & \mathbf{0} & \mathbf{0} & \mathbf{0}\\[2pt]
&&&&&&\\[-12pt]
\vdots & \vdots & \vdots & \ddots & \vdots & \vdots & \vdots\\[2pt]
&&&&&&\\[-6pt]
\mathbf{0} & \mathbf{0} & \mathbf{0} & \cdots &  \mathbbm{1} & \mathbf{0} & \mathbf{0}\\[2pt]
&&&&&&\\[-6pt]
\mathbbm{1} & \mathbf{0} & \mathbf{0} & \cdots &  \mathbf{0} & \mathbf{0} & \mathbf{0}\\[2pt]\hline
&&&&&&\\[-5pt]
\mathbf{0} & \mathbf{0} & \mathbf{0} & \cdots &  \mathbf{0} & \mathcal{P}_{M} & \mathcal{E}_1\\[2pt]\hline
&&&&&&\\[-5pt]
\mathbf{0} & \mathbf{0} & \mathbf{0} & \cdots &  \mathbf{0} & \mathbf{0} & \mathbbm{1}\\[2pt]
\end{linesmall}\right)\, ,
\end{align}
where $\mathcal{E}_1$ is defined as
\begin{align}
    \mathcal{E}_1=\begin{psmall} 0 & 0 & 0 \\[-3pt] \vdots & \vdots & \vdots \\[2pt] 0 & 0 & 0 \\ 1 & 0 & 0 \end{psmall}\in\mathbb{M}_{N-1,3}\, .
\end{align}
The symmetry matrix $\mathcal{C}^{\text{ver}}$ satisfies $(\mathcal{C}^{\text{ver}})^M=\mathbbm{1}$ and $\text{det}(\mathcal{C}^{\text{ver}})=1$.
\item rotations $\mathfrak{r}$:\\
The transformed web diagram for the case $(4,2)$ and the associated general symmetry transformation $\mathcal{R}$ are as follows
\begin{align}
&\scalebox{0.72}{\parbox{7.1cm}{\begin{tikzpicture}
\draw[ultra thick] (1.75,1.75) -- (2.25,2.25);
\draw[ultra thick] (2.75,0.75) -- (3.25,1.25);
\draw[ultra thick] (3.75,-0.25) -- (4.25,0.25);
\draw[ultra thick] (4.75,-1.25) -- (5.25,-0.75);
\draw[ultra thick] (0.75,1.75) -- (1.75,1.75) -- (1.75,0.75) -- (2.75,0.75) -- (2.75,-0.25) -- (3.75,-0.25) -- (3.75,-1.25) -- (4.75,-1.25) -- (4.75,-2.25) -- (5.75,-2.25);
\draw[ultra thick] (1,0) -- (1.75,0.75);
\draw[ultra thick] (2,-1) -- (2.75,-0.25);
\draw[ultra thick] (3,-2) -- (3.75,-1.25);
\draw[ultra thick] (4,-3) -- (4.75,-2.25);
\draw[ultra thick] (0,0) -- (1,0) -- (1,-1) -- (2,-1) -- (2,-2) -- (3,-2) -- (3,-3) -- (4,-3) -- (4,-4) -- (5,-4);
\draw[ultra thick] (1,-1) -- (0.5,-1.5);
\draw[ultra thick] (2,-2) -- (1.5,-2.5);
\draw[ultra thick] (3,-3) -- (2.5,-3.5);
\draw[ultra thick] (4,-4) -- (3.5,-4.5);
\node at (0.55,1.75) {\footnotesize $b$}; 
\node at (5.95,-2.25) {\footnotesize $b$}; 
\node at (-0.05,0.25) {\footnotesize $a$}; 
\node at (5,-4.3) {\footnotesize $a$}; 
\node at (2.4,2.4) {\footnotesize $4$};
\node at (3.4,1.4) {\footnotesize $3$};
\node at (4.4,0.4) {\footnotesize $2$};
\node at (5.4,-0.6) {\footnotesize $1$};
\node at (0.35,-1.65) {\footnotesize $4$};
\node at (1.35,-2.65) {\footnotesize $3$};
\node at (2.35,-3.65) {\footnotesize $2$};
\node at (3.35,-4.65) {\footnotesize $1$};
\draw[<->,green!50!black] (2,1.9) -- (2.9,1);
\node[green!50!black] at (2.7,1.7) {\footnotesize $\widehat{a}_1^{(1)\prime}$};
\draw[<->,green!50!black] (3,0.9) -- (3.9,0);
\node[green!50!black] at (3.7,0.7) {\footnotesize $\widehat{a}_2^{(1)\prime}$};
\draw[<->,green!50!black] (4,-0.1) -- (4.9,-1);
\node[green!50!black] at (4.7,-0.3) {\footnotesize $\widehat{a}_3^{(1)\prime}$};
\draw[<->,green!50!black] (1.3,0.2) -- (2.2,-0.7);
\node[green!50!black] at (2,0) {\footnotesize $\widehat{a}_1^{(2)\prime}$};
\draw[<->,green!50!black] (2.3,-0.8) -- (3.2,-1.7);
\node[green!50!black] at (3,-1) {\footnotesize $\widehat{a}_2^{(2)\prime}$};
\draw[<->,green!50!black] (3.3,-1.8) -- (4.2,-2.7);
\node[green!50!black] at (4,-2) {\footnotesize $\widehat{a}_3^{(2)\prime}$};
\draw[dashed] (1.75,1.75) -- (-0.325,-0.325);
\draw[dashed] (5,-4) -- (5.5,-4);
\draw[<->,green!50!black] (2.7,2.7) -- (6.7,-1.3);
\node[green!50!black] at (4.75,0.95) {\footnotesize{\bf $\rho'$}};
\draw[<->,green!50!black] (-0.25,-0.35) -- (0.65,-1.25);
\node[green!50!black] at (0.4,-0.65) {\footnotesize{\bf $S'$}};
\draw[<->,green!50!black] (0.9,0.1) -- (0.9,1.65);
\node[green!50!black] at (0.65,1.3) {\footnotesize{\bf $\tau'_1$}};
\draw[<->,green!50!black] (5.45,-0.85) -- (5.45,-3.9);
\node[green!50!black] at (5.25,-3) {\footnotesize{\bf $\tau'$}};
\draw[blue,fill=blue] (1.75,0.75) circle (0.1cm);
\node[scale=0.7,blue] at (2.1,0.95) {\footnotesize{\bf $(4,2)$}};
\draw[blue,fill=blue] (2.75,-0.25) circle (0.1cm);
\node[scale=0.7,blue] at (3.1,-0.05) {\footnotesize{\bf $(3,2)$}};
\draw[blue,fill=blue] (3.75,-1.25) circle (0.1cm);
\node[scale=0.7,blue] at (4.1,-1.05) {\footnotesize{\bf $(2,2)$}};
\draw[blue,fill=blue] (4.75,-2.25) circle (0.1cm);
\node[scale=0.7,blue] at (5.1,-2.05) {\footnotesize{\bf $(1,2)$}};
\draw[blue,fill=blue] (1,-1) circle (0.1cm);
\node[scale=0.7,blue] at (1.4,-0.8) {\footnotesize{\bf $(4,1)$}};
\draw[blue,fill=blue] (2,-2) circle (0.1cm);
\node[scale=0.7,blue] at (2.4,-1.8) {\footnotesize{\bf $(3,1)$}};
\draw[blue,fill=blue] (3,-3) circle (0.1cm);
\node[scale=0.7,blue] at (3.4,-2.8) {\footnotesize{\bf $(2,1)$}};
\draw[blue,fill=blue] (4,-4) circle (0.1cm);
\node[scale=0.7,blue] at (4.4,-3.8) {\footnotesize{\bf $(1,1)$}};
\end{tikzpicture}}}
&&\mathcal{R}=\left(\begin{linesmall}{@{}ccccc|c|c@{}}
&&&&&&\\[-6pt]
\mathcal{J}_{N} & \mathbf{0} & \cdots & \mathbf{0} & \mathbf{0} & \mathbf{0} & \mathbf{0} \\[2pt]
&&&&&&\\[-6pt]
\mathbf{0} & \mathbf{0} & \cdots & \mathbf{0}  & \mathcal{J}_{N} & \mathbf{0} & \mathbf{0}  \\[2pt]
&&&&&&\\[-6pt]
\mathbf{0} & \mathbf{0} & \cdots  & \mathcal{J}_{N} & \mathbf{0} & \mathbf{0} & \mathbf{0}  \\[2pt]
&&&&&&\\[-12pt]
\vdots & \vdots & \reflectbox{$\ddots$} & \vdots & \vdots & \vdots & \vdots  \\[2pt]
&&&&&&\\[-6pt]
\mathbf{0} & \mathcal{J}_{N} & \cdots & \mathbf{0} &  \mathbf{0} & \mathbf{0} & \mathbf{0} \\[2pt]\hline
&&&&&&\\[-5pt]
\mathbf{0} & \mathbf{0} & \cdots &  \mathbf{0} & \mathbf{0} & \mathcal{J}_{M} & \mathbf{0} \\[2pt]\hline
&&&&&&\\[-5pt]
\mathbf{0} & \mathbf{0} & \cdots & \mathbf{0} &  \mathbf{0} & \mathbf{0} &   \mathbbm{1}   \\[2pt]
\end{linesmall}\right)\,,
\end{align}

where the matrix $\mathcal{J}_{N}$ is defined as 
\begin{align}
&\mathcal{J}_{N}=\left(\begin{linesmall}{@{}cccc@{}}  
&&&\\[-6pt]
0 & \cdots & 0 & 1 \\[2pt]  0 & \cdots & 1 & 0 \\[2pt]
&&&\\[-12pt] 
\vdots & \reflectbox{$\ddots$} & \vdots & \vdots   \\[2pt]
&&&\\[-6pt]
1 & \cdots & 0 & 0 \\[2pt]
\end{linesmall}\right)\in\mathbb{M}_{N-1,N-1}&&\forall n\in\mathbb{N}\,.\label{Jdef}
\end{align}
This symmetry has the properties $\mathcal{R}^2=\mathbbm{1}$ and $\text{det}(\mathcal{R})=1$.

\end{itemize}

\item[\emph{(ii)}] \emph{Symmetries relating points in (potentially) different K\"ahler cones in the extended moduli space $\mathbb{EK}(X_{N,M})$}\\ 

\noindent
Here we consider three different similarity transformation which relate points in the extended moduli space of $\mathbb{EK}(X_{N,M})$
\begin{itemize}
\item A similarity transformation (which we shall denote $\mathfrak{a}$ in the following) that leads to a different orientation of the vertices according to the definition (\ref{OrientationVertices}) is shown in the following
\begin{align}
&\scalebox{0.72}{\parbox{7.1cm}{\begin{tikzpicture}
\draw[ultra thick] (1.75,1.75) -- (2.25,2.25);
\draw[ultra thick] (2.75,0.75) -- (3.25,1.25);
\draw[ultra thick] (3.75,-0.25) -- (4.25,0.25);
\draw[ultra thick] (4.75,-1.25) -- (5.25,-0.75);
\draw[ultra thick] (0.75,1.75) -- (1.75,1.75) -- (1.75,0.75) -- (2.75,0.75) -- (2.75,-0.25) -- (3.75,-0.25) -- (3.75,-1.25) -- (4.75,-1.25) -- (4.75,-2.25) -- (5.75,-2.25);
\draw[ultra thick] (1,0) -- (1.75,0.75);
\draw[ultra thick] (2,-1) -- (2.75,-0.25);
\draw[ultra thick] (3,-2) -- (3.75,-1.25);
\draw[ultra thick] (4,-3) -- (4.75,-2.25);
\draw[ultra thick] (0,0) -- (1,0) -- (1,-1) -- (2,-1) -- (2,-2) -- (3,-2) -- (3,-3) -- (4,-3) -- (4,-4) -- (5,-4);
\draw[ultra thick] (1,-1) -- (0.5,-1.5);
\draw[ultra thick] (2,-2) -- (1.5,-2.5);
\draw[ultra thick] (3,-3) -- (2.5,-3.5);
\draw[ultra thick] (4,-4) -- (3.5,-4.5);
\node at (0.55,1.75) {\footnotesize $b$}; 
\node at (5.95,-2.25) {\footnotesize $b$}; 
\node at (-0.05,0.25) {\footnotesize $a$}; 
\node at (5,-4.3) {\footnotesize $a$}; 
\node at (2.4,2.4) {\footnotesize $4$};
\node at (3.4,1.4) {\footnotesize $3$};
\node at (4.4,0.4) {\footnotesize $2$};
\node at (5.4,-0.6) {\footnotesize $1$};
\node at (0.35,-1.65) {\footnotesize $4$};
\node at (1.35,-2.65) {\footnotesize $3$};
\node at (2.35,-3.65) {\footnotesize $2$};
\node at (3.35,-4.65) {\footnotesize $1$};
\draw[<->,green!50!black] (2,1.9) -- (2.9,1);
\node[green!50!black] at (2.7,1.7) {\footnotesize $\widehat{a}_1^{(1)\prime}$};
\draw[<->,green!50!black] (3,0.9) -- (3.9,0);
\node[green!50!black] at (3.7,0.7) {\footnotesize $\widehat{a}_2^{(1)\prime}$};
\draw[<->,green!50!black] (4,-0.1) -- (4.9,-1);
\node[green!50!black] at (4.7,-0.3) {\footnotesize $\widehat{a}_3^{(1)\prime}$};
\draw[<->,green!50!black] (1.3,0.2) -- (2.2,-0.7);
\node[green!50!black] at (2,0) {\footnotesize $\widehat{a}_1^{(2)\prime}$};
\draw[<->,green!50!black] (2.3,-0.8) -- (3.2,-1.7);
\node[green!50!black] at (3,-1) {\footnotesize $\widehat{a}_2^{(2)\prime}$};
\draw[<->,green!50!black] (3.3,-1.8) -- (4.2,-2.7);
\node[green!50!black] at (4,-2) {\footnotesize $\widehat{a}_3^{(2)\prime}$};
\draw[dashed] (1.75,1.75) -- (-0.325,-0.325);
\draw[dashed] (5,-4) -- (5.5,-4);
\draw[<->,green!50!black] (2.7,2.7) -- (6.7,-1.3);
\node[green!50!black] at (4.75,0.95) {\footnotesize{\bf $\rho'$}};
\draw[<->,green!50!black] (-0.25,-0.35) -- (0.65,-1.25);
\node[green!50!black] at (0.4,-0.65) {\footnotesize{\bf $S'$}};
\draw[<->,green!50!black] (0.9,0.1) -- (0.9,1.65);
\node[green!50!black] at (0.65,1.3) {\footnotesize{\bf $\tau'_1$}};
\draw[<->,green!50!black] (5.45,-0.85) -- (5.45,-3.9);
\node[green!50!black] at (5.25,-3) {\footnotesize{\bf $\tau'$}};
\draw[red,fill=red] (1.75,0.75) circle (0.1cm);
\node[scale=0.7,red] at (2.1,0.95) {\footnotesize{\bf $(4,1)$}};
\draw[red,fill=red] (2.75,-0.25) circle (0.1cm);
\node[scale=0.7,red] at (3.1,-0.05) {\footnotesize{\bf $(3,2)$}};
\draw[red,fill=red] (3.75,-1.25) circle (0.1cm);
\node[scale=0.7,red] at (4.1,-1.05) {\footnotesize{\bf $(2,1)$}};
\draw[red,fill=red] (4.75,-2.25) circle (0.1cm);
\node[scale=0.7,red] at (5.1,-2.05) {\footnotesize{\bf $(1,2)$}};
\draw[red,fill=red] (1,-1) circle (0.1cm);
\node[scale=0.7,red] at (1.4,-0.8) {\footnotesize{\bf $(4,2)$}};
\draw[red,fill=red] (2,-2) circle (0.1cm);
\node[scale=0.7,red] at (2.4,-1.8) {\footnotesize{\bf $(3,1)$}};
\draw[red,fill=red] (3,-3) circle (0.1cm);
\node[scale=0.7,red] at (3.4,-2.8) {\footnotesize{\bf $(2,2)$}};
\draw[red,fill=red] (4,-4) circle (0.1cm);
\node[scale=0.7,red] at (4.4,-3.8) {\footnotesize{\bf $(1,1)$}};
\end{tikzpicture}}}
&&\mathcal{A}=\left(\begin{linesmall}{@{}ccccc|c|ccc@{}}&&&&&&&&\\[-6pt]
\mathbf{u}_1 & \mathbf{u}_2 & \cdots & \mathbf{u}_{M-1} & \mathbf{u}_M &  &  & &  \\[2pt]
&&&&&&&&\\[-6pt]
\mathbf{u}_M & \mathbf{u}_1 & \cdots & \mathbf{u}_{M-2} & \mathbf{u}_{M-1} &  &  & &  \\[2pt]
&&&&&&&&\\[-6pt]
\vdots & \vdots & \ddots & \vdots & \vdots & \mathcal{A}_{\widehat{a}\tau} &  & \mathcal{A}_{\widehat{a}\omega} &  \\[2pt]
&&&&&&&&\\[-6pt]
\mathbf{u}_3 & \mathbf{u}_4 & \cdots & \mathbf{u}_{1} & \mathbf{u}_{2} &  &  & &  \\[2pt]
&&&&&&&&\\[-6pt]
\mathbf{u}_2 & \mathbf{u}_3 & \cdots & \mathbf{u}_{M} & \mathbf{u}_{1} &  &  & &  \\[2pt]\hline
&&&&&&&&\\[-4pt]
\mathbf{0} & \mathbf{0} & \cdots & \mathbf{0} & \mathbf{0} & \mathbbm{1} & \mathbf{0}    & \mathbf{0} & \mathbf{0} \\[2pt]\hline
&&&&&&&&\\[-5pt]
\mathbf{0} & \mathbf{0} & \cdots & \mathbf{0} & \mathbf{0} & \mathbf{0} & 1    & 0 & 0 \\[2pt]
&&&&&&&&\\[-6pt]
\mathbf{0} & \mathbf{0} & \cdots & \mathbf{0} & \mathbf{0} & \mathbf{0} & 1    & -1 & 0 \\[2pt]
&&&&&&&&\\[-6pt]
\mathbf{0} & \mathbf{0} & \cdots & \mathbf{0} & \mathbf{0} & \mathbf{0} & N/M    & -2N/M & 1 \\[2pt]
\end{linesmall}\right)\,.\label{eq:Amattrans}
\end{align}
Here we do not further specify the matrices $\mathcal{A}_{\widehat{a}\tau}$, $\mathcal{A}_{\widehat{a}\omega}$ or $u_i\in\mathbb{M}_{M-1,M-1}$ (which, however, can be computed for fixed $(N,M)$ using the Mathematica package {\tt NPLSTsym} described in Appendix~\ref{App:MathematicaPackage}). Instead, we only remark that the matrix $\mathcal{A}_{\widehat{a}\widehat{a}}$ is \emph{magic}, \emph{i.e.} 
\begin{align}
&\sum_{j=1}^{M(N-1)}(\mathcal{S}_{\widehat{a}\widehat{a}})_{ij}=1=\sum_{j=1}^{M(N-1)}(\mathcal{S}_{\widehat{a}\widehat{a}})_{ji}\,,&&\forall j\in\{1,\ldots,M(N-1)\}\,,
\end{align}
which is due to the following property $\sum_{i=1}^M\mathbf{u}_i=\mathcal{J}_{N-1}$ (where $\mathcal{J}_{N-1}$ is defined in (\ref{Jdef})). Furthermore, the matrix $\mathcal{A}$ has the properties $\mathcal{A}^2=\mathbbm{1}$ and $\text{det}(\mathcal{A})=-1$.
\item A similarity transformation (which we shall call $\mathfrak{b}$ in the following) leading to a shifted web diagram with $\delta=N-M$ and different orientation of the vertices according to the definition (\ref{OrientationVertices}) is shown in the following
\begin{align}
&\scalebox{0.72}{\parbox{7.5cm}{\begin{tikzpicture}
\draw[ultra thick] (1.75,1.75) -- (2.25,2.25);
\draw[ultra thick] (2.75,0.75) -- (3.25,1.25);
\draw[ultra thick] (3.75,-0.25) -- (4.25,0.25);
\draw[ultra thick] (4.75,-1.25) -- (5.25,-0.75);
\draw[ultra thick] (0.75,1.75) -- (1.75,1.75) -- (1.75,0.75) -- (2.75,0.75) -- (2.75,-0.25) -- (3.75,-0.25) -- (3.75,-1.25) -- (4.75,-1.25) -- (4.75,-2.25) -- (5.75,-2.25);
\draw[ultra thick] (1,0) -- (1.75,0.75);
\draw[ultra thick] (2,-1) -- (2.75,-0.25);
\draw[ultra thick] (3,-2) -- (3.75,-1.25);
\draw[ultra thick] (4,-3) -- (4.75,-2.25);
\draw[ultra thick] (0,0) -- (1,0) -- (1,-1) -- (2,-1) -- (2,-2) -- (3,-2) -- (3,-3) -- (4,-3) -- (4,-4) -- (5,-4);
\draw[ultra thick] (1,-1) -- (0.5,-1.5);
\draw[ultra thick] (2,-2) -- (1.5,-2.5);
\draw[ultra thick] (3,-3) -- (2.5,-3.5);
\draw[ultra thick] (4,-4) -- (3.5,-4.5);
\node at (0.55,1.75) {\footnotesize $a$}; 
\node at (5.95,-2.25) {\footnotesize $a$}; 
\node at (0,-0.25) {\footnotesize $b$}; 
\node at (5,-4.3) {\footnotesize $b$}; 
\node at (2.4,2.45) {\scriptsize {\bf III}};
\node at (3.4,1.45) {\scriptsize {\bf IV}};
\node at (4.4,0.4) {\scriptsize {\bf I}};
\node at (5.4,-0.6) {\scriptsize {\bf II}};
\node at (0.65,-1.65) {\scriptsize {\bf I}};
\node at (1.65,-2.65) {\scriptsize {\bf II}};
\node at (2.35,-3.7) {\scriptsize {\bf III}};
\node at (3.35,-4.7) {\scriptsize {\bf IV}};
\draw[<->,green!50!black] (2,1.9) -- (2.9,1);
\node[green!50!black] at (2.7,1.7) {\footnotesize $\widehat{a}_3^{(1)\prime}$};
\draw[<->,green!50!black] (4,-0.1) -- (4.9,-1);
\node[green!50!black] at (4.7,-0.3) {\footnotesize $\widehat{a}_1^{(1)\prime}$};
\draw[<->,green!50!black] (5,-1.1) -- (5.9,-2);
\node[green!50!black] at (5.7,-1.3) {\footnotesize $\widehat{a}_2^{(1)\prime}$};
\draw[<->,green!50!black] (1.3,0.2) -- (2.2,-0.7);
\node[green!50!black] at (2,0) {\footnotesize $\widehat{a}_2^{(2)\prime}$};
\draw[<->,green!50!black] (2.3,-0.8) -- (3.2,-1.7);
\node[green!50!black] at (3.3,-1.3) {\footnotesize $\widehat{a}_3^{(2)\prime}$};
\draw[<->,green!50!black] (4.3,-2.8) -- (5.2,-3.7);
\node[green!50!black] at (5,-3) {\footnotesize $\widehat{a}_1^{(2)\prime}$};
\draw[<->,green!50!black] (2.7,2.7) -- (6.7,-1.3);
\node[green!50!black] at (4.75,0.95) {\footnotesize{\bf $\rho'$}};
\draw[dashed] (3.75,-0.25) -- (1.25,-2.75);
\draw[dashed] (0.5,-1.5) -- (0.25,-1.75);
\draw[<-,green!50!black] (0.2,-1.7) -- (-0.3,-1.2);
\draw[dotted,green!50!black]  (-0.3,-1.2) -- (-0.7,-0.8);
\draw[<-,green!50!black] (1.3,-2.8) -- (1.8,-3.3);
\draw[dotted,green!50!black] (1.8,-3.3) -- (2.2,-3.7);
\node[green!50!black] at (1.4,-3.25) {\footnotesize{\bf $S'$}};
\draw[dashed] (3.75,-0.25) -- (6.4,-0.25);
\draw[dashed] (5,-4) -- (6.4,-4);
\draw[<->,green!50!black] (6.4,-0.3) -- (6.4,-3.95);
\node[rotate=90,green!50!black,scale=0.9] at (6.65,-3) {\scriptsize{\bf $\tau'_1-\rho'+\widehat{a}_1^{(2)\prime}+\widehat{a}_2^{(2)\prime}+\widehat{a}_3^{(2)\prime}$}};
\draw[dashed] (4.25,0.25) -- (-0.6,0.25);
\draw[dashed] (0.5,-1.5) -- (-0.6,-1.5);
\draw[<-,green!50!black] (-0.5,0.3) -- (-0.5,0.8);
\draw[dotted,green!50!black] (-0.5,0.8) -- (-0.5,1.2);
\draw[<-,green!50!black] (-0.5,-1.55) -- (-0.5,-2.05);
\draw[dotted,green!50!black] (-0.5,-2.05) -- (-0.5,-2.45);
\node[green!50!black] at (0.1,0.6) {\footnotesize{\bf $\tau'-S'$}};
\draw[red,fill=red] (1.75,0.75) circle (0.1cm);
\node[scale=0.7,red] at (2.1,0.95) {\footnotesize{\bf $(3,2)$}};
\draw[red,fill=red] (2.75,-0.25) circle (0.1cm);
\node[scale=0.7,red] at (3.1,-0.05) {\footnotesize{\bf $(4,2)$}};
\draw[red,fill=red] (3.75,-1.25) circle (0.1cm);
\node[scale=0.7,red] at (4.1,-1.05) {\footnotesize{\bf $(1,2)$}};
\draw[red,fill=red] (4.75,-2.25) circle (0.1cm);
\node[scale=0.7,red] at (5.1,-2.05) {\footnotesize{\bf $(2,2)$}};
\draw[red,fill=red] (1,-1) circle (0.1cm);
\node[scale=0.7,red] at (1.4,-0.8) {\footnotesize{\bf $(4,1)$}};
\draw[red,fill=red] (2,-2) circle (0.1cm);
\node[scale=0.7,red] at (2.4,-2.2) {\footnotesize{\bf $(1,1)$}};
\draw[red,fill=red] (3,-3) circle (0.1cm);
\node[scale=0.7,red] at (3.4,-3.2) {\footnotesize{\bf $(2,1)$}};
\draw[red,fill=red] (4,-4) circle (0.1cm);
\node[scale=0.7,red] at (4.4,-3.8) {\footnotesize{\bf $(3,1)$}};
\end{tikzpicture}}}
&&\mathcal{B}=\left(\begin{linesmall}{@{}c|c|ccc@{}}&&&&\\[-6pt]
\mathcal{B}_{\widehat{a}\widehat{a}} & \mathbf{0} &  & \mathbf{B}_{\widehat{a}\omega} &  \\[2pt]\hline
&&&&\\[-5pt]
\mathcal{B}_{\tau\widehat{a}} & \mathcal{J}_{M} &  & \mathbf{B}_{\tau\omega} &  \\[2pt]\hline
&&&&\\[-5pt]
\mathbf{0} & \mathbf{0} & 1 & -2N/M & N/M  \\[2pt]
&&&&\\[-6pt]
\mathbf{0} & \mathbf{0} & 0 & -1 & 1 \\[2pt]
&&&&\\[-6pt]
\mathbf{0} & \mathbf{0} & 0 & 0 & 1 \\[2pt]
\end{linesmall}\right)\,.\label{eq:Bmattrans}
\end{align}
The dependence of $\mathcal{B}_{\widehat{a}\widehat{a}}$, $\mathcal{B}_{\tau\widehat{a}}$, $\mathcal{B}_{\widehat{a}\omega}$ and $\mathcal{B}_{\tau\omega}$ on $(N,M)$ is rather complicated and we refer the reader to the Mathematica package {\tt NPLSTsym} described in Appendix~\ref{App:MathematicaPackage}) to compute $\mathcal{B}$ for concrete values of $(N,M)$. We remark that $(\mathcal{B})^2=\mathbbm{1}$ and $\text{det}(\mathcal{B})=-1$.

\item A further similarity transformation can be obtained by first flopping all diagonal lines of the diagram in Figure~\ref{Fig:Web42}, as is shown in the example $(N,M)=(4,2)$ (where the labelling of the K\"ahler parameters follows the modified vertices as indicated)
\begin{align}
&\scalebox{0.72}{\parbox{7.1cm}{\begin{tikzpicture}
\draw[ultra thick] (1.75,1.75) -- (2.25,2.25);
\draw[ultra thick] (2.75,0.75) -- (3.25,1.25);
\draw[ultra thick] (3.75,-0.25) -- (4.25,0.25);
\draw[ultra thick] (4.75,-1.25) -- (5.25,-0.75);
\draw[ultra thick] (0.75,1.75) -- (1.75,1.75) -- (1.75,0.75) -- (2.75,0.75) -- (2.75,-0.25) -- (3.75,-0.25) -- (3.75,-1.25) -- (4.75,-1.25) -- (4.75,-2.25) -- (5.75,-2.25);
\draw[ultra thick] (1,0) -- (1.75,0.75);
\draw[ultra thick] (2,-1) -- (2.75,-0.25);
\draw[ultra thick] (3,-2) -- (3.75,-1.25);
\draw[ultra thick] (4,-3) -- (4.75,-2.25);
\draw[ultra thick] (0,0) -- (1,0) -- (1,-1) -- (2,-1) -- (2,-2) -- (3,-2) -- (3,-3) -- (4,-3) -- (4,-4) -- (5,-4);
\draw[ultra thick] (1,-1) -- (0.5,-1.5);
\draw[ultra thick] (2,-2) -- (1.5,-2.5);
\draw[ultra thick] (3,-3) -- (2.5,-3.5);
\draw[ultra thick] (4,-4) -- (3.5,-4.5);
\node at (0.55,1.75) {\footnotesize $a$}; 
\node at (5.95,-2.25) {\footnotesize $a$}; 
\node at (0,-0.25) {\footnotesize $b$}; 
\node at (5,-4.3) {\footnotesize $b$}; 
\node at (2.4,2.45) {\scriptsize {\bf I}};
\node at (3.4,1.45) {\scriptsize {\bf II}};
\node at (4.4,0.4) {\scriptsize {\bf III}};
\node at (5.4,-0.6) {\scriptsize {\bf IV}};
\node at (0.35,-1.65) {\scriptsize {\bf I}};
\node at (1.35,-2.65) {\scriptsize {\bf II}};
\node at (2.35,-3.7) {\scriptsize {\bf III}};
\node at (3.35,-4.7) {\scriptsize {\bf IV}};
\draw[<->,green!50!black] (2,1.9) -- (2.9,1);
\node[green!50!black] at (2.7,1.7) {\footnotesize $\widehat{a}_1^{(1)\prime}$};
\draw[<->,green!50!black] (3,0.9) -- (3.9,0);
\node[green!50!black] at (3.7,0.7) {\footnotesize $\widehat{a}_2^{(1)\prime}$};
\draw[<->,green!50!black] (4,-0.1) -- (4.9,-1);
\node[green!50!black] at (4.7,-0.3) {\footnotesize $\widehat{a}_3^{(1)\prime}$};
\draw[<->,green!50!black] (1.3,0.2) -- (2.2,-0.7);
\node[green!50!black] at (2,0) {\footnotesize $\widehat{a}_1^{(2)\prime}$};
\draw[<->,green!50!black] (2.3,-0.8) -- (3.2,-1.7);
\node[green!50!black] at (3,-1) {\footnotesize $\widehat{a}_2^{(2)\prime}$};
\draw[<->,green!50!black] (3.3,-1.8) -- (4.2,-2.7);
\node[green!50!black] at (4,-2) {\footnotesize $\widehat{a}_3^{(2)\prime}$};
\draw[dashed] (1.75,1.75) -- (-0.325,-0.325);
\draw[dashed] (5,-4) -- (5.5,-4);
\draw[<->,green!50!black] (2.7,2.7) -- (6.7,-1.3);
\node[green!50!black] at (4.75,0.95) {\footnotesize{\bf $\rho'$}};
\draw[<->,green!50!black] (-0.25,-0.35) -- (0.65,-1.25);
\node[green!50!black] at (0.4,-0.65) {\footnotesize{\bf $S'$}};
\draw[<->,green!50!black] (0.9,0.1) -- (0.9,1.65);
\node[green!50!black] at (0.65,1.3) {\footnotesize{\bf $\tau'_1$}};
\draw[<->,green!50!black] (5.45,-0.85) -- (5.45,-3.9);
\node[green!50!black] at (5.25,-3) {\footnotesize{\bf $\tau'$}};
\draw[orange,fill=yellow!90!green] (1.75,1.75) circle (0.1cm);
\node[scale=0.7,orange] at (1.45,1.95) {\footnotesize{\bf $(1,1)$}};
\draw[orange,fill=yellow!90!green] (2.75,0.75) circle (0.1cm);
\node[scale=0.7,orange] at (2.45,0.95) {\footnotesize{\bf $(2,1)$}};
\draw[orange,fill=yellow!90!green] (3.75,-0.25) circle (0.1cm);
\node[scale=0.7,orange] at (3.45,-0.05) {\footnotesize{\bf $(3,1)$}};
\draw[orange,fill=yellow!90!green] (4.75,-1.25) circle (0.1cm);
\node[scale=0.7,orange] at (4.45,-1.05) {\footnotesize{\bf $(4,1)$}};
\draw[orange,fill=yellow!90!green] (1,0) circle (0.1cm);
\node[scale=0.7,orange] at (0.65,-0.2) {\footnotesize{\bf $(1,2)$}};
\draw[orange,fill=yellow!90!green] (2,-1) circle (0.1cm);
\node[scale=0.7,orange] at (1.65,-1.2) {\footnotesize{\bf $(2,2)$}};
\draw[orange,fill=yellow!90!green] (3,-2) circle (0.1cm);
\node[scale=0.7,orange] at (2.65,-2.2) {\footnotesize{\bf $(3,2)$}};
\draw[orange,fill=yellow!90!green] (4,-3) circle (0.1cm);
\node[scale=0.7,orange] at (3.65,-3.2) {\footnotesize{\bf $(4,2)$}};
\end{tikzpicture}}}
&&\scalebox{1}{\parbox{4.2cm}{\begin{tikzpicture}
\draw[ultra thick] (-1,0) -- (0,0) -- (0,-1); 
\draw[ultra thick] (0,0) -- (0.75,0.75); 
\draw[orange,fill=yellow!90!green] (0,0) circle (0.1cm);
\node[scale=0.7,orange] at (0.4,-0.1) {\footnotesize{\bf $(i,j)$}};
\node at (-0.5,0.35) {\footnotesize $\tilde{h}_i^{(j)}$};
\node at (-0.3,-0.6) {\footnotesize $\tilde{v}_i^{(j)}$};
\node at (0.25,0.75) {\footnotesize $-m_i^{(j)}$};
\node at (0,-2) {\scriptsize $\tilde{h}_i^{(j)}=h_i^{(j+1)}+m_i^{(j)}+m_{i-1}^{(j+1)}$};
\node at (0,-2.5) {\scriptsize $\tilde{v}_i^{(j)}=v_i^{(j)}+m_i^{(j)}+m_{i}^{(j+1)}$};
\end{tikzpicture}}}
&\mathcal{F}=\left(\begin{linesmall}{@{}ccc|c|ccc@{}}&&&&&&\\[-6pt]
\mathcal{J}_{N} & \cdots & \mathbf{0} & \mathbf{0} &  & \mathbf{0} &  \\[2pt]
\vdots & \ddots & \vdots & \mathbf{0} &  & \mathbf{0} &  \\[2pt]
\mathbf{0} & \cdots & \mathcal{J}_{N} & \mathbf{0} &  & \mathbf{0} &  \\[2pt]\hline
&&&&&&\\[-5pt]
\mathbf{0}& \cdots& \mathbf{0} &\mathbbm{1} &  & \mathbf{0}  &  \\[2pt]\hline
&&&&&&\\[-5pt]
\mathbf{0} & \cdots&\mathbf{0} & \mathbf{0} & 1 & 0 & 0  \\[2pt]
&&&&&&\\[-6pt]
\mathbf{0}&\cdots&\mathbf{0} & \mathbf{0} & 0 & -1 & 0 \\[2pt]
&&&&&&\\[-6pt]
\mathbf{0}&\cdots&\mathbf{0} & \mathbf{0} & 0 & 0 & 1 \\[4pt]
\end{linesmall}\right)\,.\label{FlopRelDef}
\end{align}
Combining this flop transformation with transformations of the type $\mathfrak{a}$, $\mathfrak{r}$ and $\mathfrak{c}^{\text{vert}}$ described above, we find a new self-similarity transformation, which we shall denote $\mathfrak{f}$ and for which the associated matrix $\mathcal{F}$ is given in (\ref{FlopRelDef}). This matrix satisfies $(\mathcal{F})^2=\mathbbm{1}$ and $\text{det}(\mathcal{F})=-1$.
\end{itemize}

\end{itemize}

\noindent
Before providing a more group theoretical organisation of the symmetry transformations found above, we remark how they are related to the results for $M=1$ (notably to the symmetry transformations found in \cite{Bastian:2018jlf}, which are summarised in (\ref{M1Generators})): in this case the transformation $\mathfrak{c}^{\text{vert}}$ becomes trivial while $\mathfrak{r}$ and $\mathfrak{c}^{\text{hor}}$ are elements of $\widetilde{S}_N$ in (\ref{DihedralGroup}) and the matrices in (\ref{M1Generators}) are recovered by\footnote{The transformation $\mathfrak{f}$ is not part of (\ref{DihedralGroup}) since in \cite{Bastian:2018jlf} transformations that act as $S\to -S$ were not considered. Furthermore, we remark that a different basis was used in \cite{Bastian:2018jlf}, such that for a direct comparison a change of basis is required, which is induced by the following matrix
\begin{align}
\mathcal{U}=\left(\begin{linesmall}{@{}ccccc@{}}  
&&&&\\[-6pt]
 & \mathbbm{1} &  & 0 & 0 \\[2pt]  N & \cdots & N & -2N & 1 \\[2pt]
1 & \cdots & 1 & -1 & 0 \\[2pt]
1 & \cdots & 1 & 0 & 0 \\[2pt]
\end{linesmall}\right)\,.
\end{align}
}
\begin{align}
&\mathcal{G}_2(N)=\mathcal{R}\cdot \mathcal{C}^{\text{hor}}\cdot\mathcal{C}^{\text{hor}}\cdot\mathcal{B}\,,&&\text{and} &&\mathcal{G}'_2(N)=\mathcal{B}\cdot \mathcal{A}\cdot\mathcal{R}\cdot\mathcal{B}\,.\label{IdentG2}
\end{align}

\subsection{Coxeter Matrix}
In the previous Subsection we have identified the symmetry transformations $(\mathfrak{c}^{\text{hor}},\mathfrak{c}^{\text{ver}},\mathfrak{r},\mathfrak{a},\mathfrak{b},\mathfrak{f})$ which act linearly on the K\"ahler parameters of $X_{N,M}$ through the matrices $(\mathcal{C}^{\text{hor}},\mathcal{C}^{\text{ver}},\mathcal{R},\mathcal{A},\mathcal{B},\mathcal{F})$ respectively. Following the example of $M=1$ in (\ref{DihedralGroup}) and (\ref{M1Generators}), we can arrange these generators in a more group theoretical form. Indeed, generalising the form of (\ref{M1Generators}) for $M=1$, we can re-arrange the symmetries into six involutions
\begin{align}
&\mathbb{T}=\{\mathfrak{t}_i|i=1,\ldots,6\}=\{\mathfrak{r}\,,\mathfrak{r}\circ\mathfrak{c}^{\text{ver}}\,,\mathfrak{r}\circ\mathfrak{c}^{\text{hor}}\,,\mathfrak{a}\,,\mathfrak{b}\,,\mathfrak{f}\}\,,&&\text{with} &&\mathfrak{t}_i\circ\mathfrak{t}_i=\mathbbm{1}\,\hspace{0.1cm}\forall i=1,\ldots,6\,.\label{Involutions}
\end{align}
These involutions are independent of each other and allow to uniquely re-construct the original symmetry transformations. We shall denote the corresponding matrices that act on the basis of K\"ahler parameters (\ref{KaehlerBasis}) by $T_i$ respectively. Relations among these involutions are characterised by so-called braid-relations, \emph{i.e.} we can specify (minimal) integers\footnote{Here we also allow  $m_{ij}=\infty$, which means that no (finite) integer $m_{ij}$ exists such that (\ref{DefBraidRel}) is satisfied.} $m_{ij}$ such that
\begin{align}
&(\mathfrak{t}_i\circ\mathfrak{t}_j)^{\circ m_{ij}}=\mathbbm{1}\,,&&\forall i,j=1,\ldots,6\,.\label{DefBraidRel}
\end{align}
Thus, a way of characterising the group structure encoded in the symmetry transformations is through the so-called \emph{Coxeter matrix}
\begin{align}
&m_{ij}=\left(\begin{linesmall}{@{}cc@{}}  
&\\[-6pt]
m^{ss} & m^{sd} \\[2pt]  (m^{sd})^T & m^{dd}\end{linesmall}\right)_{ij}\,,&&\text{with} &&m^{ss}=\left(\begin{linesmall}{@{}ccc@{}}  
&&\\[-6pt]
1 & M & N\\[2pt]  M & 1 & N \\[2pt] N & N & 1\end{linesmall}\right)\,,&&\text{and} &&m^{dd}=\left(\begin{linesmall}{@{}ccc@{}}  
&&\\[-6pt]
1 & m_{\mathfrak{ab}} & \infty\\[2pt]  m_{\mathfrak{ab}} & 1 & \infty \\[2pt] \infty & \infty & 1\end{linesmall}\right)\,.\label{CoxeterStructure}
\end{align}
Notice that $(\mathfrak{r}\,,\mathfrak{r}\circ\mathfrak{c}^{\text{ver}}\,,\mathfrak{r}\circ\mathfrak{c}^{\text{hor}})$ constitute transformations that follow from symmetries of the webdiagram of $X_{N,M}$ (and do not require the dualities in the extended K\"ahler moduli space $\mathbb{KE}(X_{N,M})$). The structure (\ref{CoxeterStructure}) suggests that any two of these generate a dihedral group (either of order $2N$ or $2M$ respectively). Similarly, the transformations $(\mathfrak{a}\,,\mathfrak{b})$ generate the dihedral group $\text{Dih}_{m_{\mathfrak{ab}}}$, which essentially generalises $\mathbb{G}(N)$ in (\ref{M1Generators}) to $M>1$. For $m_{\mathfrak{ab}}$ we find explicitly for the first few instances
\begin{center}
    \centering
    \begin{tabular}{ c || c | c | c | c | c | c | c | c}
          $m_{\mathfrak{ab}}$ & $N=1$ & $N=2$ & $N=3$ & $N=4$ & $N=5$ & $N=6$ & $N=7$ & $N=8$ \\ \hline \hline
          $M=1$ & 3 & 2 & 3 & $\infty$ & $\infty$ & $\infty$ & $\infty$ & $\infty$ \\ \hline
          $M=2$ & 2 & 3 & $\underline{\infty}$ & 4 & $\underline{\infty}$ & $\underline{\infty}$ & $\underline{\infty}$ & $\underline{\infty}$ \\ \hline
          $M=3$ & 3 & $\underline{\infty}$ & 6 & $\underline{\infty}$ & $\underline{\infty}$ & $\infty$ & $\underline{\infty}$ & $\underline{\infty}$ \\ \hline
          $M=4$ & $\infty$ & 4 & $\underline{\infty}$ & 6 & $\underline{\infty}$ & $\underline{\infty}$ & $\underline{\infty}$ & $\infty$ \\ \hline
          $M=5$ & $\infty$ & $\underline{\infty}$ & $\underline{\infty}$ & $\underline{\infty}$ & 6 & $\underline{\infty}$ & $\underline{\infty}$ & $\underline{\infty}$ \\ \hline
          $M=6$ & $\infty$ & $\infty$ & $\infty$ & $\underline{\infty}$ & $\underline{\infty}$ & 6 & $\underline{\infty}$ & $\underline{\infty}$ \\ \hline
          $M=7$ & $\infty$ & $\underline{\infty}$ & $\underline{\infty}$ & $\underline{\infty}$ & $\underline{\infty}$ & $\underline{\infty}$ & 6 & $\underline{\infty}$ \\ \hline
          $M=8$ & $\infty$ & $\infty$ & $\underline{\infty}$ & $\infty$ & $\underline{\infty}$ & $\underline{\infty}$ & $\underline{\infty}$ & 6
    \end{tabular}
\end{center}
where underlined entries indicate that $(N,M)$ can be dualised into a system $(N',M')$ with $NM=N'M'$, $\text{gcd}(N,M)=\text{gcd}(N',M')$ but $\text{min}(M',N')<\text{min}(M,N)$. This table suggests
\begin{equation}
    m_{\mathfrak{ab}}=\begin{cases}
      6, & \text{for}\, N=M>4, \\
      \infty, & \forall N,M > 4, \,\,\text{with} \,\, N\neq M,
    \end{cases}
\end{equation}
which has been further tested up to $N,M\leq12$. Since ${\rm Dih_3}\cong S_3$ and ${\rm Dih}_6 \cong S_3 \times \mathbb Z_2$, the full symmetry group always has an $S_3$ subgroup when $N=M$: indeed, in this case there is a natural symmetry of order $3$, which exchanges the parameters $(h_i^{(j)},v_i^{(j)},m_i^{(j)})$ in Figure~\ref{Fig:General Setup}.

Finally, $m^{sd}$ in (\ref{CoxeterStructure}) has a more complicated structure that depends on the specific choice of involutions in (\ref{Involutions}). We note, however, that it can be computed explicitly using the Mathematica package {\tt NPLSTsym} as explained in Appendix~\ref{App:MathematicaPackage}.

\subsection{Particular Point in the Moduli Space and $Sp(4,\mathbb{Z})$}
Another way to characterise the symmetry transformations discussed in Section~\ref{Sect:GenSymTrans} is through their action on a particular subspace of the extended K\"ahler moduli space of $X_{N,M}$. Indeed, with respect to the basis $(\widehat{a}_i^{(j)},\tau_i,\tau,S,\rho)$ a three-dimensional subspace can be defined by
\begin{align}
\mathbb{K}_3(X_{N,M})=\big\{x\in \mathbb{K}(X_{N,M})\mid x(\mathfrak{v})=\big(\overbrace{\tfrac{\rho}{N},\ldots,\tfrac{\rho}{N}}^{M(N-1)\text{-times}},\overbrace{\tfrac{\tau}{M},\ldots,\tfrac{\tau}{M}}^{M\text{-times}},\tau,S,\rho\big)\text{ with } (\tau,S,\rho)\in\mathbb{R}^3\big\}\,.\label{SubsetSpaceK3}
\end{align}
The symmetry transformations $(\mathfrak{c}^{\text{hor}},\mathfrak{c}^{\text{ver}},\mathfrak{r},\mathfrak{a},\mathfrak{b},\mathfrak{f})$ derived in the previous Section, map $\mathbb{K}_3$ into itself. Indeed, from the perspective of the basis of K\"ahler parameters, the corresponding matrices satisfy
\begin{align}
&\forall \vec{v}\in \mathbb{V}:\,\mathcal{S}\cdot \vec{v}\in\mathbb{V}\,,&&\text{with} &&\left\{\begin{array}{l}\mathcal{S}\in\{\mathcal{C}^{\text{hor}},\mathcal{C}^{\text{ver}},\mathcal{R},\mathcal{A},\mathcal{B},\mathcal{F}\}\,, \\[4pt] \mathbb{V}=\big\{\big(\underbrace{\tfrac{\rho}{N},\ldots,\tfrac{\rho}{N}}_{M(N-1)\text{-times}},\underbrace{\tfrac{\tau}{M},\ldots,\tfrac{\tau}{M}}_{M\text{-times}},\tau,S,\rho\big)\mid(\tau,S,\rho)\in\mathbb{R}^3\big\}\,.\end{array}\right.\nonumber
\end{align}
Concretely, the transformations act on the remaining K\"ahler parameters $(\tau,S,\rho)$ of $\mathbb{V}$ as
\begin{align}
&\mathfrak{c}^{\text{hor}}\,,\mathfrak{c}^{\text{ver}}\,,\mathfrak{r}\,:\hspace{0.5cm}(\tau,S,\rho)\longrightarrow (\tau,S,\rho)\,,\nonumber\\
&\mathfrak{a}\,:\hspace{2.2cm}(\tau,S,\rho)\longrightarrow \left(\tau,\tau-S,\tfrac{N(\tau-2S)}{M}+\rho\right)\,,\nonumber\\
&\mathfrak{b}\,:\hspace{2.2cm}(\tau,S,\rho)\longrightarrow \left(\tau+\tfrac{N(\rho-2S)}{M},\rho-S,\rho\right)\,,\nonumber\\
&\mathfrak{f}\,:\hspace{2.25cm}(\tau,S,\rho)\longrightarrow \left(\tau,-S,\rho\right)\,.\label{Sp4Trafos}
\end{align}
Furthermore, it was argued in \cite{Hohenegger:2016eqy} that for any point in $\mathbb{K}_3(X_{N,M})$ the Nekrasov-Shatashvili limit of the free energy can be related to the free energy of the theory labelled by $(N,M)=(1,1)$, upon rescaling the three remaining K\"ahler parameters $(\tau,S,\rho)$. Since the partition function $\mathcal{Z}_{1,1}(\tau,S,\rho,\epsilon_{1,2})$ was argued to be invariant under an $Sp(4,\mathbb{Z})$ symmetry \cite{Dijkgraaf:1996xw,Hohenegger:2013ala}, we expect such a symmetry to be also present for $\mathcal{Z}_{N,M}$ for all points in $\mathbb{K}_3(X_{N,M})$ (at least in the NS-limit) and the transformations (\ref{Sp4Trafos}) to naturally fit into it. To show that this is the case, we first define a map from $\mathbb{K}_3(X_{N,M})$ to a genus two Riemann surface, by arranging the three parameters $(\tau,S,\rho)$ in $\mathbb{V}$ into the following period matrix 
\begin{align}
&\Omega=\left(\begin{linesmall}{@{}cc@{}}\tau/M & S/M \\ S/M & \rho/N
\end{linesmall}\right)\,.
\end{align}
An $Sp(4,\mathbb{Z})$ action\footnote{In fact, in \cite{Bender} a representation of $Sp(4,\mathbb{Z})$ has been given in terms of 2 generators that satisfy 8 relations. The former can be given in terms of the following 2 matrices
\begin{align}
&K=\left(\begin{linesmall}{@{}cccc@{}} 1 & 0 & 0 & 0  \\ 1 & -1 & 0 & 0 \\ 0 & 0 & 1 & 1 \\ 0 & 0 & 0 & -1\end{linesmall}\right)\,,&&\text{and} &&L=\left(\begin{linesmall}{@{}cccc@{}} 0 & 0 & -1 & 0  \\ 0 & 0 & 0 & -1 \\ 1 & 0 & 1 & 0 \\ 0 & 1 & 0 & 0\end{linesmall}\right)\,.\label{Sp4Presentation}
\end{align}}  on $\Omega$ can then be defined as
\begin{align}
&\left(\begin{linesmall}{@{}cc@{}} A & B \\ C & D
\end{linesmall}\right):\,\Omega\longmapsto (A\,\Omega+B)(C\,\Omega+D)^{-1}\,,&&\text{with}&&\begin{array}{l}A^T D-C^T B=\mathbbm{1}_{2\times 2}\,,\\ 
DA^T-CB^T=\mathbbm{1}_{2\times 2}\,,\end{array}&&\text{and} &&\begin{array}{l}A^T C=C^T A\,,\\
B^T D=D^T B\,,\end{array}\nonumber
\end{align}
where $A,B,C,D$ are $2\times 2$ matrices with integer entries.
 The transformations (\ref{Sp4Trafos}) can indeed be arranged in this form by\footnote{These $Sp(4,\mathbb{Z})$ transformations are not unique, since, for example, changing the sign of $A$ also reproduces the transformations (\ref{Sp4Trafos}).}
\begin{align}
&B=C=\left(\begin{linesmall}{@{}cc@{}} 0 & 0 \\ 0 & 0
\end{linesmall}\right)\,,&&(D^{-1})^T=A\,,&&A=\left\{\begin{array}{lcl} \left(\begin{linesmall}{@{}cc@{}} 1 & 0 \\ 0 & 1
\end{linesmall}\right) & \text{for} & \mathfrak{c}^{\text{hor}}\,,\mathfrak{c}^{\text{ver}}\,,\mathfrak{r}\,,\\[4pt]
\left(\begin{linesmall}{@{}cc@{}} 1 & 0 \\ 1 & -1 \end{linesmall}\right) & \text{for} & \mathfrak{a}\,, \\[4pt]
\left(\begin{linesmall}{@{}cc@{}} 1 & -N/M \\ 0 & -1 \end{linesmall}\right) & \text{for} & \mathfrak{b}\,, \\[4pt] 
\left(\begin{linesmall}{@{}cc@{}} 1 & 0 \\ 0 & -1 \end{linesmall}\right) & \text{for} & \mathfrak{f}\,. 
\end{array}\right.
\end{align} 
From the perspective of this $Sp(4,\mathbb{Z})$, the transformations $\mathfrak{c}^{\text{hor}}\,,\mathfrak{c}^{\text{ver}}\,,\mathfrak{r}$ therefore act trivially, while $\mathfrak{f}$ can be identified with the element $L^6$ (with $L$ given in (\ref{Sp4Presentation})). The remaining transformations can naturally be related to generators of a $PSL(2,\mathbb{Z})\subset Sp(4,\mathbb{Z})$: indeed, the $Sp(4,\mathbb{Z})$ action of $\mathfrak{a}\circ \mathfrak{f}$ and $\mathfrak{f}\circ \mathfrak{b}$ can be written as
\begin{align}
&B=C=\left(\begin{linesmall}{@{}cc@{}} 0 & 0 \\ 0 & 0
\end{linesmall}\right)\,,&&(D^{-1})^T=A\,,&&A=\left\{\begin{array}{lcl} S\, T\, S & \text{for} & \mathfrak{a}\circ \mathfrak{f}\,,\\[4pt] T^{N/M} & \text{for} &  \mathfrak{b}\circ \mathfrak{f}\,, 
\end{array}\right.
\end{align} 
where $T=\left(\begin{linesmall}{@{}cc@{}} 1 & 1  \\ 0 & 1 \end{linesmall}\right)$ and $S=\left(\begin{linesmall}{@{}cc@{}} 0 & 1  \\ -1 & 0 \end{linesmall}\right)$ are the generators of $PSL(2,\mathbb{Z})$. We note, that $\mathfrak{a}\circ \mathfrak{f}$ and $\mathfrak{b}\circ \mathfrak{f}$ are not involutions (and indeed the orders of $\mathcal{A} \mathcal{F}$ and $\mathcal{B} \mathcal{F}$ are infinite.)






\section{Little String Theories and Quiver Algebras}\label{Sec:ZLSTVOA}
The symmetries discussed in the previous Section act as linear transformations on the K\"ahler parameters of the manifold $X_{N,M}$. In Section~\ref{SubSect:KahlerParameters} we have interpreted these parameters in terms of a LST with specific gauge structure and matter content. As we shall discuss in more detail in the subsequent Section~\ref{Sect:ActionSymmetries}, $(\mathfrak{c}^{\text{hor}},\mathfrak{c}^{\text{ver}},\mathfrak{r},\mathfrak{a},\mathfrak{b},\mathfrak{f})$ are thus (non-perturbative) symmetries of this theory, which notably leave the partition function $\mathcal{Z}_{N,M}$ invariant. However, due to an extensive network of dualities, this theory admits various other descriptions exhibiting additional algebraic structures. In this Section we shall review one particular such dual description \cite{Kimura:2015rgi,Kimura:2016dys,Kimura:2017hez} that enjoys an underlying  {\it affine quiver algebra},\footnote{In the original papers~\cite{Kimura:2015rgi,Kimura:2016dys,Kimura:2017hez}, it has been called the (affine) quiver W-algebra. Since we do not focus on the W-algebraic aspects in this paper, we simply call it the affine quiver algebra.} while in the subsequent Section we shall exhibit how the non-perturbative symmetries found above act on this algebraic structure.

Concretely, in the formulation provided in \cite{Kimura:2015rgi,Kimura:2016dys,Kimura:2017hez} (see also~\cite{Kimura:2020jxl} for a review), the partition function $\mathcal{Z}_{N,M}$ (or more generally the partition function of a quiver gauge theory in five and six dimensions with eight supercharges) can be written as the correlation function of specific vertex operators (called screening currents) that are generated by an infinite-dimensional Heisenberg algebra. In the following, we shall review this construction highlighting in particular the underlying affine quiver algebra. In order to keep the discussion as simple as possible (and to focus on the algebraic structures and symmetries, which are independent of this choice), we shall work in the unrefined limit $\epsilon_1=-\epsilon_2=\epsilon$ (with the notation $q=e^{2\pi i \epsilon}$): the generalisation to the refined case is straight-forward. 


\subsection{Affine Quiver Algebra}
Let $\Gamma$ be a quiver with $M$ nodes (labelled by $i,j=1,\ldots,M$) that are connected by oriented edges $e: i\to j$. We define the $m$-th mass-deformed Cartan matrix of $\Gamma$ as (we recall $q=e^{2\pi i \epsilon_1}$ and $t=e^{-2\pi i \epsilon_2}$)
\begin{align}
    c_{ij}^{[m]} = (1 + (t/q)^m) \delta_{ij} - \sum_{e: i\to j} \mu_e^m (t/q)^m - \sum_{e:j\to i} \mu_e^{-m}\,,&& \forall m\in\mathbb{Z}\,, \label{DefCartanRef}
\end{align}
where $\mu_e$ is the mass deformation parameter associated to the edge $e$.
In the unrefined limit $\epsilon_1=-\epsilon_2=\epsilon$ (such that $t=q$), this deformed Cartan matrix becomes independent of $\epsilon$,
\begin{align}
&c_{ij}^{[m]} = 2 \delta_{ij} - \sum_{e: i\to j} \mu_e^m - \sum_{e:j\to i} \mu_e^{-m}\,,&& \forall m\in\mathbb{Z}\, . \label{DefCartan}
\end{align}
As explained in \cite{Kimura:2015rgi,Kimura:2016dys,Kimura:2017hez}, this matrix describes an algebraic structure that is encoded in the form of the quiver $\Gamma$ itself. 
For the theories with eight supercharges, every edge appears as a pair of $i \to j$ and $j \to i$ that forms a bifundamental hypermultiplet. Hence, the change of orientation of each edge $i \to j$ and $j \to i$ gives rise to the following symmetry in general,
\begin{align}
    \mu_e (t/q) \leftrightarrow \mu_e^{-1} \quad \xrightarrow{q = t} \quad \mu_e \leftrightarrow \mu_e^{-1} . \label{QuiverOrientationSym}
\end{align}
Therefore, we do not need to take care of orientations of the edges, and thus such a quiver can be identified with the Dynkin diagram encoding a Lie algebra structure. Moreover, it is also instrumental in writing the partition function of the six dimensional gauge theory with bifundamental matter associated with $\Gamma$, which can be written in terms of correlation functions of so-called screening currents \cite{Kimura:2016dys}\footnote{In full rigour, the screening current comprises an additional component corresponding to the Chern-Simons term found in 5d gauge theory. This term is also essential in the context of 6d theories to ensure the matching between the algebraic formalism and the gauge theory, as discussed in \cite{Kimura:2016dys}. We do not address this additional term, as it cancels out in the specific case under examination.}: 
\begin{align}
&S_{i,x} = {: \exp \left( s_{i,0} \log x + \Tilde{s}_{i,0} +\sum_{m\in\mathbb Z^*} s_{i,m}^{(+)} x^{-m} + s_{i,m}^{(-)} x^m \right) :}\,,&&\forall i\in \{1,\ldots,M\}\,.\label{eq:screeningcurrents}
\end{align}
Here $s_{i,m}^{(\pm)}, \Tilde{s}_{i,0}$ are free field modes (which can be constructed from a Heisenberg algebra \cite{Kimura:2015rgi}). The symbol $:-:$ means the normal ordering where all the creation operators $(s_{i,m}^{(\pm)})_{m\le0}$ are placed to the left of the annihilation operators $(s_{i,m}^{(\pm)})_{m > 0}$ and $\tilde{s}_{i,0}$. We denote the vacuum state of the free field Fock space by $|0\rangle$ obeying $s_{i,m > 0}^{(\pm)} |0\rangle = \langle 0 | s_{i,m \le 0}^{(\pm)} = 0$. The vacuum expectation value of any normal ordered operators becomes zero, $\left< 0 | :\mathcal{O}_1 \cdots \mathcal{O}_n: | 0 \right> = 0$. The free field modes are defined by the following commutation relations governed by the deformed Cartan matrix~(\ref{DefCartan}):
\begin{align}
&\left[s_{i,m}^{(\pm)},s_{j,m^\prime}^{(\pm)} \right] = \mp \frac{c_{ji}^{[\pm m]}}{m(1-\Qr^{\pm m})} \delta_{m+m^\prime,0}\,,&& \left[\Tilde s_{i,0}, s_{j,m}^{(\pm)} \right]= - \delta_{m,0}\, c_{ji}^{[0]}\,.\label{CommutationRelations}
\end{align}
Here $\rho$ is an elliptic parameter and $x$ in (\ref{eq:screeningcurrents}) is the (formal) position of the vertex operator on an elliptic curve $\mathscr{E}=\mathbb{C}/(\mathbb{Z}+\rho\mathbb{Z})$. Realising the web diagram discussed in Section~\ref{Sect:OverviewPartitionFunction} using the five-brane web, $x$ specifies the position of D5 branes~\cite{Kimura:2022zsm,Kimura:2023bxy}. To compute the partition function $\mathcal{Z}_{N,M}$ of the LSTs discussed in Section~\ref{Sect:OverviewPartitionFunction}, $\rho$ is identified with the corresponding K\"ahler parameter in (\ref{KaehlerBasis}), while $x$ is taken to be different combinations of the (exponential of) $\widehat{b}_i^{(j)}$ appearing in (\ref{DefBParameters}). From this point of view, it seems to be natural to regard the K\"ahler parameters as complex variables in this formalism. In fact, this complexification plays an important role in the computation of the partition function in the vertex operator formalism as explained below. We shall then write the partition function $\mathcal{Z}_{N,M}$ in (\ref{DefPartitionFunction}) as a combination of elliptic two-point correlation functions of the screening currents of the following form
\begin{equation}\label{corSS}
    \bigg\langle 0 \, \bigg| S_{i,Q_{\widehat b_k^{(i)}}} S_{j,Q_{\widehat b_l^{(j)}}} \bigg| \, 0 \bigg\rangle = \exp \left( -\sum_{n \in \mathbb Z^*} \frac{c_{ji}^{[n]}}{n (1-\Qr^n)} \frac{Q_{\widehat b_l^{(j)}}^n}{Q_{\widehat b_k^{(i)}}^n} \right)\,,
\end{equation}
To make this construction more concrete, we first consider $M=1$ in the following Subsection, before moving on to the general case in Section~\ref{Sec:QuiverGenM}.

\subsection{The case $(N,1)$ or the $\widehat{A}_0$ Quiver}\label{N1QuiverAlg}
We first consider the $\widehat A_0$ quiver with one $U(N)$ node and matter in the adjoint representation

\begin{wrapfigure}{r}{0.21\textwidth}
${}$\\[0.3cm]
\centering
\scalebox{0.9}{\parbox{1.3cm}{\begin{tikzpicture}
\draw[ultra thick] (2,0.6) circle (0.6cm);
\draw[ultra thick,fill=white] (2,0) circle (0.5cm);
\node at (2,0) {\scriptsize$U(N)$};
\end{tikzpicture}}}\caption{\emph{$\widehat{A_0}$ quiver with one $U(N)$ node and adjoint matter}}
\label{Fig:A0Quiver}
\end{wrapfigure}

\noindent
as shown in Figure~\ref{Fig:A0Quiver}. The $n$-th mass-deformed Cartan matrix, as defined in (\ref{DefCartanRef}), for this quiver reads explicitly
\begin{align}\label{cartana0}
&c^{[n]} = [1 + (t/q)^n - (t/q)^n \Qs^n - \Qs^{-n}] \xrightarrow{q = t} [2 - \Qs^n - \Qs^{-n}]\,.
\end{align}
As discussed in Section~\ref{Sect:OverviewPartitionFunction}, the UV completion of the gauge theory associated with this quiver is a Little String Theory (indeed, Figure~\ref{Fig:A0Quiver} is a particular case of the quiver shown in Figure~\ref{Fig:Quiver}). Using the topological vertex formalism (\emph{i.e.} the formalism reviewed in Section~\ref{subsec:partitionfunction}) the partition function of this LST was written in \cite{Filoche:2022qxk} as
\begin{align}
\mathcal{Z}_{N,1}=W_N(\emptyset)\,&\sum_{\alpha_1,\ldots,\alpha_N}Q_\tau^{|\alpha_1|+\ldots+|\alpha_N|}\,\left(\prod_{k=1}^N \frac{\vartheta_{\alpha_k\alpha_k}(Q_S;\rho)}{\vartheta_{\alpha_k\alpha_k}(1;\rho)}\right) \prod_{1\leq i<j\leq N}\mathcal{T}_{\alpha_j\alpha_i}(\rho,S,\widehat{a}^{(1)}_{1,\ldots,N-1},\epsilon)\,,\label{N1PartitionFunction}
\end{align}
where we use the same notation as in Section~\ref{Sect:OverviewPartitionFunction}. Specifically, the Nekrasov subfunctions
\begin{align}
\mathcal{T}_{\alpha_j \alpha_i} &= \left(- \frac{\phi_{-2,1}(S;\rho)}{4\pi^2} \right)^{|\alpha_i| + |\alpha_j|}\, \widehat{\mathcal{T}}_{\alpha_j\alpha_i}(\rho,S,\widehat{a}_{1,\ldots,N}^{(1)},\epsilon)\,,\label{DefNekrasovSubfun}
\end{align}
can be written in terms of Kronecker-Eisenstein series (see \eqref{DefOmega} for the definition), \emph{i.e.}
\begin{align}
\widehat{\mathcal{T}}_{\alpha_j\alpha_i}&=\prod_{(r,s)\in \alpha_j} \Omega(\widehat b_j^{(1)} - \widehat b_i^{(1)} + n_{r,s}^{\alpha_j,\alpha_i} \epsilon,S;\rho)\,\Omega(\widehat b_j^{(1)} - \widehat b_i^{(1)} + n_{r,s}^{\alpha_j,\alpha_i} \epsilon,-S;\rho) \nonumber\\
    & \hspace{0.6cm}\times \prod_{(r,s)\in \alpha_i} \Omega(\widehat b_j^{(1)} - \widehat b_i^{(1)} - n_{r,s}^{\alpha_i,\alpha_j}\epsilon,S;\rho)\,\Omega(\widehat b_j^{(1)} - \widehat b_i^{(1)} - n_{r,s}^{\alpha_i,\alpha_j}\epsilon,-S;\rho)\,.\label{FormNekSub}
\end{align}
In (\ref{DefNekrasovSubfun}) $\phi_{-2,1}$ is a standard Jacobi form of weight $-2$ and index $1$, which is defined in (\ref{DefPhi21}), and $n_{r,s}^{\alpha_i,\alpha_j}$ in (\ref{FormNekSub}) are combinatoric numbers which are explicitly given in \eqref{eq:nalphalpha}. 

 Using the definition of Kronecker-Eisenstein series in term of Jacobi theta functions~\eqref{DefOmega} and the exponential form of the Jacobi theta functions~\eqref{eq:thetafunction}, pairs of Kronecker-Eisenstein series can be written as correlation functions (\ref{corSS}) of screening currents. In fact, the full Nekrasov subfunctions (\ref{DefNekrasovSubfun}) can be re-written in this fashion \cite{Kimura:2016dys}: to this end, we perform the products over $n\in\{n_{r,s}^{\alpha_j,\alpha_i},\,\,(r,s) \in \alpha_j\}\cup \{-n_{r,s}^{\alpha_i,\alpha_j},\,\,(r,s)\in \alpha_i \}$
\begin{equation}
    \mathcal{T}_{\alpha_j \alpha_i} = \exp \left(-\sum_{n\in \mathbb Z^*} \frac{\Qs^n + \Qs^{-n} - 2}{n(1-\Qr^n)} \frac{Q_{\widehat b_j^{(1)}}^n}{Q_{\widehat b_i^{(1)}}^n} f_{\alpha_j \alpha_i}(q^n) \right)\,.\label{FormNekSubRed}
\end{equation}
The functions $f_{\alpha_j \alpha_i}$ were first introduced in the context of Chern-Simon theory in \cite{Iqbal:2003ix} and are defined in Appendix~\eqref{eq:fCSalphalpha}. 
The product over all such contributions that appears in the partition function (\ref{N1PartitionFunction}), can in turn be re-written \cite{Kimura:2016dys} in terms of correlation functions of radially ordered products of elliptic screening currents
\begin{equation}
    \Qt^{\sum_{i=1}^N \left|\alpha_i\right|} \prod_{1\leq i<j\leq N} {\mathcal T}_{\alpha_j \alpha_i} = \bigg\langle 0 \, \bigg| \prod_{x\in \mathcal{X}_{\alpha_1, \ldots \alpha_N}}^\prec S_{0,x} \, \bigg| \, 0 \bigg\rangle \Big/ \bigg\langle 0 \, \bigg| \prod_{x\in \mathcal{X}_{\emptyset, \ldots \emptyset}}^\prec S_{0,x} \, \bigg| \, 0 \bigg\rangle\,.
\end{equation}
In this context, $\mathcal{X}_{\alpha_1, \ldots ,\alpha_N}$ represents the collection of Chern characters corresponding to the fixed point (identified by the partitions $\alpha_1, \ldots, \alpha_N$) of the moduli space of instantons under its automorphism group. This group is the direct product of the $U(1)$ transformations generated by the regularisation parameters $\epsilon_{1,2}$ (with $\epsilon_1 + \epsilon_2 = 0$; $q = t$), as well as the Cartan subgroups of the $U(N)$ gauge group. The set $\mathcal{X}_{\alpha_1, \ldots ,\alpha_N}$ can be expressed in terms of the K\"ahler parameters of $X_{N,1}$ introduced in the previous Section: 
\begin{equation}
    \mathcal{X}_{\alpha_1,\ldots,\alpha_N} = \bigsqcup_{i \in \{ 1, \ldots,N \}} \mathcal{X}_{\alpha_i}, \quad \mathcal{X}_{\alpha_i} = \left\lbrace Q_{\widehat b_i^{(1)}} q^{-\alpha_{i,k}+k-1} \right\rbrace_{k \in \{1,\ldots,\infty\}}.
\end{equation}
We can thus also write
\begin{equation}
    \Qt^{|\alpha_j|}\, \mathcal{T}_{\alpha_j \alpha_i} = \bigg\langle 0 \, \bigg| \prod_{x\in \mathcal{X}_{\alpha_i}\bigsqcup \mathcal{X}_{\alpha_j} }^\prec S_{0,x} \, \bigg| \, 0 \bigg\rangle \Big/ \bigg\langle 0 \, \bigg| \prod_{x\in \mathcal{X}_{\emptyset} \bigsqcup \mathcal{X}_\emptyset}^\prec S_{0,x} \, \bigg| \, 0 \bigg\rangle, \quad \quad \forall j>i \, ,
    \label{eq:S-correlaotr_ratio}
\end{equation}
where we assume $|Q_{\widehat b_j^{(1)}}/Q_{\widehat b_i^{(1)}}| \gg 1 $ for $j > i$ and $|q|<1$ to guarantee the radial ordering in the computation:\footnote{Such an assumption is typically applied to discuss the Seiberg--Witten geometry from the instanton partition function~\cite{Nekrasov:2003rj}.} The sequences $\{\alpha_{i,k} - k + 1\}_{i = 1,\ldots,N,k=1,\ldots,\infty}$ are strictly decreasing as $\alpha_{i,k} - k + 1 > \alpha_{i,k+1} - (k+1) + 1$, while the partitions are non-increasing sequences, $\alpha_{i,k} \ge \alpha_{i,k+1}$.
Although the correlation function of the screening currents involve infinite products (due to the one-loop contribution), the ratio of them appearing in \eqref{eq:S-correlaotr_ratio} gives rise to the Nekrasov subfunction, which is a finite product of theta function factors, and thus we can take all the K\"ahler parameters real again after the computation.

\subsection{The case $(N,M)$ or the $\widehat{A}_{M-1}$ Quiver}\label{Sec:QuiverGenM}
The discussion above can be generalised to an $\widehat A_{M-1}$ quiver, \emph{i.e.} the LST associated to the circular quiver shown in Figure~\ref{Fig:Quiver} with $M$ gauge nodes of type $U(N)$ and bifundamental hypermultiplet matter. In this case, the partition function $\mathcal{Z}_{N,M}$ in (\ref{DefPartitionFunction}) does not admit a straight-forward simple decomposition in terms of Kronecker-Eisenstein series along the lines of (\ref{FormNekSub}). However, it can still be formulated in terms of correlation functions of screening currents exhibiting a non-trivial $\widehat{\mathfrak{a}}_{M-1}$ affine quiver algebra~\cite{Kimura:2016dys}. To make the latter manifest, we start from the $n$-th mass deformed Cartan matrix (\ref{DefCartan}), which for the quiver in Figure~\ref{Fig:Quiver} reads explicitly
\begin{equation}\label{eq:genMCartan}
    c_{ij}^{[n]} = 2 \delta_{ij} - \left(Q_{\widehat S}^{-n} \,\delta_{i,j+1} + Q_{\widehat S}^{n}\,\delta_{i+1,j}\right) \, , \quad \begin{array}{l}\forall n \in \mathbb Z \hspace{0.2cm}\text{and} \hspace{0.2cm}i,j\in \{1,\ldots,M\} \text{ mod }M\, ,\\
    \widehat{S}=S/M\,.\end{array}
\end{equation}
and which factorises as follows
\begin{align}
&c^{[n]}=2\mathbbm{1} - \mathbf{Q}_{\widehat S} - \mathbf{Q}_{\widehat S}^{-1} = (\mathbbm{1}  - \mathbf{Q}_{\widehat S})(\mathbbm{1}  - \mathbf{Q}_{\widehat S}^{-1})\,,&&\text{with} &&\mathbf{Q}_{\widehat S}=Q^n_{\widehat{S}}\begin{psmall}0 & 1 & 0 & \ldots  & 0 \\ 0 & 0 & 1 &  \ldots & 0 \\[-4pt] \vdots & & \vdots & \ddots & \vdots & \\[1pt] 0 & 0 & 0 &  \ldots &  1 \\ 1 & 0 & 0 &  \ldots &  0\end{psmall}\,.
\end{align}
The fact that the Cartan matrix (and thus also the LST orbifold theory) admits only one mass deformation parameter is compatible with the constraints imposed by the Calabi-Yau condition of the manifold $X_{N,M}$ discussed in Section~\ref{SubSect:KahlerParameters}. The partition function (\ref{DefPartitionFunction}) can then be rewritten in terms of correlators of the form (\ref{corSS}), indeed we have  \cite{Kimura:2016dys}:
\begin{align}
    \Qt^{\sum_{i=1}^N \left| \alpha_i^{(1)}\right|} &\prod_{j=1}^{M-1} Q_{\tau_j}^{\sum_{i=1}^N \left|\alpha_i^{(j+1)}\right|-\left|\alpha_i^{(1)}\right|} \,\mathcal{P}_{\vec\alpha^{(1)},\ldots,\vec\alpha^{(M)}} \nonumber\\
  &= \bigg\langle 0 \, \bigg| \prod_{x\in \mathcal{X}_{\vec\alpha^{(1)}, \ldots, \vec\alpha^{(M)}}}^\prec S_{{\bf i}(x),x} \, \bigg| \, 0 \bigg\rangle \Big/ \bigg\langle 0 \, \bigg| \prod_{x\in \mathcal{X}_{\emptyset, \ldots, \emptyset}}^\prec S_{{\bf i}(x),x} \, \bigg| \, 0 \bigg\rangle \, ,\label{PartitionFunctionCorrelators}
\end{align}
with ${\bf i}: \mathcal{X} \to \{1, \ldots, M \}$ a map such that $\forall x \in \mathcal{X}_{\vec \alpha^{(j)}}:\,\, {\bf i}(x)=j$. Moreover, the collection of Chern characters, denoted as $\mathcal{X}_{\vec \alpha^{(1)},\ldots,\vec \alpha^{(M)}}$ and defined in accordance with Subsection~\ref{N1QuiverAlg}, can be expressed as a disjoint union over $NM$ partitions:
\begin{equation}\label{eq:genMfixpoints}
    \mathcal{X}_{\vec \alpha^{(1)},\ldots,\vec \alpha^{(M)}} = \bigsqcup_{j\in \{1,\ldots,M\}} \bigsqcup_{i \in \{1,\ldots, N\}} \mathcal{X}_{\alpha_i^{(j)}}, \quad \mathcal{X}_{\alpha_i^{(j)}} = \{\nu_i^{(j)} q^{-\alpha_{i,k}^{(j)}+k-1} \}_{k \in \{1,\ldots,\infty \}}\,.
\end{equation}
Here $\nu_i^{(j)}$ is a set of $NM$ Coulomb moduli of the quiver gauge theory, which can be written as combinations of the $(N-1)M$ K\"ahler parameters $\widehat{a}_i^{(j)}$ (with $i \in \{1,\ldots,N-1\}$ and $j \in \{1,\ldots,M\}$) of $X_{N,M}$, which were introduced in Figure~\ref{Fig:General Setup}. Concretely, the $\nu_i^{(j)}$ correspond to solutions of the following matching conditions:
\begin{equation}
    \frac{\nu_i^{(s)}}{\nu_j^{(s+1)}} = \widehat Q_{i,i-j}^{(s)} Q_{\widehat S}^{-1}, \quad \frac{\nu_i^{(s)}}{\nu_j^{(s)}} = \overline{Q}_{i,i-j}^{(s)}, \quad \forall i,j,s,\quad 1 \leq i <j \leq N, \quad 1 \leq s \leq M, 
\end{equation}
where $\widehat Q$ and $\overline{Q}$ are defined in~\eqref{eq:Qvariables}. These matching conditions correspond to the doubly elliptic version of the ones found in~\cite{Bao:2011rc}.
In order to impose the radial ordering properly, we assume the relation $|\widehat Q_{i,i-j}^{(s)} Q_{\widehat S}^{-1}| \gg |\overline{Q}_{i,i-j}^{(s)}| \gg 1$ for $i<j$ in the computation of the correlation function. After obtaining the partition function, we can then relax this condition via analytic continuation.




\section{Action of the Symmetries on the Quiver Algebra}\label{Sect:ActionSymmetries}
The re-organisation (\ref{PartitionFunctionCorrelators}) allows to re-write the partition functions $\mathcal{Z}_{N,M}$ in (\ref{DefPartitionFunction}) in a way, which makes the affine quiver algebra (\emph{i.e.} the affine algebra encoded in the form of the quiver in Figure~\ref{Fig:Quiver}) manifest: the key element in this construction is the (deformed) Cartan matrix (\ref{DefCartan}) (concretely (\ref{eq:genMCartan}) for the LSTs discussed in Section~\ref{Sect:OverviewPartitionFunction}). Since the latter explicitly depends on the mass parameters of the LST, which transform in a non-trivial manner under the symmetries found in Section~\ref{Sect:ReviewSyms}, the latter also act on the affine quiver algebra. This Section is dedicated to making this action more precise: for concreteness, we shall first start with the case $M=1$ (for which the action on the LST free energy has previously been discussed in \cite{Bastian:2018jlf}), before discussing the general case.

\subsection{Physical Symmetries for $M=1$}
For $M=1$, the symmetries\footnote{For $M=1$, we do not include the symmetry $\mathfrak{c}^{\text{ver}}$, which is trivial in this case.} $\mathfrak{r}$, $\mathfrak{c}^{\text{hor}}$, $\mathfrak{f}$, $\mathfrak{a}$ and $\mathfrak{b}$ act through multiplication by the corresponding matrices on the K\"ahler parameters $\vec{v}=(\widehat{a}_1^{(1)},\ldots,\widehat{a}_{N-1}^{(1)},\tau,S,\rho)$. Such a simple action was found in \cite{Bastian:2018jlf} also on the Fourier coefficients of the free energy associated with the partition function $\mathcal{Z}_{N,1}$. Indeed, upon introducing\footnote{Here and in \cite{Bastian:2018jlf} the free energy is defined using the plethystic logarithm, which counts only single particle BPS states. The result can be generalised to also include multiparticle states, \emph{i.e.} the full BPS free energy.}
\begin{align}
\mathcal{F}_{N,1}=\text{PLog} \mathcal{Z}_{N,1}(\vec{v};\epsilon_{1,2}):=\sum_{k=1}^\infty \frac{\mu(k)}{k}\,\ln\mathcal{Z}_{N,1}(k\vec{v};\epsilon_{1,2})\,,
\end{align}
where $\text{PLog}$ denotes the plethystic logarithm and $\mu$ the M\"obius function, we can (at least formally) write the following Fourier series
\begin{align}
\mathcal{F}_{N,1}=\sum_{n,m=0}^\infty Q_\tau^n Q_\rho^m \sum_{\ell_1,\ldots,\ell_{N-1}\in\mathbb{Z}} e^{2\pi i \sum_{r=1}^{N-1}\ell_r\,\widehat{a}_r^{(1)}}\,\sum_{k\in\mathbb{Z}}\,Q_S^k\,f_{\ell_1,\ldots,\ell_{N-1},n,k,m}(\epsilon_{1,2})\,,
\end{align}
where $f_{\ell_1,\ldots,\ell_{N-1},n,k,m}$ are ($\epsilon$-dependent) Fourier coefficients. The fact that the symmetries $\mathfrak{r}$, $\mathfrak{c}^{\text{hor}}$, $\mathfrak{f}$, $\mathfrak{a}$ and $\mathfrak{b}$ leave the partition function invariant then leads to the following non-trivial relations among the latter
\begin{align}\label{fouriercoeftrans}
&f_{\ell_1,\ldots,\ell_{N-1},n,k,m}=f_{\ell'_1,\ldots,\ell'_{N-1},n',k',m'}\,,&&\text{for} &&(\ell'_1,\ldots,\ell'_{N-1},n',k',m')^T=\mathcal{S}^T\cdot (\ell_1,\ldots,\ell_{N-1},n,k,m)^T\,,
\end{align}
with $\mathcal{S}$ any of the matrices in $\{\mathcal{R},\mathcal{C}^{\text{hor}},\mathcal{F},\mathcal{A},\mathcal{B}\}$. These relations were extensively tested in \cite{Bastian:2018jlf}, also in the case of generic $\epsilon_{1,2}$. 

Invariance of the free energy a priori also suggests invariance of the partition function itself. For certain symmetries, the latter can in fact also be verified directly. To understand this, it is instructive to consider some examples 
\begin{itemize}
\item {\bf Cyclic permutation $\mathfrak{c}^{\text{hor}}$:}\\
The symmetry $\mathfrak{c}^{\text{hor}}$ acts in the following fashion on $\vec{v}$
\begin{align}
&\mathfrak{c}^{\text{hor}}:&& (\widehat{a}^{(1)}_1, \widehat{a}^{(1)}_2 \ldots, \widehat{a}^{(1)}_{N-1} , \rho , S ,\tau) \longrightarrow \left(\widehat{a}^{(1)}_2,\widehat{a}^{(1)}_3, \ldots , \widehat{a}^{(1)}_{N-1}, \rho - \sum_{i=1}^{N-1} \widehat{a}^{(1)}_i , \rho , S ,\tau\right)\,.\nonumber
\end{align}
which therefore does not mix different instanton levels in the partition function (since $\tau$ is left invariant). Moreover, the fact that this transformation is a symmetry of $\mathcal{Z}_{N,1}$ can be directly understood from the form (\ref{DefPartitionFunction}) of the partition function, which in the unrefined limit can be rewritten as in (\ref{N1PartitionFunction}). In this form, the symmetry $\mathfrak{c}^{\text{hor}}$ acts as a simple cyclic permutation on the argument of the Nekrasov subfunction
\begin{equation}\label{mapChorN1moduli}
    \mathfrak{a}_{ij}=\sum_{k=i}^{j-1} \widehat{a}_k=\widehat{b}_j^{(1)}-\widehat{b}_i^{(1)} \xrightarrow{\quad \mathfrak{c}^{\text{hor}} \quad} \begin{cases}
      \mathfrak{a}_{(i+1)(j+1)}, & \text{if}\ 1 < j < N, \\
      \rho - \mathfrak{a}_{1(i+1)}, & \text{if}\ j = N.
    \end{cases}
\end{equation}
Furthermore, since 
\begin{align}
    \mathcal{T}_{\alpha_N\alpha_i}(\rho-\mathfrak{a}_{1(i+1)},\rho,S;\epsilon)&=\exp \left(-\sum_{m\in \mathbb{Z}^*} \frac{1}{m} \frac{\Qs^m + \Qs^{-m} -2}{1-\Qr^m} f_{\alpha_N \alpha_i}(q^m) \Qr^m Q_{\mathfrak{a}_{1(i+1)}}^{-m} \right)\nonumber\\
    &= \mathcal{T}_{\alpha_i \alpha_N}(\mathfrak{a}_{1(i+1)},\rho,S;\epsilon)\,,
\end{align}
where we have used $f_{\alpha_j\alpha_i}(q)=f_{\alpha_i\alpha_j}(q)$ (see (\ref{Deffcoefs})) along with a change of the summation variable $m\to -m$, the partition function $\mathcal{Z}_{N,1}$ in (\ref{N1PartitionFunction}) is invariant under $\mathfrak{c}^{\text{hor}}$ (up to a renaming of the partitions $\alpha_{1,\ldots,N}$).

\item {\bf Inversion of the mass parameter}:\\
The composition of symmetries can lead to simple transformations, for example
\begin{align}
&\mathfrak{f}\circ \mathfrak{r}:&& (\widehat{a}^{(1)}_1, \widehat{a}^{(1)}_2 \ldots, \widehat{a}^{(1)}_{N-1} , \rho , S ,\tau) \longrightarrow \left(\widehat{a}^{(1)}_1, \widehat{a}^{(1)}_2 \ldots, \widehat{a}^{(1)}_{N-1} , \rho , -S ,\tau\right)\,,\label{DefCombinedRS}
\end{align}
\emph{i.e.} it only flips the sign of the mass parameter.\footnote{This sign flip symmetry holds even in the refined case. In this case, moreover, there exists another symmetry $S \to \epsilon_1 + \epsilon_2 - S$, which corresponds to \eqref{QuiverOrientationSym}.} This is indeed a symmetry of the Nekrasov subfunctions, which can be manifestly seen from (\ref{FormNekSubRed}) or from (\ref{DefNekrasovSubfun}) and (\ref{FormNekSub}) (upon using $\phi_{-2,1}(-S;\rho)=\phi_{-2,1}(S,\rho)$). Furthermore, upon using properties of the $\vartheta_{\alpha\alpha}$ defined in (\ref{DefCurlyTheta}), one can show that the full partition function $\mathcal{Z}_{N,1}$ is invariant under the transformation (\ref{DefCombinedRS}) (see also \cite{Hohenegger:2015btj}).

\item {\bf Shift of the mass parameter:}\\
An example of a non-trivial symmetry that leaves the set of gauge parameters $(\widehat{a}_1^{(1)},\ldots,\widehat{a}_{N-1}^{(1)})$ invariant but also does not mix different instanton orders in the partition function is
\begin{align}
&\mathfrak{r}\circ\mathfrak{c}^{\text{hor}}\circ\mathfrak{c}^{\text{hor}}\circ\mathfrak{b}:&&  (\widehat{a}^{(1)}_1, \widehat{a}^{(1)}_2 \ldots, \widehat{a}^{(1)}_{N-1} , \rho , S ,\tau)\nonumber\\
& &&\hspace{1cm}\longrightarrow (\widehat{a}^{(1)}_1, \widehat{a}^{(1)}_2 \ldots, \widehat{a}^{(1)}_{N-1} , \rho , \rho-S ,\tau-2N S+N\rho)\,.\label{ShiftTrans}
\end{align}
which is in fact the transformation that was described in \cite{Bastian:2018jlf} by the matrix $\mathcal{G}_2(N)$ (see equation (\ref{IdentG2})). To see this explicitly, we use the fact that the Kronecker-Eisenstein series (for arbitrary arguments $x$ and $\kappa=\pm 1$) and $\phi_{-2,1}$ in (\ref{FormNekSub}) behave in the following manner under this transformation
\begin{align}
&\Omega(x,\kappa S;\rho)\to e^{2\pi i\kappa x}\,\Omega(x,\kappa S;\rho)\,,&&\phi_{-2,1}(S,\rho)\to \Qr^{-1}\Qs^2\,\phi_{-2,1}(S,\rho)\,.
\end{align}
Thus, the building blocks of the partition function (\ref{N1PartitionFunction}) transform in the following manner
\begin{align}
\prod_{k=1}^N \frac{\vartheta_{\alpha_k\alpha_k}(Q_S;\rho)}{\vartheta_{\alpha_k\alpha_k}(1;\rho)}&\longrightarrow \Qr^{-\sum_{k=1}^N |\alpha_k|}\,Q_S^{2\sum_{k=1}^N|\alpha_k|}\,\prod_{k=1}^N \frac{\vartheta_{\alpha_k\alpha_k}(Q_S;\rho)}{\vartheta_{\alpha_k\alpha_k}(1;\rho)} \\
\mathcal{T}_{\alpha_j\alpha_i}&\longrightarrow \Qr^{-|\alpha_i|-|\alpha_j|}\Qs^{2|\alpha_i|+2|\alpha_j|}\,\mathcal{T}_{\alpha_2\alpha_1}\,.
\end{align}
Together with the transformation of $\Qt\to \Qt\,\Qr^N\,\Qs^{-2N}$, this implies that the partition function (\ref{N1PartitionFunction}) is invariant under (\ref{ShiftTrans}) order by order in the instanton expansion.

\end{itemize}
We remark that these have only been examples of 'simple' symmetries, which leave the partition function invariant order by order in the instanton expansion. More general symmetries (which mix $\tau$ with the remaining parameters in a non-trivial fashion) will act more trivially, but are nevertheless symmetries of $\mathcal{Z}_{N,1}$, as discussed above.



\subsection{Action on Quiver Algebra}

In the general case $M>1$, the symmetries cannot be systematically understood as concrete transformations of the building blocks of the partition function $\mathcal{Z}_{N,M}$ (in particular due to contributions in $\Qr^{1/M}$ to the $\vartheta_{\alpha \beta}$ functions entering the partition function~\eqref{Pexternalpartition}). Since these transformations are self-similarities of the web diagram, it is expected that they relate Fourier coefficients of the free energy similarly to~\eqref{fouriercoeftrans}, and should therefore leave the partition function invariant. Furthermore, there exists a well defined action of these symmetries on the affine quiver algebra and the quiver moduli, which we exhibit here explicitly for a few interesting examples:

\begin{itemize}
\item {\bf Cyclic permutations $\mathfrak c^{\text{hor}},\mathfrak c^{\text{ver}}$ and rotations $\mathfrak r$:}\\
A first simple example is given by all symmetries relating points in the same Kähler cone. This set of symmetries leave the mass-deformed Cartan matrix and by extension the quiver algebra invariant but corresponds to re-organisations of the sets of Chern characters of a given fixed point~\eqref{eq:genMfixpoints}:
\begin{align}\label{eq:cveractionquiver}
    &\mathfrak c^{\rm hor}:\,\mathcal{X}_{\alpha_i^{(j)}}=\{\nu_{i}^{(j)} q^{-\alpha_{i,k}^{(j)}+k -1} \}_{k \in \{1,\ldots,\infty\}} \overset{\frak c^{\rm hor}}{\longrightarrow} \{\nu_{i+1}^{(j)} q^{-\alpha_{i,k}^{(j)}+k -1} \}_{k \in \{1,\ldots,\infty\}}\, , \nonumber \\
    &\mathfrak c^{\rm ver}:\,\mathcal{X}_{\alpha_i^{(j)}}=\{\nu_{i}^{(j)} q^{-\alpha_{i,k}^{(j)}+k -1} \}_{k \in \{1,\ldots,\infty\}} \overset{\frak c^{\rm ver}}{\longrightarrow} \{\nu_{i}^{(j+1)} q^{-\alpha_{i,k}^{(j)}+k -1} \}_{k \in \{1,\ldots,\infty\}}\, ,\nonumber \\
    &\mathfrak r: \, \mathcal{X}_{\alpha_i^{(j)}}=\{\nu_{i}^{(j)} q^{-\alpha_{i,k}^{(j)}+k -1} \}_{k \in \{1,\ldots,\infty\}} \overset{\frak r}{\longrightarrow} \{\nu_{N-i}^{(M-j+2)} q^{-\alpha_{i,k}^{(j)}+k -1} \}_{k \in \{1,\ldots,\infty\}}\,,
\end{align}
where the lower indices are understood $\rm mod$ $N$ with value in $\{1,\ldots,N\}$ and the upper indices $\rm mod$ $M$ with value in $\{1,\ldots,M\}$. $\mathfrak c^{\text{hor}}$ is interesting in this context: although it is a symmetry of the partition function, this is not immediately manifest. Indeed, (contrary to $\mathfrak c^{\text{ver}}$ and $\mathfrak r$) it is not a one to one map over the set of arguments of the $\vartheta$-functions in (\ref{Pexternalpartition}): $ \{\nu_i^{(s)}/\nu_j^{(s+1)},\, \nu_i^{(s)}/\nu_{j}^{(s)}| \, 1 \leq i <j \leq N,\, 1\leq s \leq M\}=: {\bf N}$. Explicitly, we have $\nu_{N-1}^{(s)}/\nu_{N}^{(s)}\xrightarrow[]{\mathfrak c^{\text{hor}}}\nu_{N}^{(s)}/\nu_{1}^{(s)} \notin {\bf N}$, which is manifest in the $(N,1)$ case (see~\eqref{mapChorN1moduli}). As explained in Subsection~\ref{Sect:GenSymTrans}, this is a consequence of the choice of parametrisation for the web diagram and $\mathfrak c^{\text{hor}}$ corresponds to a map between the possible choices of parametrisation. 

\item {\bf Inversion of the mass parameter:}\\
Another simple example is given by $\mathfrak f$ that acts on $S$ as $S \overset{\frak f}{\longrightarrow} -S$, which corresponds an invariance of the mass-deformed Cartan matrix for quiver gauge theories with bifundamental matter \eqref{QuiverOrientationSym}. Hence, it becomes $S \to \epsilon_1 + \epsilon_2 - S$ in the refined case. This is tied with the following re-organisation of the sets $\mathcal{X}_{\alpha_i^{(j)}}$:
\begin{equation}
    \mathfrak f: \, \mathcal{X}_{\alpha_i^{(j)}}=\{\nu_{i}^{(j)} q^{-\alpha_{i,k}^{(j)}+k -1} \}_{k \in \{1,\ldots,\infty\}} \overset{\frak f}{\longrightarrow} \{\nu_{N-i}^{(j)} q^{-\alpha_{i,k}^{(j)}+k -1} \}_{k \in \{1,\ldots,\infty\}}\,,
\end{equation}
which is a one to one map over the set of parameters in the partition function~\eqref{DefPartitionFunction}.

\item {\bf Shift of the mass parameter in $\rho$:}\\
Non-trivial examples of symmetries acting of the elliptic quiver algebra are given by $\mathfrak b$ type symmetries that act on $S$ like $S \overset{\frak b}{\longrightarrow} \rho - S$. In order to understand the action of this class of symmetries on the affine quiver algebra we examine three examples:

\begin{itemize}
    \item First, let us re-examine the transformation~\eqref{ShiftTrans} corresponding to the action of $\mathfrak r \circ \mathfrak c^{\rm hor}\circ \mathfrak c^{\rm hor} \circ\mathfrak b$ over the moduli space of $(N,1)$ web diagrams. The mass deformed Cartan matrix for the $\widehat A_{0}$ quiver gauge theory~\eqref{eq:genMCartan} transforms simply under $\mathfrak r \circ \mathfrak c^{\rm hor}\circ \mathfrak c^{\rm hor} \circ\mathfrak b$:
\begin{equation}
    c^{[m]} \overset{}{\longrightarrow} 2 - (Q_{S}^{-m}\Qr^{m}+Q_{S}^{m}\Qr^{-m}).
\end{equation}
Using~\eqref{CommutationRelations}, one obtains the transformation rule for the commutation relations under $\mathfrak r \circ \mathfrak c^{\rm hor}\circ \mathfrak c^{\rm hor} \circ\mathfrak b$:
\begin{equation}
    \left[s_{m}^{(\pm)},s_{m^\prime}^{(\pm)}\right] \overset{}{\longrightarrow} \left[s_{m}^{(\pm)},s_{m^\prime}^{(\pm)}\right] \pm \frac{(\Qr^{\mp m} Q_{S}^{\pm m} - Q_{S}^{\mp m})}{m}\delta_{m+m^\prime,0}.
\end{equation}
This modification of the free field algebra is then absorbed by the full coupling $\tau$:
\begin{equation}
    \tau \longrightarrow \tau - 2 N S + N\rho.
\end{equation}
\item We next consider the action of $\mathfrak b$ over the basis of K\"ahler parameters in the $(N,M)$ case. In this case, the mass deformed Cartan matrix for $\widehat{A}_{M-1}$ quiver gauge theories transforms as:
\begin{equation}
    c_{ij}^{[n]} \overset{\frak b}{\longrightarrow} 2 \delta_{ij} - \left(Q_{\widehat S}^n \Qr^{-n/M}\delta_{i,j+1} +Q_{\widehat S}^{-n} \Qr^{n/M}\delta_{i+1,j}\right),\quad \forall n \in \mathbb{Z}\,.
\end{equation}
This transformation cannot be formulated as a simple deformation of the free field algebra, it is however still compensated by a non-trivial transformation of the coupling constants $(\tau_1,\ldots,\tau_{M-1},\tau)$ and K\"ahler parameters $\widehat a_i^{(j)}$ (see~\eqref{eq:Bmattrans}). For general $M>1$, $\mathfrak b$ type symmetries are interesting examples of symmetries that are natural from the geometrical perspective but do not correspond to simple transformations of the quiver algebra. From the geometric perspective, identities over Kronecker-Eisenstein series with such a shift in $\Qr^{1/M}$ have been found in~\cite{Mikosz2021}.
\item In addition, for $M>1$, interesting examples arise from combinations of $\mathfrak f$ and $\mathfrak b$. For instance, one can consider $\mathfrak{b}^{(M)}:=((\mathfrak{c}^{\rm hor})^{-M-1} \circ \mathfrak{f} \circ \mathfrak{b})^M$. In this case, the bifundamental mass parameters undergoes $S \to S + M \rho$ and the mass deformed Cartan matrix changes as:
\begin{equation}
    c_{ij}^{[n]} \overset{\mathfrak{b}^{(M)}}{\longrightarrow} 2 \delta_{ij} - \left(Q_{\widehat S}^{n} \Qr^{n} \delta_{i+1,j} + Q_{\widehat S}^{-n} \Qr^{-n} \delta_{i,j+1} \right), \quad \forall n \in \mathbb{Z}\,.
\end{equation}
Similar to the $\widehat A_0$ case, we can derive the transformation rule for the algebra:
\begin{equation}
    [s_{i,m}^{(\pm)},s_{j,m^\prime}^{(\pm)}] \overset{\mathfrak{b}^{(M)}}{\longrightarrow} [s_{i,m}^{(\pm)},s_{j,m^\prime}^{(\pm)}] \pm \frac{\Qr^{\pm m} Q_{\widehat S}^{\pm m} \delta_{i,j+1} - Q_{\widehat S}^{\pm m} \delta_{i+1,j}}{m} \delta_{m+m^\prime,0}.
\end{equation}
Once again extra contributions are absorbed in the $M$ gauge couplings $(\tau_1, \ldots, \tau_{M-1},\tau)$:
\begin{equation}
\begin{split}
    &\tau \overset{\mathfrak{b}^{(M)}}{\longrightarrow} \tau - 2NS + NM \rho,\\
    &\tau_j \overset{\mathfrak{b}^{(M)}}{\longrightarrow} \tau_j - 2\frac{N}{M} S + c_{j,M,N}\, \rho, \quad \forall j \in \{1,\ldots,M-1\},
\end{split}
\end{equation}
where $c_{j,M,N}$ is given by the following table:
\begin{center}
\begin{tabular}{c||c|c}
        $c_{j,M,N}$ & $M$ even & $M$ odd \\
        \hline
        $j$ even & $M + 2(N/M -1)(M-2j)$ & $M + 2(N/M -1)((N-j)+3)$ \\
        $j$ odd & $M$ & $M+ (N/M -1)(N-2(j-1))$ 
\end{tabular}
\end{center}
These asymmetric coefficients for $(\tau_1,\ldots,\tau_{M-1})$ are a consequence of the choice of basis for the couplings. The $\widehat a_i^{(j)}$ K\"ahler parameters are left unchanged under $\mathfrak b^{(M)}$.
\end{itemize}

\item {\bf Shift of the mass parameter in $\tau$:}\\
We finish this list of examples with $\frak a$ type symmetries that act on $S$ like $S \overset{\frak a}{\longrightarrow} \tau - S$. This symmetry does not respect the expansion in $(\tau_1,\ldots,\tau_{M-1},\tau)$ as can be seen from the general form of the $\mathcal{A}$ matrix~\eqref{eq:Amattrans} (due to the non vanishing $\mathcal{A}_{\widehat a \tau}$ component). However, if we consider the $(\tau,S,\rho)$ triplet, this transformation is symmetric to $\frak b$ upon exchange of $\tau$ and $\rho$. In the doubly elliptic situation considered here, the partition function ${\cal Z}_{N,M}$ exhibits a triality structure~\cite{Bastian:2017ary} (see Subsection~\ref{subsec:partitionfunction}). Consequently, the partition function can equivalently be expanded in powers of $\Qr$ with $\tau$ playing the role of the elliptic parameter entering the definition of the screening currents, this corresponds to an expansion around the subspace where $\int_{\bf h} \omega \to \infty$. In this alternative expansion, ${\frak a}$ type transformations can be understood as a modification of the free field algebra compensated by a shift of $\Qr$.

\end{itemize}



\section{Conclusions and Outlook}\label{Sect:Conclusions}
In this paper we have studied non-perturbative symmetries of a class of Little String orbifolds. These theories can be engineered by $N$ parallel M5-branes spread out along a circle that probe a transverse $\mathbb{Z}_M$ orbifold background. The moduli space of these theories contains various different regions that describe dual supersymmetric gauge theories on $\mathbb{R}^4\times T^2$ with different gauge structure and matter content \cite{Bastian:2018dfu,Bastian:2017ary}. In this work we have focused on a circular quiver gauge theory with $M$ nodes of the gauge group $U(N)$ and hypermultiplet matter in the bifundamental representation (or adjoint in the case of $M=1$). A sketch of the quiver is shown in Figure~\ref{Fig:Quiver}.  

LSTs of this type permit numerous dual descriptions \cite{Haghighat:2013gba,Haghighat:2013tka,Hohenegger:2013ala,Hohenegger:2015cba,Kimura:2015rgi,Kimura:2016dys,Kimura:2017hez} which can be used to obtain explicit expressions for the full instanton BPS-partition function $\mathcal{Z}_{N,M}$ (see eq.~(\ref{DefPartitionFunction})). One such description is F-theory compactified on a class of toric Calabi-Yau threefolds $X_{N,M}$, which allows to compute $\mathcal{Z}_{N,M}$ as the topological string partition function (for which efficient building blocks were developed in \cite{Bastian:2017ing}). Dualities of $X_{N,M}$ (in the sense of \cite{Hohenegger:2016yuv}) give rise to symmetries of the LST: in this paper we have established for general $X_{N,M}$ explicitly the duality maps for various classes of such symmetries in the form of linear transformations on the K\"ahler parameters. Although linear, concrete realisations of these symmetry transformations in the form of matrices can become quite complicated. We have therefore provided the Mathematica package {\tt NPLSTsym} that allows to compute the latter explicitly for any integer numbers $N,M$ (see Appendix~\ref{App:MathematicaPackage} for the documentation). From the perspective of the quiver gauge theory mentioned previously, these symmetries generally mix coupling constants with the remaining mass- and Coulomb-branch parameters and thus act inherently non-perturbatively. We have furthermore shown that these symmetries can be generated by a set of involutions for which we have worked out the Coxeter matrix. We have in particular identified two such involutions that form a dihedral group and generalise a similar structure found in \cite{Bastian:2018jlf} for $M=1$. Furthermore, we have characterised these symmetries through their action on the particular subspace (\ref{SubsetSpaceK3}) of the (extended) K\"ahler moduli space of $X_{N,M}$ and have shown that they are characterised through elements of a specific $SL(2,\mathbb{Z}) \subset Sp(4,\mathbb{Z})$. The symmetry transformations found in this paper certainly do not constitute all (non-perturbative) symmetries of the LSTs: indeed, due to the structure of $X_{N,M}$ as a double elliptic fibration \cite{Kanazawa:2016tnt,Hohenegger:2016yuv,Hohenegger:2015btj} there are modular symmetries acting on $\tau$ and $\rho$ (see also \cite{Haghighat:2018gqf}). While such symmetries are not captured in our approach (indeed, per construction all transformations act linearly on the K\"ahler parameters), the set of symmetries discussed here has the benefit of being very concretely realised. This has already been exploited in the case of $M=1$ to establish further properties and symmetries of the LSTs (see \emph{e.g.} \cite{Bastian:2019hpx,Bastian:2019wpx,Hohenegger:2019tii,Hohenegger:2020slq,Hohenegger:2020gio}). We therefore foresee further applications of our results in exploring algebraic properties of the LSTs

As one concrete such application, we have here considered the implications of our results on other descriptions of the LSTs. Indeed, the non-perturbative symmetry transformations are valid also in other dual formulations of the LST, even though they are no longer manifest. As an example, we have studied in this paper a description of the quiver gauge theory in terms of a vertex operator algebra \cite{Kimura:2015rgi,Kimura:2016dys,Kimura:2017hez}. In this formalism, the fundamental building blocks of the partition function $\mathcal{Z}_{N,M}$ can be written as correlation functions of so-called screening currents, which encode algebraic properties of the underlying quiver. The commutator algebra of these currents is governed by (a deformed version of the) Cartan matrix of the affine quiver algebra (which in the current case is simply the affine Lie algebra $\widehat{\mathfrak{a}}_{M-1}$. Since the deformation is due to the mass parameters of the gauge theory (which are identified with some of the K\"ahler parameters of $X_{N,M}$), the symmetry transformations derived previously act in a non-trivial fashion on this Cartan matrix, thus acting non-trivially on the quiver algebra. While some symmetry generators also leave the correlation functions invariant (or transform them in a simple fashion), general transformations act in a much more complicated manner. The duality therefore provides us with a novel type of automorphism on the affine quiver algebra, that would be difficult to obtain from other means. 

In the future we foresee further applications and generalisations of the ideas and results presented in this work: the LSTs discussed in this work constitute a particular class, for which computations (for example of the partition function) can be performed very concretely. More general types of LSTs \cite{Heckman:2013pva,Heckman:2015bfa,Bhardwaj:2015xxa} (see \cite{Bhardwaj:2015oru} for a classification from F-theory) exists. While a description in terms of brane webs might not be available, the quiver algebra formalism can be extended in a more general fashion. In the future it will therefore be interesting to analyse possible automorphisms of the latter and their implications for gauge theory. Furthermore, the mass parameters (which act as the deformations of the Cartan matrix of the quiver algebra) have recently been interpreted as deformations in higher-dimensional generalisations of the Nekrasov $\Omega$-brackground \cite{Nekrasov:2009JJM,Nekrasov:2015wsu,Nekrasov:2017cih}. It would be interesting to understand how the non-perturbative symmetries found in this work can be understood from this perspective.

\section*{Acknowledgements}
BF and SH would like to thank Oliver Schlotterer for invaluable discussions on Kronecker-Eisenstein series and related modular objects. The work of TK was in part supported by EIPHI Graduate School (No.~ANR-17-EURE-0002) and Bourgogne-Franche-Comté region.

\appendix
\section{Notations and definitions}
This appendix is dedicated to the different notations and definitions used to express the partition function $\mathcal{Z}_{N,M}$ that appears in many places in the main body of this work. 

We first start by discussing modular functions that are relevant in this work: we use two reprensentions for the Jacobi theta function $\theta_1$ (with modular parameter $\rho$ such that $|\Qr| < 1$):
\begin{equation}
    \theta_1(\A;\rho) = -i \Qr^{1/8} \Qa{}^{1/2} \prod_{n=1}^\infty (1-\Qr^n) (1-\Qa{} \Qr^n)(1-\Qa{}^{-1} \Qr^{n-1})
\end{equation}
and its exponential form:
\begin{equation}\label{eq:thetafunction}
    \theta_1(\A;\rho) = i \Qr^{1/12} \eta(\Qr) \Qa{}^{-1/2} \exp \left(-\sum_{n\in\mathbb Z^*} \frac{\Qa{}^n}{n(1-\Qr^n)} \right),
\end{equation}
where $\eta$ is the Dedekind eta function:
\begin{equation}
    \eta(\rho) = \Qr^{1/24} \prod_{n=1}^\infty (1-\Qr^n).
\end{equation}
The Jacobi theta function $\theta_1$ also verifies the following shift symmetry
\begin{align}
\theta_1(\pm S+x\mp\rho,\rho)=\Qr^{-1/2}\,\Qs\,e^{\pm 2\pi i x}\,\theta_1(\pm S+x,\rho)\,.
\end{align}
As argued in \cite{Filoche:2022qxk}, the partition function $\mathcal{Z}_{N,1}$ (\emph{i.e.} for $M=1$) can be written using modular building blocks, in particular the Kronecker-Eisenstein series:
\begin{align}
&\Omega(u,v;\rho) = \text{exp}\left( 2\pi i v\frac{{\rm Im}(u)}{{\rm Im}(\rho)}\right) \frac{\theta_1(u+v;\rho) \theta_1^{\prime}(0;\rho)}{\theta_1(u;\rho)\theta_1(v;\rho)}\,,&&\forall u,v\in\mathbb{C}\,,\label{DefOmega}
\end{align}
and the standard weight $-2$, index $1$ Jacobi form:
\begin{equation}
    \phi_{-2,1}(u,\rho) = \frac{\theta_1^2(u;\rho)}{\eta(\rho)^6}, \quad\quad\quad \forall u \in \mathbb C\,.\label{DefPhi21}
\end{equation}
To write $\mathcal{Z}_{N,1}$ in terms of the exponential building blocks (\ref{FormNekSubRed}) (labelled by integer partitions $\alpha_{1,2}$), a set of combinatorial functions appear, which are defined as: 
\begin{equation}\label{eq:fCSalphalpha}
    f_{\alpha_1,\alpha_2}(q)= \sum_{(r,s)\in \alpha_1} q^{n_{r,s}^{\alpha_1,\alpha_2}} + \sum_{(r,s)\in \alpha_2} q^{-n_{r,s}^{\alpha_2,\alpha_1}},
\end{equation}
with the integers
\begin{equation}\label{eq:nalphalpha}
    n_{r,s}^{\alpha_1,\alpha_2} = \alpha_{2,s}^t + \alpha_{1,r} - r - s + 1, \quad \quad \forall (r,s) \in \alpha_1 \, .
\end{equation}
These can also be written alternatively in the context of Chern-Simon theory in \cite{Iqbal:2003ix}:
\begin{align}
&f_{\alpha_1\alpha_2}(q)=(q-2+q^{-1}) f_{\alpha_1}(q) f_{\alpha_2}(q)+f_{\alpha_1}(q)+f_{\alpha_2}(q)\,,&&\text{with} &&f_{\beta}(q)=\sum_{j=1}^{\ell(\beta)}\sum_{i=1}^{\beta_j}q^{i-j}\,.\label{Deffcoefs}
\end{align}
These functions in particular satisfy
\begin{align}
f_{\alpha_1^t\alpha_2^t}(q^{-1})=f_{\alpha_1\alpha_2}(q)\,.\label{RelTransposeF}
\end{align}

In Subsection~\ref{subsec:partitionfunction}, we review the topological vertex computation for the $(N,M)$ partition function. The basic building blocks (\ref{WbuildingBlock}) are written in terms of the following $\mathcal J_{\mu\nu}$ function, where $\mu,\nu$ are integer partitions and $\Qr<1$:
\begin{align}\label{Jfunction}
    &\mathcal{J}_{\mu \nu}(x;q,t) = \prod_{k=1}^\infty J_{\mu \nu}(\Qr^{k-1} x;q,t)\, , \nonumber\\ 
    &J_{\mu\nu}(x;q,t) = \prod_{(i,j)\in \mu} \left(1-x q^{\nu_j^t-i+\frac{1}{2}} t^{\mu_i -j + \frac{1}{2}} \right) \prod_{(i,j)\in \nu} \left(1-x q^{-\mu_j^t + i - \frac{1}{2}}t^{-\nu_i +j -\frac{1}{2}} \right).
\end{align}
The full partition function $\mathcal{Z}_{N,M}$ can systematically be reformulated using $\vartheta_{\mu\nu}$ functions which are defined as: 
\begin{align}
\vartheta_{\mu\nu}(x;\rho)&=\prod_{(i,j)\in\mu}\vartheta\left(x^{-1}q^{-\nu_j^t+i-\frac{1}{2}}t^{-\mu_i+j-\frac{1}{2}};\rho\right)\prod_{(i,j)\in\nu}\vartheta\left(x^{-1}q^{\mu_j^t-i+\frac{1}{2}}t^{\nu_i-j+\frac{1}{2}};\rho\right)\,,\nonumber\\
\vartheta(x;\rho)&=\left(x^{1/2}-x^{-1/2}\right)\prod_{k=1}^\infty(1-x \Qr^k)(1-x^{-1}\Qr^k)=\frac{i\Qr^{-1/8}\theta_1(z;\rho)}{\prod_{k=1}^\infty(1-\Qr^k)}\,,\label{DefCurlyTheta}
\end{align}
where $\mu,\nu$ are integer partitions and $\Qr<1$.

\section{{\tt NPLSTsym} Mathematica Package}\label{App:MathematicaPackage}
In Section~\ref{Sect:GenSymTrans} we have established symmetry transformations that act linearly on the K\"ahler parameters of $X_{N,M}$. In this Appendix we describe the Mathematica package {\tt NPLSTsym}\footnote{The package has been written using Mathematica 13 and has been tested with Mathematica 12 and 13.} that allows to compute the associated matrices. It also allows to study their group theoretical relations by computing automatically the Coxeter matrix of a given set of involutions. 
\paragraph{Package import}
In order to use the {\tt NPLSTsym} Mathematica package, the {\tt NPLSTsym.wl} file must be downloaded from the supplementary material to the arXiv preprint. The functions {\tt Cvermat, Chormat, Rmat, Amat, Bmat, Fmat, SymOrder} and {\tt Coxetermat} can be imported using the following command:
\begin{mmaCell}{Input}
<<\mmaDef{PATH/NPLSTsym.wl}
\end{mmaCell}
where PATH must be replaced by the location of the {\tt NPLSTsym.wl} file.
\paragraph{Symmetry matrices calculation} The functions {\tt Cvermat, Chormat, Rmat, Amat, Bmat} and {\tt Fmat} are dedicated respectively to the computation of the $\mathcal{C}^{\rm ver}, \mathcal{C}^{\rm hor}, \mathcal{R}, \mathcal{A}, \mathcal{B}$ and $\mathcal{F}$ matrices described in Subsection~\ref{Sect:GenSymTrans}. The matrix is given in the $(\vec{a}^{(1)},\ldots,\vec{a}^{(M)},\tau_1,\ldots,\tau_{M-1},\tau,S,\rho)$ basis defined~\eqref{VectorKaehlerModuli}. 
\paragraph{}- {\tt Cvermat} takes two arguments: $N\in\mathbb N$ and $M\in\mathbb N$ and returns the $\mathcal{C}^{\rm ver}$ matrix. Example for $N=4, M=2$:
\begin{mmaCell}{Input}
\mmaDef{Cvermat}[4,2] // MatrixForm
\end{mmaCell}
\begin{mmaCell}[verbatimenv=]{Output}
\(\begin{pmatrix}
{\tt 0} & \tt 0 & \tt 0 & \tt 1 & \tt0 & \tt0 &\tt 0 & \tt0 &\tt 0 & \tt0 \\
\tt0 & \tt0 & \tt0 & \tt0 & \tt1 & \tt0 & \tt0 & \tt0 & \tt0 & \tt0 \\
\tt0 &\tt 0 &\tt 0 &\tt 0 &\tt 0 &\tt 1 &\tt 0 &\tt 0 & \tt0 & \tt0 \\
\tt1 & \tt0 & \tt0 & \tt0 &\tt 0 &\tt 0 &\tt 0 &\tt 0 &\tt 0 &\tt 0 \\
\tt0 & \tt1 &\tt 0 &\tt 0 &\tt 0 &\tt 0 &\tt 0 &\tt 0 &\tt 0 &\tt 0 \\
\tt0 &\tt 0 &\tt 1 &\tt 0 &\tt 0 &\tt 0 &\tt 0 &\tt 0 &\tt 0 &\tt 0 \\
\tt0 &\tt 0 &\tt 0 &\tt 0 &\tt 0 &\tt 0 &\tt -1 &\tt 1 &\tt 0 &\tt 0 \\
\tt0 &\tt 0 &\tt 0 &\tt 0 &\tt 0 &\tt 0 &\tt 0 &\tt 1 &\tt 0 &\tt 0 \\
\tt0 & \tt0 & \tt0 & \tt0 &\tt 0 &\tt 0 &\tt 0 &\tt 0 &\tt 1 &\tt 0 \\
\tt0 & \tt0 & \tt0 & \tt0 & \tt0 & \tt0 & \tt0 & \tt0 & \tt0 &\tt 1 \\
\end{pmatrix}\)
\end{mmaCell}
\paragraph{} - {\tt Chormat} takes two arguments: $N\in\mathbb N$ and $M\in\mathbb N$ and returns the $\mathcal{C}^{\rm hor}$ matrix. Example for $N=4, M=2$:
\begin{mmaCell}{Input}
\mmaDef{Chormat}[4,2] // MatrixForm
\end{mmaCell}
\begin{mmaCell}[verbatimenv=]{Output}
\(\begin{pmatrix}
\tt 0 &\tt 1 &\tt 0 &\tt 0 &\tt 0 &\tt 0 &\tt 0 &\tt 0 &\tt 0 &\tt 0 \\
 \tt0 & \tt0 & \tt1 & \tt0 &\tt 0 &\tt 0 &\tt 0 &\tt 0 &\tt 0 &\tt 0 \\
 \tt-1 & \tt-1 &\tt -1 &\tt 0 &\tt 0 &\tt 0 &\tt 0 &\tt 0 &\tt 0 &\tt 1 \\
\tt 0 &\tt 0 &\tt 0 &\tt 0 &\tt 1 &\tt 0 &\tt 0 &\tt 0 &\tt 0 &\tt 0 \\
 \tt0 & \tt0 & \tt0 & \tt0 &\tt 0 &\tt 1 &\tt 0 &\tt 0 &\tt 0 &\tt 0 \\
 \tt0 & \tt0 &\tt 0 &\tt -1 &\tt -1 &\tt -1 &\tt 0 &\tt 0 &\tt 0 &\tt 1 \\
 \tt-\frac{3}{2} &\tt -1 & \tt-\frac{1}{2} &\tt \frac{3}{2} &\tt 1 &\tt \frac{1}{2} &\tt 1 &\tt 0 &\tt 0 & \tt0 \\
 \tt0 & \tt0 & \tt0 & \tt0 &\tt 0 &\tt 0 &\tt 0 &\tt 1 &\tt 0 &\tt 0 \\
 \tt0 & \tt0 & \tt0 & \tt0 & \tt0 &\tt 0 &\tt 0 &\tt 0 &\tt 1 &\tt 0 \\
 \tt0 & \tt0 & \tt0 & \tt0 &\tt 0 &\tt 0 &\tt 0 &\tt 0 &\tt 0 &\tt 1 \\
\end{pmatrix}\)
\end{mmaCell}
\paragraph{} - {\tt Rmat} takes two arguments: $N\in\mathbb N$ and $M\in\mathbb N$ and returns the $\mathcal{R}$ matrix. Example for $N=4, M=2$:
\begin{mmaCell}{Input}
\mmaDef{Rmat}[4,2] // MatrixForm
\end{mmaCell}
\begin{mmaCell}[verbatimenv=]{Output}
\(\begin{pmatrix}
\tt0 &\tt 0 &\tt 1 & \tt0 &\tt 0 & \tt0 &\tt 0 &\tt 0 & \tt0 &\tt 0 \\
 \tt0 & \tt1 &\tt 0 &\tt 0 &\tt 0 &\tt 0 &\tt 0 &\tt 0 &\tt 0 &\tt 0 \\
\tt 1 &\tt 0 & \tt0 &\tt 0 &\tt 0 &\tt 0 &\tt 0 & \tt0 & \tt0 & \tt0 \\
 \tt0 &\tt 0 &\tt 0 &\tt 0 & \tt0 & \tt1 & \tt0 & \tt0 &\tt 0 &\tt 0 \\
\tt 0 &\tt 0 & \tt0 & \tt0 &\tt 1 &\tt 0 &\tt 0 &\tt 0 &\tt 0 &\tt 0 \\
\tt 0 & \tt0 & \tt0 & \tt1 & \tt0 & \tt0 &\tt 0 &\tt 0 &\tt 0 &\tt 0 \\
 \tt0 & \tt0 &\tt 0 & \tt0 &\tt 0 &\tt 0 &\tt 1 &\tt 0 &\tt 0 &\tt 0 \\
 \tt0 &\tt 0 & \tt0 & \tt0 &\tt 0 &\tt 0 &\tt 0 &\tt 1 &\tt 0 &\tt 0 \\
\tt 0 &\tt 0 &\tt 0 &\tt 0 &\tt 0 &\tt 0 &\tt 0 &\tt 0 &\tt 1 & \tt0 \\
 \tt0 & \tt0 & \tt0 & \tt0 &\tt 0 & \tt0 &\tt 0 & \tt0 & \tt0 &\tt 1 \\
\end{pmatrix}\)
\end{mmaCell}
\paragraph{} - {\tt Amat} takes two arguments: $N\in\mathbb N$, $M\in\mathbb N$ with $M|N$ and returns the $\mathcal{A}$ matrix. Example for $N=4, M=2$:
\begin{mmaCell}{Input}
\mmaDef{Amat}[4,2] // MatrixForm
\end{mmaCell}
\begin{mmaCell}[verbatimenv=]{Output}
\(\begin{pmatrix}
\tt-\frac{1}{2} &\tt -1 &\tt -\frac{1}{2} &\tt \frac{1}{2} &\tt 1 &\tt \frac{3}{2} &\tt 1 &\tt 0 & \tt-1 &\tt 0 \\
\tt 1 &\tt 2 &\tt 1 &\tt -1 &\tt -1 &\tt -1 &\tt -1 &\tt 1 &\tt -1 &\tt 0 \\
 \tt-\frac{1}{2} &\tt -1 &\tt -\frac{1}{2} & \tt\frac{3}{2} &\tt 1 & \tt\frac{1}{2} &\tt 1 & \tt0 & \tt-1 &\tt 0 \\
\tt \frac{1}{2} &\tt 1 & \tt\frac{3}{2} &\tt -\frac{1}{2} &\tt -1 & \tt-\frac{1}{2} & \tt-1 &\tt 1 & \tt-1 &\tt 0 \\
\tt -1 &\tt -1 &\tt -1 &\tt 1 &\tt 2 & \tt1 &\tt 1 & \tt0 &\tt -1 &\tt 0 \\
 \tt\frac{3}{2} &\tt 1 &\tt \frac{1}{2} &\tt -\frac{1}{2} &\tt -1 &\tt -\frac{1}{2} &\tt -1 &\tt 1 &\tt -1 &\tt 0 \\
\tt 0 & \tt0 &\tt 0& \tt0 & \tt0 &\tt 0 &\tt 1 &\tt 0 &\tt 0 &\tt 0 \\
 \tt0 & \tt0 & \tt0 & \tt0 &\tt 0 & \tt0 &\tt 0 &\tt 1 &\tt 0 &\tt 0 \\
 \tt0 &\tt 0 &\tt 0 &\tt 0 &\tt 0 &\tt 0 &\tt 0 &\tt 1 &\tt -1 &\tt 0 \\
 \tt0 &\tt 0 &\tt 0 &\tt 0 &\tt 0 &\tt 0 &\tt 0 &\tt 2 &\tt -4 &\tt 1 \\
\end{pmatrix}\)
\end{mmaCell}

\paragraph{} - {\tt Bmat} takes two arguments: $N\in\mathbb N$, $M\in\mathbb N$ with $M|N$ and returns the $\mathcal{B}$ matrix. Example for $N=4, M=2$:
\begin{mmaCell}{Input}
\mmaDef{Bmat}[4,2] // MatrixForm
\end{mmaCell}
\begin{mmaCell}[verbatimenv=]{Output}
\(\begin{pmatrix}
 \tt0 &\tt 1 &\tt 0 &\tt 0 &\tt 0 &\tt 0 &\tt 0 &\tt 0 & \tt0 &\tt 0 \\
\tt 1 &\tt 0 & \tt0 &\tt 0 & \tt0 &\tt 0 &\tt 0 &\tt 0 &\tt 0 &\tt 0 \\
\tt -1 &\tt -1 &\tt -1 &\tt 0 &\tt 0 &\tt 0 &\tt 0 &\tt 0 &\tt 0 &\tt 1 \\
 \tt0 &\tt 0 &\tt 0 &\tt -1 &\tt -1 &\tt -1 & \tt0 & \tt0 & \tt0 &\tt 1 \\
 \tt0 &\tt 0 & \tt0 & \tt0 &\tt 0 &\tt 1 &\tt 0 &\tt 0 &\tt 0 &\tt 0 \\
\tt 0 & \tt0 & \tt0 &\tt 0 & \tt1 &\tt 0 &\tt 0 &\tt 0 &\tt 0 &\tt 0 \\
\tt -\frac{1}{2} &\tt -1 &\tt -\frac{3}{2} &\tt \frac{3}{2} &\tt 1 & \tt\frac{1}{2} &\tt 1 & \tt0 &\tt -1 &\tt \frac{1}{2} \\
\tt 0 &\tt 0 &\tt 0 &\tt 0 & \tt0 &\tt 0 &\tt 0 &\tt 1 &\tt -4 &\tt 2 \\
\tt 0 &\tt 0 &\tt 0 &\tt 0 &\tt 0 & \tt0 &\tt 0 & \tt0 & \tt-1 &\tt 1 \\
\tt 0 &\tt 0 &\tt 0 &\tt 0 &\tt 0 &\tt 0 &\tt 0 &\tt 0 &\tt 0 &\tt 1 \\
\end{pmatrix}\)
\end{mmaCell}
\paragraph{} - {\tt Fmat} takes two arguments: $N$ and $M$ and returns the $\mathcal{F}$ matrix. Example for $N=4, M=2$:
\begin{mmaCell}{Input}
\mmaDef{Fmat}[4,2] // MatrixForm
\end{mmaCell}
\begin{mmaCell}[verbatimenv=]{Output}
\(\begin{pmatrix}
\tt0 & \tt0 & \tt1 & \tt0 & \tt0 & \tt0 & \tt0 & \tt0 &\tt 0 & \tt0 \\
\tt 0 &\tt 1 &\tt 0 & \tt0 & \tt0 & \tt0 & \tt0 &\tt 0 &\tt 0 &\tt 0 \\
 \tt1 & \tt0 & \tt0 & \tt0 & \tt0 & \tt0 & \tt0 & \tt0 &\tt 0 &\tt 0 \\
 \tt0 & \tt0 & \tt0 & \tt0 &\tt 0 &\tt 1 &\tt 0 &\tt 0 &\tt 0 &\tt 0 \\
 \tt0 &\tt 0 & \tt0 & \tt0 &\tt 1 & \tt0 &\tt 0 & \tt0 & \tt0 & \tt0 \\
 \tt0 & \tt0 & \tt0 & \tt1 & \tt0 & \tt0 & \tt0 & \tt0 &\tt 0 & \tt0 \\
 \tt0 & \tt0 & \tt0 & \tt0 & \tt0 & \tt0 & \tt1 & \tt0 & \tt0 & \tt0 \\
 \tt0 & \tt0 &\tt 0 & \tt0 &\tt 0 &\tt 0 & \tt0 & \tt1 &\tt 0 &\tt 0 \\
 \tt0 &\tt 0 &\tt 0 &\tt 0 &\tt 0 &\tt 0 &\tt 0 &\tt 0 &\tt -1 &\tt 0 \\
 \tt0 & \tt0 &\tt 0 &\tt 0 &\tt 0 &\tt 0 &\tt 0 &\tt 0 &\tt 0 &\tt 1 \\
\end{pmatrix}\)
\end{mmaCell}

\paragraph{Symmetry order function} The {\tt SymOrder} function takes two arguments: the matrix associated with a symmetry $\mathcal M$ and the depth which corresponds to an integer $n_{\rm max}$. {\tt SymOrder} returns the smallest integer $k$ such that $\mathcal M^k = \mathds 1$. If $\forall n \leq n_{\rm max}$, $\mathcal M^n \neq \mathds 1$, {\tt SymOrder} returns $\infty$. Example:

\begin{mmaCell}{Input}
\mmaDef{SymOrder}[\mmaDef{Chormat}[4,2],10]
\end{mmaCell}
\begin{mmaCell}{Output}
4
\end{mmaCell}

\paragraph{Automatic Coxeter matrix calculation} The {\tt Coxetermat} function takes two arguments: a set of involutions under the form of a Mathematica list of matrices and the depth which corresponds to an integer $n_{\rm max}$. If $a,b$ are two involutions such that $(ab)^n \neq \mathbbm{1}$, $\forall n \leq n_{\rm max}$ then the braiding coefficient will be set to $\infty$. Example:
\begin{mmaCell}{Code}
\mmaDef{Coxetermat}[{\mmaDef{Amat}[4,2],\mmaDef{Bmat}[4,2],\mmaDef{Fmat}[4,2]},10]
\end{mmaCell}
\begin{mmaCell}{Output}
\{\{1,4,∞\},\{4,1,∞\},\{∞,∞,1\}\}
\end{mmaCell}

\begingroup
\sloppy
\printbibliography
\endgroup

\end{document}